\documentclass[10pt, twocolumn,aps,pra,superscriptaddress]{revtex4-2}

\usepackage{xcolor,graphicx}%,tikz}

\usepackage[colorlinks=true]{hyperref}
\usepackage{soul}
\usepackage{colortbl}
\usepackage[normalem]{ulem}

\usepackage{dcolumn}% Align table columns on decimal point
\usepackage{bm}% bold math
\usepackage{comment}
\usepackage{nccmath,amsmath,lipsum}
\usepackage{svg}

\usepackage{physics}
\usepackage{qcircuit}
\usepackage[mathlines]{lineno}

\newcommand{\cnn}{Universit\'e Paris-Saclay, Centre de Nanosciences et de Nanotechnologies, CNRS, 10 Boulevard Thomas Gobert, 91120, Palaiseau, France}

\newcommand{\UD}{University of Delaware, Newark, DE 19716}

\newcommand{\quandela}{Quandela SAS, 7 Rue L\'eonard de Vinci, 91300 Massy, France}

\newcommand{\UTS}{School of Mathematical and Physical Sciences, University of Technology Sydney, Ultimo, New South Wales
2007, Australia}

\newcommand{\ParisCite}{Universit\'e Paris Cit\'e, Centre de Nanosciences et de Nanotechnologies, CNRS, 10 Boulevard Thomas Gobert, 91120, Palaiseau, France}

\begin{document}

\title{Deterministic and reconfigurable graph state generation with a single solid-state quantum emitter}% Force line breaks with \\

\author{H.~Huet}
%\email{helio.huet@universite-paris-saclay.fr}
\affiliation{\cnn}

\author{P.~R.~Ramesh}
\affiliation{\cnn}
\affiliation{\UD}

\author{S.~C.~Wein}
\affiliation{\quandela}

\author{N.~Coste}
\affiliation{\cnn}
\affiliation{\UTS}

\author{P.~Hilaire}
\author{N.~Somaschi}
\affiliation{\quandela}

\author{M.~Morassi}
\author{A.~Lemaître}
\author{I.~Sagnes}
\affiliation{\cnn}

\author{M.~F.~Doty}
\affiliation{\UD}

\author{O.~Krebs}
\affiliation{\cnn}

\author{L.~Lanco}
\affiliation{\cnn}
\affiliation{\ParisCite}

\author{D.~A.~Fioretto}
\affiliation{\cnn}
\affiliation{\quandela}

\author{P.~Senellart}
\affiliation{\cnn}

\begin{abstract}

Measurement-based quantum computing offers a promising route towards scalable, universal photonic quantum computation. This approach relies on the deterministic and efficient generation of photonic graph states in which many photons are mutually entangled with various topologies. Recently, deterministic sources of graph states have been demonstrated with quantum emitters in both the optical and microwave domains. In this work, we demonstrate deterministic and reconfigurable graph state generation with optical solid-state integrated quantum emitters. Specifically, we use a single semiconductor quantum dot in a cavity to generate caterpillar graph states, the most general type of graph state that can be produced with a single emitter. By using fast detuned optical pulses, we achieve full control over the spin state, enabling us to vary the entanglement topology at will. We perform quantum state tomography of two successive photons, measuring Bell state fidelities up to 0.80$\pm$0.04 and concurrences up to 0.69$\pm$0.09, while maintaining high photon indistinguishability. This simple optical scheme, compatible with commercially available quantum dot-based single photon sources, brings us a step closer to fault-tolerant quantum computing with spins and photons.

\end{abstract}

\maketitle
%\linenumbers

Realizing universal, fault-tolerant quantum computation is a long sought-after objective. Measurement-based quantum computation offers a possible path toward more rapid scaling of computational resources to achieve this aim~\cite{raussendorfOneWayQuantumComputer2001a,raussendorfTopologicalFaulttoleranceCluster2007a,paesaniHighThresholdQuantumComputing2023a}. This paradigm relies on a class of entangled states known as graph states, of which linear cluster states and GHZ states (locally equivalent to star graph states) are prominent examples~\cite{GHZ,Raussendorf2003}. In this regard, photonic graph states are ideal candidates due to their limited sensitivity to decoherence~\cite{zhangLosstolerantAllphotonicQuantum2022}. 

Photonic graph states were first generated using linear optics gates~\cite{kokLinearOpticalQuantum2007,liMultiphotonGraphStates2020} and parametric down-conversion sources~\cite{luExperimentalEntanglementSix2007}, but these approaches have severe scaling limitations inherent to the probabilistic nature of the gates and sources. Recently, better scaling was obtained using efficient deterministic single-photon sources based on semiconductor quantum dots (QDs)~\cite{istratiSequentialGenerationLinear2020a,Pont2024,Chen2024}. However, the most efficient way to generate such graph states relies on deterministic entanglement mediated by the spin of a quantum emitter~\cite{reisererCavitybasedQuantumNetworks2015,lindnerPhotonicClusterState2009}. Such schemes have recently been demonstrated in the optical domain  with trapped atoms and QDs~\cite{yangSequentialGenerationMultiphoton2022,thomasEfficientGenerationEntangled2022,thomasFusionDeterministicallyGenerated2024b,coganDeterministicGenerationIndistinguishable2023,costeHighrateEntanglementSemiconductor2023a,suContinuousDeterministicAllphotonic2024a,mengDeterministicPhotonSource2024}, and in the microwave domain with superconducting  qubits~\cite{Ferreira2024,OSullivan2024}.  In this regard,
QD deterministic sources of graph states are highly promising because they offer emission in the optical domain  for long distance propagation, solid-state integration, and record single photon generation rates~\cite{Somaschi2016,Tomm2021,dingHighefficiencySinglephotonSource2025}. However, to date, only limited topology of entanglement has been generated~\cite{coganDeterministicGenerationIndistinguishable2023,costeHighrateEntanglementSemiconductor2023a,suContinuousDeterministicAllphotonic2024a,mengDeterministicPhotonSource2024}, without at-will reconfigurability.

In this work we demonstrate the generation of 4-partite entanglement with arbitrary topology by producing a class of entangled graph states called caterpillar graph states~\cite{petterssonDeterministicGenerationConcatenated2024}. These states are the most general type of graph state that can be generated with a single emitter and include linear cluster, GHZ, and redundantly encoded linear cluster states. Moreover, these caterpillar states can be used for efficient fusion operations, which are crucial for generating multi-dimensional graph states and implementing quantum error correction protocols~\cite{hilaireNeardeterministicHybridGeneration2023,paesaniHighThresholdQuantumComputing2023a}. We achieve this result using an optical method that provides full control of the spin state of a single electron trapped in a QD while retaining compatibility with the entanglement generation scheme~\cite{lindnerPhotonicClusterState2009}. In addition, we demonstrate that our protocol allows for on-demand reconfigurability of the entanglement generation.

\section*{\label{Main}Controllable platform for spin-photon entanglement}

\begin{figure*}
\includegraphics[scale=0.6]{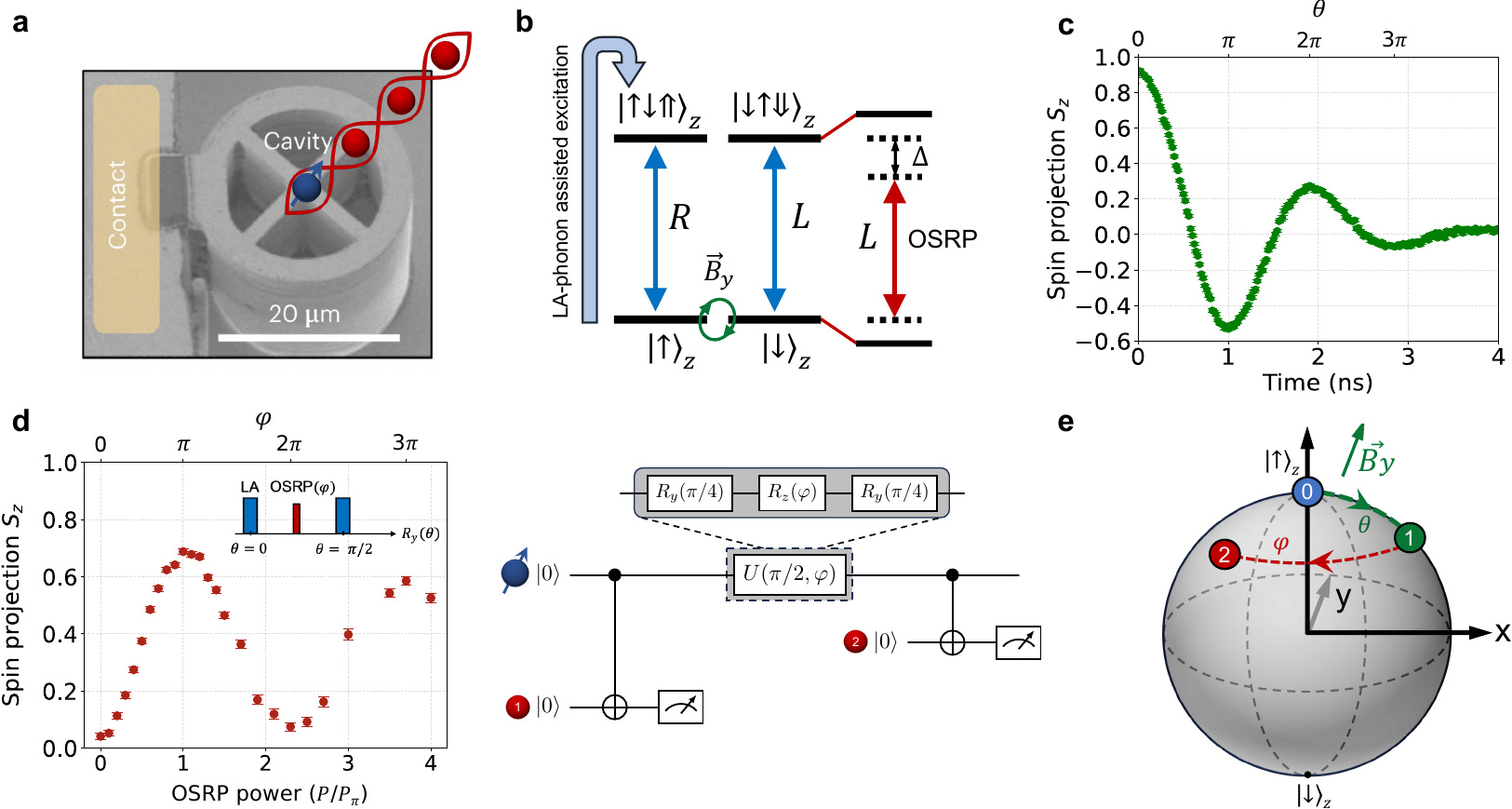}
\caption{\textbf{Complete spin control for versatile spin-photon entanglement.} \textbf{a}, Scanning electron micrograph of an electrically-contacted QD-micropillar cavity device and schematic representation of entanglement between QD spin and emitted photons. \textbf{b}, Optical selection rules of a negatively charged QD under small ($<100$ mT) transverse magnetic field $\vec{B}_y$. LA-phonon assisted excitation is used to excite the QD with a blue-detuning of 0.8 nm. The fast (4ps), red-detuned and circularly polarized optical spin rotation pulse (OSRP) induces an AC Stark shift that imprints a phase shift between the $\ket{\uparrow_z}$ and $\ket{\downarrow_z}$ states, which is equivalent to a coherent rotation about the \textit{z}-axis. \textbf{c}, Spin projection along the \textit{z}-axis ($S_z$) as a function of time (and equivalent rotation angle $\theta)$, illustrating the coherent Larmor precession undergone by the electron spin for B = 60 mT. \textbf{d}, (Left) Spin projection $S_z$ as a function of OSRP power (and equivalent rotation angle $\varphi$), demonstrating rotation of the electron spin about the \textit{z}-axis. The 3-pulse sequence (inset) used to measure $S_z$ is composed of two excitation pulses (labeled LA) and one OSRP with variable power. (Right) Equivalent quantum circuit diagram which features the unitary gate $U(\theta,\varphi)$ we can perform by combining Larmor precession and OSRP. \textbf{e}, Representation of spin control in the Bloch sphere. Starting from a measurement of a photon in the \textit{R} polarization basis which heralds the spin state in up $\ket{\uparrow}_z$ (marked 0), a 60 mT transverse magnetic field induces a Larmor precession in the $xz$-plane. After an arbitrary rotation by angle $\theta$, an OSRP rotates the spin about the \textit{z}-axis with an angle $\varphi$.}
\label{fig:device}
\end{figure*}

Our platform for generating graph states is based on spin-photon entanglement using an InGaAs semiconductor QD in an optical cavity, as shown in Fig.~\ref{fig:device}\textbf{a}. The QD is deterministically embedded in the center of a micropillar cavity using the in-situ lithography technique~\cite{Dousse2008}. The QD-cavity coupling  provides a significant and unpolarized Purcell enhancement of the single photon emission, with a corresponding photon radiative lifetime of 200 ps, as well as a high collection efficiency. The cavity is electrically contacted, allowing us to apply an electrical bias to tune the QD energy~\cite{Somaschi2016}.

A single electron is trapped in the QD, serving as a host spin that can be optically addressed. For such a charged QD, the two excited states are trion transitions consisting of two electrons and one hole ($\ket{\uparrow\downarrow\Uparrow}$ or $\ket{\uparrow\downarrow\Downarrow}$). 
The optical selection rules for this system (depicted in Fig.~\ref{fig:device}\textbf{b}) couple the spin of the electron in the ground state, either $\ket{\uparrow}$ or $\ket{\downarrow}$, to the polarization of the emitted photon, circular right ($R$) or circular left ($L$), respectively. In this work, we make use of these optical selection rules to successively entangle the polarization degree of freedom of emitted photons with the state of the single spin.

To take advantage of the mapping between the spin state and photon polarization, we use longitudinal-acoustic (LA) phonon-assisted excitation. Laser pulses that are blue-detuned from the QD transition by approximately 0.8 nm populate the trion state. This gives us high occupation probability, high photon indistinguishability~\cite{Thomas2021}, and access to the polarization degree of freedom of the emitted photons~\cite{costeProbingDynamicsCoherence2023}, as opposed to resonant excitation scheme \cite{Somaschi2016}.

In order to fully harness the spin-photon interface for versatile entangled state generation, we require multi-axis control over the electron spin state in the Bloch sphere. We use a 60 mT external magnetic field, along the $y$ direction perpendicular to the growth direction $z$. This enables the coherent Larmor precession of the electron spin  about the $y$-axis, effectively implementing the rotation gate $R_y(\theta)$, where $\theta$ represents the rotation angle. We measure the Larmor precession using polarization-resolved time-correlations by exciting the QD with a linearly-polarized continuous wave laser and monitoring the evolution over time of the spin projection along the \textit{z}-axis $S_z$. This is accomplished with a two-photon correlation measurement. The detection of the first photon in the $R$ polarization basis heralds the spin in the $\ket{\uparrow}$ state, due to the optical selection rules. We then measure the polarization of the second emitted photon as a function of time (or equivalently, as a function of the rotation angle $\theta$) to quantify the spin projection along the \textit{z}-axis, defined as $S_{z} = \dfrac{I_R-I_L}{I_R+I_L}$, where $I_R$ ($I_L$) are the conditional detection counts in $R$ ($L$) polarization. The observed oscillations evidence a Larmor period of~$\approx$ 1.85 ns and are damped by the electron coherence time of approximately 2 ns, as shown in Fig.~\ref{fig:device}\textbf{c}.

For additional control over the spin, we use a fast (4 ps) optical spin rotation pulse (OSRP) to deterministically rotate the electron spin about the optical \textit{z}-axis~\cite{greilichUltrafastOpticalRotations2009,Press2008,Berezovsky2008,Stockill2016}. It consists of a circularly-left polarized, red-detuned ($\Delta = 1.2$ nm) laser pulse that couples to one of the two trion transitions. This induces an AC Stark shift (represented in Fig.~\ref{fig:device}\textbf{b}) and leads to a rotation in the $xy$-plane of the Bloch sphere for the duration of the pulse. This rotation is described by the gate $R_z(\varphi)$, where $\varphi$ depends on the OSRP power. We measure this rotation using a three-pulse sequence, sketched in the inset of Fig.~\ref{fig:device}\textbf{d}. The first linearly-polarized pulse is used to excite the QD, with photon detection in $R$ again, heralding the spin in the $\ket{\uparrow}$ state. We then let the spin precess about the magnetic field axis for a time corresponding to the precession $\theta = \pi/4$. Following this, we apply an OSRP with variable power. After another $\theta = \pi/4$ precession, we then finally measure the spin projection $S_z$ through a polarization measurement of the second emitted photon, which is obtained using an additional linearly-polarized excitation pulse. The entire sequence is summarized as a quantum circuit in Fig.~\ref{fig:device}\textbf{d} (right) where we define a unitary gate $U(\theta,\varphi)= R_y(\theta/2)R_z(\varphi)R_y(\theta/2)$. We find $S_z$ to oscillate as a function of the OSRP power (see Fig.~\ref{fig:device}\textbf{d}, left), indicating a control over the spin about the \textit{z}-axis.  

By combining the $R_y(\theta)$ and $R_z(\varphi)$ rotation gates, controlled by the external magnetic field and the OSRP, respectively, we demonstrate full control over the spin within the Bloch sphere, as depicted in Fig.~\ref{fig:device}\textbf{e}. This allows the implementation of arbitrary quantum gates, which we use to generate various spin-photon graph states in the following section.

\section*{Versatile graph state generation}

\begin{figure*}
\includegraphics[scale=0.35]{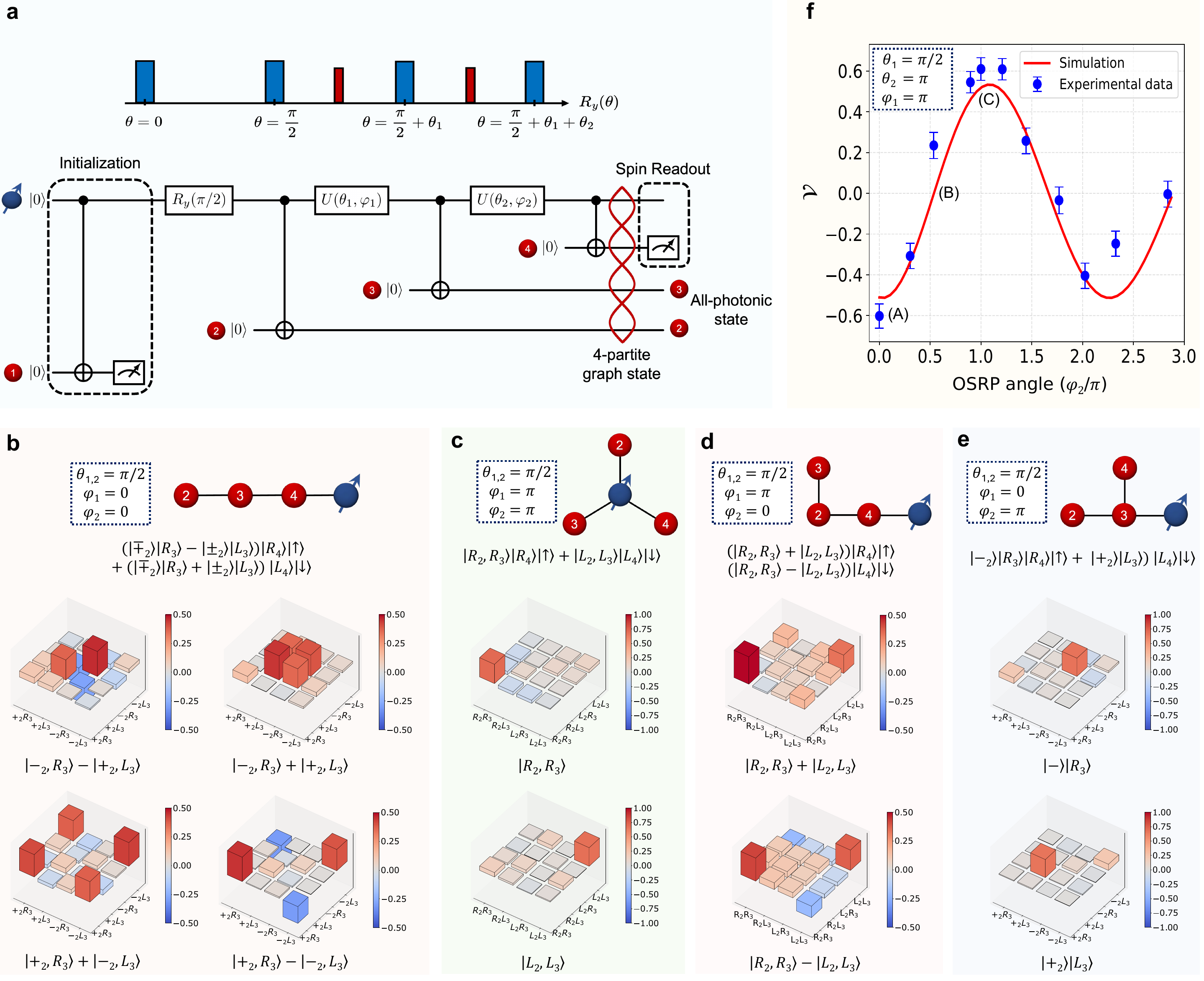}
\caption{\textbf{Reconfigurable generation of caterpillar graph states.} \textbf{a}, Optical excitation sequence and corresponding quantum circuit diagram used to generate arbitrary 4-partite caterpillar graph state. The labeled numbers in the circuit denote the order of photon emission. The first and last emitted photons allow for initialization and readout of the spin state, respectively. The Larmor precession of the spin acts as a $R_y(\theta)$ gate, while the OSRP serves as a $R_z(\varphi)$ gate, together forming an unitary gate $U(\theta,\varphi)$. \textbf{b}, 4-partite spin-photon linear cluster state generated with $\theta_{1,2}=0$ and $\varphi_{1,2}=0$, along with the measured real part (see imaginary part in Supplementary Fig.~\ref{fig:measured cluster imag}) of the density matrix of photon pair \#2 and \#3, conditioned on photons \#1 and \#4 being measured in $R/R$ (top left), $R/L$ (top right), $L/R$ (bottom right) or $L/L$ (bottom left). \textbf{c-e}, Graph representation, unitary gate parameters ($\theta_{1,2}$ and $\varphi_{1,2}$), and corresponding real part (see imaginary part in Supplementary Fig.~\ref{fig:measured cluster imag}) of the two-photon density matrix, conditioned on photons \#1 and \#4 being measured in $R/R$ (top) or $R/L$ (bottom), for multiple 4-partite states generated by this protocol: GHZ state (\textbf{c}), linear cluster states with redundant encoding between photons \#2 and \#3 (\textbf{d}) and linear cluster states with redundant encoding between photons \#3 and \#4 (\textbf{e}). \textbf{f}, Measured (symbols) and simulated (solid line) visibility $\mathcal{V}$ of four-photon correlations as a function of the angle $\varphi_2$ of the second unitary gate in the sequence, demonstrating continuous variation in the generated state. The states corresponding to points (A), (B), and (C) are defined in the main text. The simulated sinusoidal fit is obtained using model parameters extracted from
fitting the two-photon density matrices of \textbf{b-e}.}

\label{fig:cluster}
\end{figure*}

To generate 4-partite spin-photon entanglement we use four linearly-polarized excitation pulses with equal delays $t=600$ ps between them, leading to the emission of four successive photons, as sketched in Fig.~\ref{fig:cluster}\textbf{a}. The time $t$ between pulses is set to match a quarter of the spin precession period for a 60 mT magnetic field, while accounting for spontaneous emission time that effectively delays the spin precession. This ensures the spin undergoes an effective $R_y(\pi/2)$ rotation during that time. The 60 mT magnetic field amplitude is chosen so that the spin undergoes multiple $\pi/2$ precession periods within its coherence time, while preserving still mostly circular optical selection rules. Additionally, this field amplitude sets the $\pi/2$ precession time to exceed the spontaneous emission time, thereby preventing significant degradation of the state fidelity.
Because the $\approx2$ ns spin coherence time of the spin is significantly smaller than the 12 ns repetition period of our scheme, we assume the spin begins the sequence in a mixed state with equal probability of $\ket{\uparrow}$ and $\ket{\downarrow}$. The first emitted photon is measured in the $R$ ($L$) polarization basis, heralding the spin state in $\ket{\uparrow}$ ($\ket{\downarrow}$). After a time $t$, in the absence of decoherence, the spin is in the superposition state  $\dfrac{1}{\sqrt{2}}\left(\pm\ket{\uparrow} +\ket{\downarrow}\right)$, where the sign depends on the spin heralding. The photon emission triggered by laser pulse \#2 then leaves the system in the spin-photon entangled state $\ket{\Psi_2}= \dfrac{1}{\sqrt{2}}\left(\pm\ket{R_2,\uparrow} +\ket{L_2,\downarrow}\right)$, assuming instantaneous photon emission lifetime. The indices refer to the order of photon emission (i.e. $R_2$ refers to photon \#2). We now apply two unitary gates, $U(\theta_1,\varphi_{1})$ and $U(\theta_2,\varphi_{2})$, acting on the spin and each followed by an excitation pulse that leads to photon emission \#3 and \#4. By controlling $\theta_{1,2}$ and $\varphi_{1,2}$ through adjustments in the time delay between the excitation pulses, and the OSRP power, we generate various 4-partite spin-photon entangled graph states. Fig.~\ref{fig:cluster}\textbf{a} illustrates the experimental sequence and the corresponding quantum circuit diagram.

For $\theta_{1,2}=\pi/2$ and $\varphi_{1,2} = 0$, i.e. with no OSRP as in~\cite{costeHighrateEntanglementSemiconductor2023a,coganDeterministicGenerationIndistinguishable2023,
suContinuousDeterministicAllphotonic2024a}, we generate a state locally equivalent to a four-qubit linear cluster (4LC)
state (see Supplementary Information for detailed calculations),

\begin{equation} 
\begin{split}
\ket{\Psi_{4-\text{LC}}}= \dfrac{1}{2}(\ket{\mp_{2},R_{3}}-\ket{\pm_{2},L_{3}})\ket{R_{4}}\ket{\uparrow} + \\
(\ket{\mp_{2},R_{3}}+\ket{\pm_{2},L_{3}})\ket{L_{4}}\ket{\downarrow}
\end{split}
\end{equation}
where $\ket{+}$ and $\ket{-}$ are respectively defined as $\ket{+}=\dfrac{1}{\sqrt{2}}\left(\ket{R}+\ket{L}\right)$ and $\ket{-}=\dfrac{1}{\sqrt{2}}\left(\ket{R}-\ket{L}\right)$. 

We then disentangle the spin from the photonic chain to minimize additional decoherence. This is done by measuring the last emitted photon in the $R/L$ polarization basis, as its polarization state is directly mapped to the spin state. This ideally leaves the system in one of the four fully photonic entangled Bell states, depending on the polarization state measured for the first and last photon:

\begin{equation*}
\begin{split}
&\ket{\tilde{\phi}_{\pm}}_{R} = \dfrac{1}{\sqrt{2}}\left(\ket{-,R_3} \pm \ket{+_2,L_3}\right)  \\
&\ket{\tilde{\psi}_{\pm}}_{L} = \dfrac{1}{\sqrt{2}}\left(\ket{+_2,R_3} \pm \ket{-_2,L_3}\right) 
\end{split}
\end{equation*}
where the indices indicate the  first photon polarization measurement outcome. Fig.~\ref{fig:cluster}\textbf{b} presents the measured polarization density matrices of the photon pair conditioned on the measure of the first and last photon in $R/L$ basis. We find fidelities $F_{\tilde{\phi}_+}=0.78\pm0.04$, $F_{\tilde{\phi}_-}=0.69\pm0.02$, $F_{\tilde{\psi}_+}=0.80\pm0.04$ and $F_{\tilde{\psi}_-}=0.73\pm0.01$  to the ideal corresponding Bell states, and concurrences $ C_{\tilde{\phi}_+}=0.69\pm0.09$, $C_{\tilde{\phi}_-}=0.44 \pm 0.05$, $C_{\tilde{\psi}_-}=0.65\pm0.08$ and $C_{\tilde{\psi}_+}=0.49\pm0.03$. The variations in concurrence and fidelity are likely due to a small polarized Purcell effect which leads to a residual ($\approx 4\%$) polarized single photon emission. Uncertainties are obtained assuming a shot noise limited error on the total number of 4-photon coincidences. Both the fidelity and concurrence are fundamentally limited by the spin coherence time of $\approx$~2~ns, as well as the 200~ps trion radiative lifetime. The latter limitation is caused by the spontaneous emission time jitter interrupting the spin evolution, resulting in an effective emission-induced spin dephasing. This jitter is minimized when the trion \textit{g}-factor is much smaller than the electron \textit{g}-factor~\cite{rameshImpactHole$g$factor2025}. We thus chose a 60 mT magnetic field, corresponding to a 600 ps $\pi/2$ precession period, to compromise between emission-induced and nuclear-induced spin dephasing. A shorter trion radiative lifetime can further mitigate this issue by reducing the emission jitter on the spin gate. Nonetheless, to the best of our knowledge, this is the highest reported fidelity for an entangled pair within a multipartite QD spin-photon cluster state ~\cite{suContinuousDeterministicAllphotonic2024a}. Additionally, by using a numerical model to simulate these measurements we can evaluate all the relevant parameters of our experiment (summed up in Supplementary Table~\ref{tab:parameter table}) and then estimate a fidelity for the 4-partite linear cluster state. Our model exploits the zero-photon-generator method~\cite{weinSimulatingPhotonCounting2024}, allowing direct access to the conditional polarization photon coincidences of the 4-partite state without requiring the computation of multi-time integrated correlation functions. Due to spin decoherence, the state fidelity degrades over time and we thus chose to evaluate it at 600 ps after the final excitation pulse, as it corresponds to one more $\pi/2$-rotation of the spin. We find $F^{sim}_{4} = 0.66 \pm 0.05$. We note that for our experimental parameters, we estimate an upper bound on the 4-partite fidelity of 0.79,  due only to the emission-induced spin dephasing. This upper bound increases to 0.95 for a trion lifetime of 50 ps, which can be achieved using a micropillar cavity with a Purcell factor of around 24. From our simulations, we predict that this improvement would bring the 4-partite fidelity to 0.73, then primarily limited by electron-nuclear interactions. More details about the simulation model can be found in the Supplementary Information. 

Another class of entangled states that are fundamental resources for photonic quantum computing are the so-called GHZ states~\cite{GHZ}, locally equivalent to star graph states. With our protocol, GHZ states are generated by setting $\theta_{1,2}=\pi/2$ and $\varphi_{1,2} = \pi$. In this configuration, the unitary gate $U(\theta,\varphi)$ becomes a $Z$ gate which fully flips the spin about the \textit{z}-axis by the time the following photon is emitted. This leads to the generation of a 4-partite GHZ state (in the following, we only consider the heralded $\ket{\uparrow}$ case) : 

\begin{equation} 
\ket{\Psi_{4-\text{GHZ}}}= \dfrac{1}{\sqrt{2}}\left(\ket{R_2,R_3,R_4}\ket{\uparrow} + \ket{L_2,L_3,L_4}\ket{\downarrow} \right)
\label{GHZ eq}
\end{equation}
Now, when measuring the last photon in the $R/L$ basis to disentangle the spin, the remaining photon pair is projected onto a fully separable state, $\ket{R_2,R_3}$ or $\ket{L_2,L_3}$. The measured density matrices for these two states, obtained through 4-photon correlations, are shown in Fig.~\ref{fig:cluster}\textbf{c}. We find a fidelity to the ideal state of  $0.71 \pm 0.02$ and $0.68 \pm 0.02$, respectively. The 4-partite GHZ state fidelity is estimated to be $F^{sim}_{4} = 0.45 \pm 0.05$. This reduced fidelity compared to the linear cluster state is primarily attributed to optically-induced dephasing from the OSRP.

It is worth noting that these GHZ states can be generated without using OSRP ($\varphi=0$) by letting the spin undergo a $\theta = 2\pi$ (or $\pi$ for a locally equivalent state) precession between excitation pulses. However, this approach increases the spin precession time, which leads to a reduced state fidelity due to the limited spin coherence time. Optical spin control circumvents this problem as we can use arbitrary time delays between successive excitation pulses, limited only by the photon radiative lifetime. In the following, we continue to use a 600 ps delay as it provides an optimal working point with an overlap between successive photons of only 5\%.

When now setting $\theta_{1,2}=\pi/2$, $\varphi_{1} = \pi$ and $\varphi_{2} = 0$, we generate a redundantly encoded linear cluster state (RLC). Redundantly encoded qubits are crucial resources for quantum computation, as they can be used to perform efficient fusion operations with a higher success rate than ancilla-assisted fusions~\cite{browneResourceEfficientLinearOptical2005,herrera-martiPhotonicImplementationTopological2010,hilaireNeardeterministicHybridGeneration2023}. We represent these redundantly encoded states as a horizontal chain of qubits while the redundancy is introduced by attaching additional qubits vertically, thus creating a state locally equivalent to a caterpillar graph state (see Supplementary Information), as represented in Fig.~\ref{fig:cluster}\textbf{d}.
More specifically, the generated state is a 4-partite linear cluster state with photon \#2 and \#3 being redundantly encoded: 

\begin{equation} 
\begin{split}
\ket{\psi_{4-\text{RLC}1}}= \dfrac{1}{2}((\ket{R_2,R_3} +  \ket{L_2,L_3})\ket{R_4}\ket{\uparrow} + \\
(\ket{R_2,R_3} - \ket{L_2,L_3})\ket{L_4}\ket{\downarrow})
\end{split}
\label{eq:redundant1}
\end{equation}
One can see that for this state, when disentangling the spin by measuring the last photon in the $R/L$ basis, we are left with a photonic Bell state of the form:

\begin{equation*}
\ket{\phi_{\pm}} = \dfrac{1}{\sqrt{2}}\left(\ket{R_2,R_3} \pm\ket{L_2,L_3}\right)
\end{equation*}
The measured density matrix corresponding to $\ket{\phi_+}$ and $\ket{\phi_-}$ are shown in Fig.~\ref{fig:cluster}\textbf{d}. We find a fidelity to the target Bell state of $0.58 \pm 0.03$ and $0.61 \pm 0.02$ and a concurrence of $0.41 \pm 0.06$ and $0.45 \pm 0.04$, respectively, while the 4-partite entanglement is estimated to be $F^{sim}_{4} = 0.53 \pm 0.05$.\\

We finally generate yet another redundantly encoded 4-partite linear cluster by setting $\theta_{1,2}=\pi/2$, $\varphi_{1} = 0$ and $\varphi_{1} = \pi$. This yields the state:

\begin{equation}
\begin{split}
\ket{\psi_{4-\text{RLC}2}} =  \dfrac{1}{\sqrt2} \left(- \ket{-_2, R_{3}} \ket{R_4} \ket{\uparrow} + \ket{+_2, L_{3}} \ket{L_4} \ket{\downarrow} \right)
\end{split}
\end{equation}
for which now photon \#3 and \#4 are redundantly encoded.
When disentangling the spin by measuring the last photon in $R$ or $L$ we are now left with a separable state $\ket{-_2,R_3}$ or $\ket{+_2,L_3}$. The measured density matrices for these two states are shown in Fig.~\ref{fig:cluster}\textbf{e} for which we extract a fidelity to the ideal two-photon state of $0.67 \pm 0.02$ and $0.66 \pm 0.02$ and a simulated 4-partite fidelity of $F^{sim}_{4}=0.53 \pm 0.05$.
All the experimental data are reproduced using our simulation model, and are shown in the Supplementary information (Fig.~\ref{fig:simulated all}) along with a summary table that details all measured and simulated fidelities (Supplementary Table~\ref{tab:fidelity table}). We find that our simulations show excellent agreement with the experimental results, with an average absolute difference of only $3 \pm 2\%$.

\begin{figure}   
\includegraphics[scale=0.43]{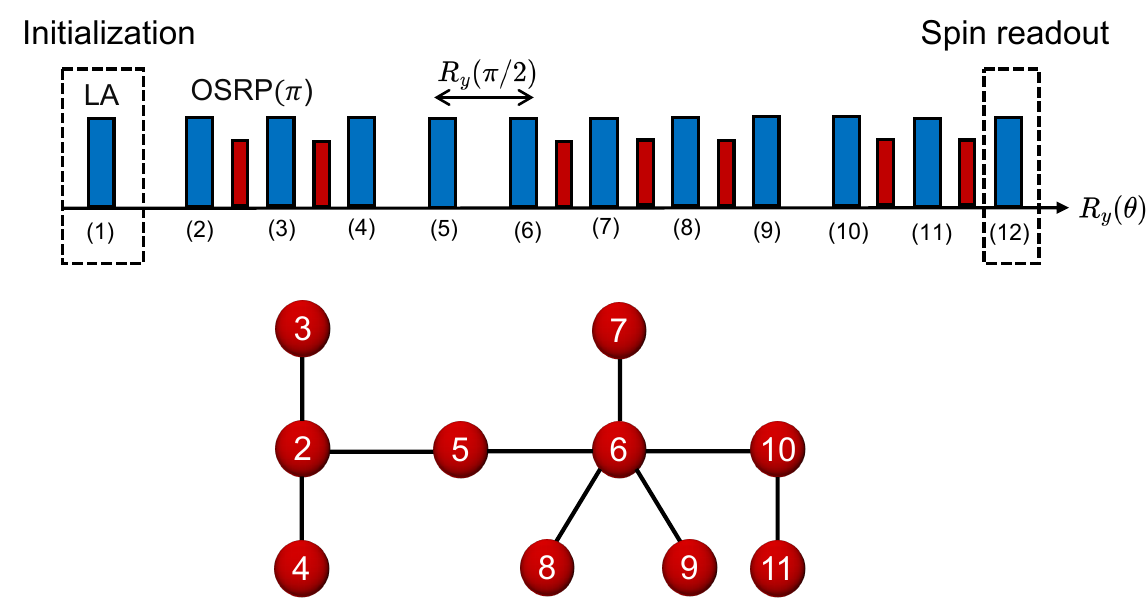}
\caption{\textbf{Near-future example of all-photonic arbitrary caterpillar graph state generation.} Pulse sequence combining excitation pulses (LA) and OSRP ($\varphi = \pi$, equivalent to a $Z$ spin gate) for the generation of all-photonic arbitrary caterpillar graph state that can be generated with our protocol. Each photon emitted following a $R_y(\pi/2)$ gate will be encoded in a new node of the caterpillar graph state, whereas photons emitted after a $Z$ gate will be redundantly encoded with the previous one, within the same node. The simulated fidelity of this state is estimated to be $0.80 \pm 0.01$ using a realistic near-term positive trion source whose parameters are described in Supplementary Fig.~\ref{fig:simulated fidelity prediction table}\textbf{b}.}
\label{fig:caterpillar}
\end{figure}

As a final illustration of the versatility of our approach, we now fix $\theta_{1}=\pi/2$, $\theta_{2}=\pi$, $\varphi_{1} = \pi$ and we scan $\varphi_{2}$ between 0 and $3\pi$. By doing so, we continuously change the output 4-partite spin-photon entangled state. In order to quantify this effect we measure oscillations in the visibility  $\mathcal{V}$ defined as :
\begin{equation*}
\resizebox{1\hsize}{!}{$\mathcal{V} = \dfrac{C_{R_1R_2R_3R_4}+C_{R_1L_2L_3L_4}-C_{R_1R_2R_3L_4}-C_{R_1L_2L_3R_4}}{C_{R_1R_2R_3R_4}+C_{R_1L_2L_3L_4}+C_{R_1R_2R_3L_4}+C_{R_1L_2L_3R_4}}$}
\label{S}
\end{equation*}
where $C_{R_1R_2R_3R_4}$, $C_{R_1L_2L_3L_4}$, $C_{R_1R_2R_3L_4}$, and $C_{R_1L_2L_3R_4}$ are four-photon coincidences measured in the $R$ or $L$ polarization basis. When $\varphi_{2}$ is set to zero, we ideally generate the state (A) = $\dfrac{1}{\sqrt{2}}\left(-\ket{R_2,R_3,L_4}\ket{\uparrow} + \ket{L_2,L_3,R_4}\ket{\downarrow} \right)$, which corresponds to $\mathcal{V}$ = -1. The polarization of the last photon is inverted relative to the GHZ state defined in Eq~\ref{GHZ eq}, as the spin undergoes a $\pi$ precession between the last two excitation pulses. At a rotation angle $\varphi_2=\pi/2$, we obtain the state (B) = $-i(\ket{R_2,R_3}+\ket{L_2,L_3})\ket{R_4}\ket{\uparrow} -(\ket{R_2,R_3}-\ket{L_2,L_3})\ket{L_4}\ket{\downarrow}$, for which $\mathcal{V} = 0$. Finally, when $\varphi_2$ reaches $\pi$, it fully flips the spin about the \textit{z}-axis, and we recover the GHZ state (C) defined in Eq~\ref{GHZ eq} ($\mathcal{V}$ = +1). Fig.~\ref{fig:cluster}\textbf{f} presents both the measured and simulated visibility $\mathcal{V}$ as a function of OSRP angle $\varphi_2$. We find $\mathcal{V}$ to oscillate as described above, demonstrating continuous control over the state. The simulation, performed using parameters obtained from fitting the two-photon density matrices of Fig.~\ref{fig:cluster}, shows good agreement with the experimental data, further validating our model. We attribute the reduced amplitude visibility from -0.6 to 0.6 to spin decoherence and the imperfect spin rotation gate fidelity of $0.87\pm0.05$, which we extract from our simulation model.

The protocol described in this work can be extended to generate fully photonic caterpillar graph states of arbitrary topology. Fig.~\ref{fig:caterpillar} shows an example of a 10-photon caterpillar graph state, along with the pulse sequence that combines excitation pulses and OSRPs. The time interval between excitation pulses corresponds to a $R_y(\pi/2)$ rotation of the spin, ensuring that each photon emitted following only a $R_y(\pi/2)$ gate is encoded as a new node of the caterpillar graph state. However, applying an OSRP with $\varphi=\pi$ between consecutive excitation pulses effectively performs a $Z$ gate, causing the newly emitted photon to be redundantly encoded with the previous one. This creates highly redundant nodes that locally resemble GHZ or star graphs, within the caterpillar graph state.

\section*{Conclusion \& Perspectives}

In conclusion, we have demonstrated a versatile approach to the on-demand generation of multipartite entangled states consisting of a solid-state spin and single photons. We have shown that with an optical pulse we can rotate the spin to continuously vary the type of entangled state that we produce. With this approach, we report for the first time with a solid-state spin the versatile generation of 4-partite linear cluster states, GHZ states, and redundantly-encoded cluster states with two-photon entanglement fidelities of up to 0.80. Notably, the emitted photons maintain high indistinguishability ($M > 82\%$, see Supplementary Fig.~\ref{fig:HOM}) across all protocols described in this work. This is a crucial requirement for generating higher-dimensional graph states using fusion operations~\cite{browneResourceEfficientLinearOptical2005}. Indeed, photon distinguishability induces a fusion measurement error with probability $(1-M)/2$, which remains below $10\%$ for our system. Owing to the inherent qubit redundancy of the caterpillar graph generation scheme, we can  combine these fusion operations with a repetition code to further mitigate logical fusion measurement errors (below $3\%$ and $1\%$ with respectively 3 and 5 successful fusion operations)~\cite{chanTailoringFusionbasedPhotonic2024}. Additionally, using a semiconductor source of indistinguishable single photons in a weak magnetic field makes this approach compatible with commercial integration to obtain a plug-and-play source of multiphoton entanglement~\cite{margariaEfficientFiberpigtailedSource2024,Maring2024} with, for instance, a permanent magnet in a compact cryostat~\cite{steindlResonantTwoLaserSpinState2023}.

We do acknowledge several areas for improvement. In particular, the device used does not show the state-of-the-art brightness that has been achieved by an InGaAs QD-cavity platform~\cite{dingHighefficiencySinglephotonSource2025}. Improved Purcell enhancement will allow for more photon emission during the spin coherence time and improve fidelity owing to the reduced trion excited state lifetime. We could also extend the coherence time of the spin through the use of a hole spin~\cite{costeProbingDynamicsCoherence2023}, or well-documented nuclear spin cooling~\cite{Greilich2007,Ethier-Majcher2017a,Gangloff2019,Prechtel2016a} and dynamical decoupling techniques~\cite{Hahn1950,Press2010}. 

\begin{comment}

\end{comment}

With realistic near-term improvements to the source — specifically a positive trion with a 100 ps radiative lifetime — and improved spin rotation gate fidelity (0.995), our simulation model predicts the generation of entangled states with up to 30 photons (see Supplementary Fig.~\ref{fig:simulated fidelity prediction table}). Notably, the entanglement fidelity remains above 80\% for the 10-photon caterpillar graph state shown in Fig.~\ref{fig:caterpillar}, which represents a realistic aim for the near future. 

The building blocks of caterpillar state generation demonstrated in this work have direct implications for scalable quantum architectures. For instance, in the fusion-based quantum computation framework, a 14-photon caterpillar state source could enable the efficient production of 24-photon Shor-encoded (2,2) 6-ring resource states~\cite{weinMinimizingResourceOverhead2024}. This approach significantly reduces the number of required sources while relaxing loss tolerance constraints in fusion and switch networks. Furthermore, the techniques developed here are essential for remote spin-spin entanglement protocols within the spin-optical quantum computing (SPOQC) architecture~\cite{gliniastySpinOpticalQuantumComputing2024,hilaireEnhancedFaulttolerancePhotonic2024,chanTailoringFusionbasedPhotonic2024}. As a result, the versatility of our method paves the way for generating more complex entanglement resources, offering a promising path towards practical fault-tolerant quantum computation.

\noindent \textbf{Acknowledgements.}
The authors thank Petr Steindl for feedback on the manuscript. This work was partially supported by the Paris Ile-de-France Région in the framework of DIM SIRTEQ, the European Union’s Horizon 2020 FET OPEN project QLUSTER (Grant ID 862035), Horizon CL4 program under the grant agreement 101135288 for EPIQUE project, by the European Commission as part of the EIC accelerator program under the grant agreement 190188855 for SEPOQC project, the Plan France 2030 through the projects ANR22-PETQ-0011, ANR-22-PETQ-0006 and ANR-22-PETQ-0013, the French National Research Agency (ANR) project SPIQE (ANR\-14\-CE32\-0012), and a public grant overseen by the French National Research Agency as part of the ”Investissements d’Avenir” programme (Labex NanoSaclay, reference: ANR\-10\-LABX\-0035). P.R.R. acknowledges the financial support of the Fulbright-Universit\'{e} Paris-Saclay Doctoral Research Award. This work was done within the C2N micro nanotechnologies platforms and partly supported by the RENATECH network and the General Council of Essonne.

\section*{\label{Methods}Methods}
\noindent \textbf{Device and experimental procedure.}
The device used here consists of a self-assembled InGaAs QD grown by molecular beam epitaxy in a GaAs matrix. The QDs are positioned in the center of a $\lambda$ GaAs cavity between two sets of distributed Bragg reflectors (DBRs). Each set consists of alternating pairs of GaAs and $\text{Al}_{0.9}\text{Ga}_{0.1}\text{As}$, with 14 pairs on top and 28 on the bottom. The cavity is etched into a micropillar structure with radial support arms to allow connecting to a planar mesa with an electrical contact. The QD and cavity are inside of a p-i-n diode with a 20 nm $\text{Al}_{x}\text{Ga}_{1-x}\text{As}$ tunneling barrier to aid in trapping charge carriers.

The QD-cavity device is operated at 4K in a closed-cycle Montana cryostat. Superconducting magnetic coils allow application of up to 500 mT magnetic field in the in-plane \textit{y}-direction. Both the LA phonon-assisted excitation ($\Delta=-0.8$nm) and OSRP ($\Delta=+1.2$nm) pulses come from a femtosecond (110 fs) Ti:Sapphire laser with a repetition rate of 81 MHz. The pulses are spectrally filtered using a 4-\textit{f} pulse-shaping line with a spatial light modulator to obtain 15 ps and 4 ps pulses, respectively. The beams are then spatially separated with a band-pass filter, with their power and polarization independently controlled using variable neutral density filters and waveplates, before being recombined on a second band-pass filter. A combination of fiber and free-space delays are used to produce a sequence, consisting of up to four excitation pulses separated by 600 ps and up to two OSRPs, which repeats every 12.2 ns. A second laser set to 860nm is used in continuous wave at very low power to stabilize the charge environment of the QD and reduce blinking. The emitted photons are collected through a lens with a numerical aperture of 0.7 in a confocal microscope configuration. The laser is separated from the single photons with narrow band-pass filters (0.8 nm bandwidth). 

To quantify entanglement we measure four-photon polarization correlation measurements. Due to detector dead-time ($\approx$ 30 ns), the polarization measurements are performed in a time-to-spatial demultiplexed tomography setup composed of three 50:50 beamsplitters that split the emitted photons into four paths with equal probability. Each path consists of a quarter and half waveplate with a polarizing beamsplitter and a pair of superconducting nanowire single photon detectors (SNSPDs). Two of the paths are set to measure first and last photons in the $R/L$ polarization basis for spin initialization and readout. The other two are set to perform full quantum state tomography of the pair of photon \#2 and \#3 in order to reconstruct their polarization states. The overall photon detection efficiency before the tomography setup is approximately 4\%, resulting in a 4-photon event rate of around 200 Hz with an 81 MHz laser repetition rate. With photon demultiplexing in the tomography setup and accounting for additional optical losses, the measured useful 4-photon event rate is reduced to about 0.5 Hz. A full schematic of the experimental setup is shown in Supplementary Fig.~\ref{fig:setup}.

\noindent \textbf{Simulation.}
Estimation of experimental parameters and state fidelity are based on a four-level trion system evolving following a Markovian master equation to describe the impact of spontaneous emission. The hyperfine interaction between the electron spin and the nuclei is captured by an additional Zeeman Hamiltonian to model the fluctuating Overhauser (OH) field with an isotropic Gaussian distribution. The excitation pulses and OSRPs are modeled as instantaneous unitary rotations of the trion system, each followed by a possible pure dephasing channel to capture optically-induced decoherence. To obtain entangled light-matter states, the master equation model is used to compute the light-matter process matrix that maps a single-qubit electron spin state to a two-qubit spin-polarization state at a later point in time. This map is used repeatedly in conjunction with single-qubit rotations to construct the desired multipartite density matrices corresponding to each studied pulse configuration. Additional details about the simulation model can be found in the Supplementary information.

\noindent \textbf{Author contribution.}
H.H. and P.R.R. conducted the experimental investigation, data analysis, methodology, and visualization. S.C.W. developed the numerical model and carried out simulations. N.C., P.H. and D.A.F. contributed to the conceptualization, methodology, and formal analysis. Device design, growth, and fabrication were performed by N.S., I.S., P.S., L.L. M.M. and A.L. All authors participated to scientific discussions. D.A.F. and P.S. supervised the project. H.H., P.R.R., S.C.W., P.S. wrote the paper with feedback from all authors.

\noindent \textbf{Data and materials availability.}
All data acquired and used in this work is property of the Centre for Nanoscience and Nanotechnology and is available upon reasonable request to pascale.senellart-mardon@cnrs.fr or helio.huet@universite-paris-saclay.fr.

\noindent \textbf{Competing interest.}
N.S. is a co-founder and P.S. is a scientific advisor and co-founder of the company Quandela. The other authors declare no competing interests.

\bibliography{Biblio}

%apsrev4-2.bst 2019-01-14 (MD) hand-edited version of apsrev4-1.bst
%Control: key (0)
%Control: author (8) initials jnrlst
%Control: editor formatted (1) identically to author
%Control: production of article title (0) allowed
%Control: page (0) single
%Control: year (1) truncated
%Control: production of eprint (0) enabled
\begin{thebibliography}{55}%
\makeatletter
\providecommand \@ifxundefined [1]{%
 \@ifx{#1\undefined}
}%
\providecommand \@ifnum [1]{%
 \ifnum #1\expandafter \@firstoftwo
 \else \expandafter \@secondoftwo
 \fi
}%
\providecommand \@ifx [1]{%
 \ifx #1\expandafter \@firstoftwo
 \else \expandafter \@secondoftwo
 \fi
}%
\providecommand \natexlab [1]{#1}%
\providecommand \enquote  [1]{``#1''}%
\providecommand \bibnamefont  [1]{#1}%
\providecommand \bibfnamefont [1]{#1}%
\providecommand \citenamefont [1]{#1}%
\providecommand \href@noop [0]{\@secondoftwo}%
\providecommand \href [0]{\begingroup \@sanitize@url \@href}%
\providecommand \@href[1]{\@@startlink{#1}\@@href}%
\providecommand \@@href[1]{\endgroup#1\@@endlink}%
\providecommand \@sanitize@url [0]{\catcode `\\12\catcode `\$12\catcode `\&12\catcode `\#12\catcode `\^12\catcode `\_12\catcode `\%12\relax}%
\providecommand \@@startlink[1]{}%
\providecommand \@@endlink[0]{}%
\providecommand \url  [0]{\begingroup\@sanitize@url \@url }%
\providecommand \@url [1]{\endgroup\@href {#1}{\urlprefix }}%
\providecommand \urlprefix  [0]{URL }%
\providecommand \Eprint [0]{\href }%
\providecommand \doibase [0]{https://doi.org/}%
\providecommand \selectlanguage [0]{\@gobble}%
\providecommand \bibinfo  [0]{\@secondoftwo}%
\providecommand \bibfield  [0]{\@secondoftwo}%
\providecommand \translation [1]{[#1]}%
\providecommand \BibitemOpen [0]{}%
\providecommand \bibitemStop [0]{}%
\providecommand \bibitemNoStop [0]{.\EOS\space}%
\providecommand \EOS [0]{\spacefactor3000\relax}%
\providecommand \BibitemShut  [1]{\csname bibitem#1\endcsname}%
\let\auto@bib@innerbib\@empty
%</preamble>
\bibitem [{\citenamefont {Raussendorf}\ and\ \citenamefont {Briegel}(2001)}]{raussendorfOneWayQuantumComputer2001a}%
  \BibitemOpen
  \bibfield  {author} {\bibinfo {author} {\bibfnamefont {R.}~\bibnamefont {Raussendorf}}\ and\ \bibinfo {author} {\bibfnamefont {H.~J.}\ \bibnamefont {Briegel}},\ }\bibfield  {title} {\bibinfo {title} {A {One}-{Way} {Quantum} {Computer}},\ }\href {https://doi.org/10.1103/PhysRevLett.86.5188} {\bibfield  {journal} {\bibinfo  {journal} {Physical Review Letters}\ }\textbf {\bibinfo {volume} {86}},\ \bibinfo {pages} {5188} (\bibinfo {year} {2001})}\BibitemShut {NoStop}%
\bibitem [{\citenamefont {Raussendorf}\ \emph {et~al.}(2007)\citenamefont {Raussendorf}, \citenamefont {Harrington},\ and\ \citenamefont {Goyal}}]{raussendorfTopologicalFaulttoleranceCluster2007a}%
  \BibitemOpen
  \bibfield  {author} {\bibinfo {author} {\bibfnamefont {R.}~\bibnamefont {Raussendorf}}, \bibinfo {author} {\bibfnamefont {J.}~\bibnamefont {Harrington}},\ and\ \bibinfo {author} {\bibfnamefont {K.}~\bibnamefont {Goyal}},\ }\bibfield  {title} {\bibinfo {title} {Topological fault-tolerance in cluster state quantum computation},\ }\href {https://doi.org/10.1088/1367-2630/9/6/199} {\bibfield  {journal} {\bibinfo  {journal} {New Journal of Physics}\ }\textbf {\bibinfo {volume} {9}},\ \bibinfo {pages} {199} (\bibinfo {year} {2007})}\BibitemShut {NoStop}%
\bibitem [{\citenamefont {Paesani}\ and\ \citenamefont {Brown}(2023)}]{paesaniHighThresholdQuantumComputing2023a}%
  \BibitemOpen
  \bibfield  {author} {\bibinfo {author} {\bibfnamefont {S.}~\bibnamefont {Paesani}}\ and\ \bibinfo {author} {\bibfnamefont {B.~J.}\ \bibnamefont {Brown}},\ }\bibfield  {title} {\bibinfo {title} {High-{{Threshold Quantum Computing}} by {{Fusing One-Dimensional Cluster States}}},\ }\href {https://doi.org/10.1103/PhysRevLett.131.120603} {\bibfield  {journal} {\bibinfo  {journal} {Physical Review Letters}\ }\textbf {\bibinfo {volume} {131}},\ \bibinfo {pages} {120603} (\bibinfo {year} {2023})}\BibitemShut {NoStop}%
\bibitem [{\citenamefont {Greenberger}\ \emph {et~al.}(1989)\citenamefont {Greenberger}, \citenamefont {Horne},\ and\ \citenamefont {Zeilinger}}]{GHZ}%
  \BibitemOpen
  \bibfield  {author} {\bibinfo {author} {\bibfnamefont {D.~M.}\ \bibnamefont {Greenberger}}, \bibinfo {author} {\bibfnamefont {M.}~\bibnamefont {Horne}},\ and\ \bibinfo {author} {\bibfnamefont {A.}~\bibnamefont {Zeilinger}},\ }\href@noop {} {\emph {\bibinfo {title} {Bell’s theorem, quantum theory and conceptions of the universe}}}\ (\bibinfo  {publisher} {Springer},\ \bibinfo {year} {1989})\BibitemShut {NoStop}%
\bibitem [{\citenamefont {Raussendorf}\ \emph {et~al.}(2003)\citenamefont {Raussendorf}, \citenamefont {Browne},\ and\ \citenamefont {Briegel}}]{Raussendorf2003}%
  \BibitemOpen
  \bibfield  {author} {\bibinfo {author} {\bibfnamefont {R.}~\bibnamefont {Raussendorf}}, \bibinfo {author} {\bibfnamefont {D.~E.}\ \bibnamefont {Browne}},\ and\ \bibinfo {author} {\bibfnamefont {H.~J.}\ \bibnamefont {Briegel}},\ }\bibfield  {title} {\bibinfo {title} {{Measurement-based quantum computation on cluster states}},\ }\href {https://doi.org/10.1103/PhysRevA.68.022312} {\bibfield  {journal} {\bibinfo  {journal} {Physical Review A}\ }\textbf {\bibinfo {volume} {68}},\ \bibinfo {pages} {32} (\bibinfo {year} {2003})}\BibitemShut {NoStop}%
\bibitem [{\citenamefont {Zhang}\ \emph {et~al.}(2022)\citenamefont {Zhang}, \citenamefont {Liu}, \citenamefont {Li}, \citenamefont {Fei}, \citenamefont {Yin}, \citenamefont {Li}, \citenamefont {Liu}, \citenamefont {Mao}, \citenamefont {Chen},\ and\ \citenamefont {Pan}}]{zhangLosstolerantAllphotonicQuantum2022}%
  \BibitemOpen
  \bibfield  {author} {\bibinfo {author} {\bibfnamefont {R.}~\bibnamefont {Zhang}}, \bibinfo {author} {\bibfnamefont {L.-Z.}\ \bibnamefont {Liu}}, \bibinfo {author} {\bibfnamefont {Z.-D.}\ \bibnamefont {Li}}, \bibinfo {author} {\bibfnamefont {Y.-Y.}\ \bibnamefont {Fei}}, \bibinfo {author} {\bibfnamefont {X.-F.}\ \bibnamefont {Yin}}, \bibinfo {author} {\bibfnamefont {L.}~\bibnamefont {Li}}, \bibinfo {author} {\bibfnamefont {N.-L.}\ \bibnamefont {Liu}}, \bibinfo {author} {\bibfnamefont {Y.}~\bibnamefont {Mao}}, \bibinfo {author} {\bibfnamefont {Y.-A.}\ \bibnamefont {Chen}},\ and\ \bibinfo {author} {\bibfnamefont {J.-W.}\ \bibnamefont {Pan}},\ }\bibfield  {title} {\bibinfo {title} {Loss-tolerant all-photonic quantum repeater with generalized {{Shor}} code},\ }\href {https://doi.org/10.1364/OPTICA.439170} {\bibfield  {journal} {\bibinfo  {journal} {Optica, OPTICA}\ }\textbf {\bibinfo {volume} {9}},\ \bibinfo {pages} {152} (\bibinfo {year} {2022})}\BibitemShut {NoStop}%
\bibitem [{\citenamefont {Kok}\ \emph {et~al.}(2007)\citenamefont {Kok}, \citenamefont {Munro}, \citenamefont {Nemoto}, \citenamefont {Ralph}, \citenamefont {Dowling},\ and\ \citenamefont {Milburn}}]{kokLinearOpticalQuantum2007}%
  \BibitemOpen
  \bibfield  {author} {\bibinfo {author} {\bibfnamefont {P.}~\bibnamefont {Kok}}, \bibinfo {author} {\bibfnamefont {W.~J.}\ \bibnamefont {Munro}}, \bibinfo {author} {\bibfnamefont {K.}~\bibnamefont {Nemoto}}, \bibinfo {author} {\bibfnamefont {T.~C.}\ \bibnamefont {Ralph}}, \bibinfo {author} {\bibfnamefont {J.~P.}\ \bibnamefont {Dowling}},\ and\ \bibinfo {author} {\bibfnamefont {G.~J.}\ \bibnamefont {Milburn}},\ }\bibfield  {title} {\bibinfo {title} {Linear optical quantum computing with photonic qubits},\ }\href {https://doi.org/10.1103/RevModPhys.79.135} {\bibfield  {journal} {\bibinfo  {journal} {Reviews of Modern Physics}\ }\textbf {\bibinfo {volume} {79}},\ \bibinfo {pages} {135} (\bibinfo {year} {2007})}\BibitemShut {NoStop}%
\bibitem [{\citenamefont {Li}\ \emph {et~al.}(2020)\citenamefont {Li}, \citenamefont {Qin}, \citenamefont {Chen}, \citenamefont {Duan}, \citenamefont {Yu}, \citenamefont {Huo}, \citenamefont {H{\"o}fling}, \citenamefont {Lu}, \citenamefont {Chen},\ and\ \citenamefont {Pan}}]{liMultiphotonGraphStates2020}%
  \BibitemOpen
  \bibfield  {author} {\bibinfo {author} {\bibfnamefont {J.-P.}\ \bibnamefont {Li}}, \bibinfo {author} {\bibfnamefont {J.}~\bibnamefont {Qin}}, \bibinfo {author} {\bibfnamefont {A.}~\bibnamefont {Chen}}, \bibinfo {author} {\bibfnamefont {Z.-C.}\ \bibnamefont {Duan}}, \bibinfo {author} {\bibfnamefont {Y.}~\bibnamefont {Yu}}, \bibinfo {author} {\bibfnamefont {Y.}~\bibnamefont {Huo}}, \bibinfo {author} {\bibfnamefont {S.}~\bibnamefont {H{\"o}fling}}, \bibinfo {author} {\bibfnamefont {C.-Y.}\ \bibnamefont {Lu}}, \bibinfo {author} {\bibfnamefont {K.}~\bibnamefont {Chen}},\ and\ \bibinfo {author} {\bibfnamefont {J.-W.}\ \bibnamefont {Pan}},\ }\bibfield  {title} {\bibinfo {title} {Multiphoton {{Graph States}} from a {{Solid-State Single-Photon Source}}},\ }\href {https://doi.org/10.1021/acsphotonics.0c00192} {\bibfield  {journal} {\bibinfo  {journal} {ACS Photonics}\ }\textbf {\bibinfo {volume} {7}},\ \bibinfo {pages} {1603} (\bibinfo {year} {2020})}\BibitemShut {NoStop}%
\bibitem [{\citenamefont {Lu}\ \emph {et~al.}(2007)\citenamefont {Lu}, \citenamefont {Zhou}, \citenamefont {G{\"u}hne}, \citenamefont {Gao}, \citenamefont {Zhang}, \citenamefont {Yuan}, \citenamefont {Goebel}, \citenamefont {Yang},\ and\ \citenamefont {Pan}}]{luExperimentalEntanglementSix2007}%
  \BibitemOpen
  \bibfield  {author} {\bibinfo {author} {\bibfnamefont {C.-Y.}\ \bibnamefont {Lu}}, \bibinfo {author} {\bibfnamefont {X.-Q.}\ \bibnamefont {Zhou}}, \bibinfo {author} {\bibfnamefont {O.}~\bibnamefont {G{\"u}hne}}, \bibinfo {author} {\bibfnamefont {W.-B.}\ \bibnamefont {Gao}}, \bibinfo {author} {\bibfnamefont {J.}~\bibnamefont {Zhang}}, \bibinfo {author} {\bibfnamefont {Z.-S.}\ \bibnamefont {Yuan}}, \bibinfo {author} {\bibfnamefont {A.}~\bibnamefont {Goebel}}, \bibinfo {author} {\bibfnamefont {T.}~\bibnamefont {Yang}},\ and\ \bibinfo {author} {\bibfnamefont {J.-W.}\ \bibnamefont {Pan}},\ }\bibfield  {title} {\bibinfo {title} {Experimental entanglement of six photons in graph states},\ }\href {https://doi.org/10.1038/nphys507} {\bibfield  {journal} {\bibinfo  {journal} {Nature Physics}\ }\textbf {\bibinfo {volume} {3}},\ \bibinfo {pages} {91} (\bibinfo {year} {2007})}\BibitemShut {NoStop}%
\bibitem [{\citenamefont {Istrati}\ \emph {et~al.}(2020)\citenamefont {Istrati}, \citenamefont {Pilnyak}, \citenamefont {Loredo}, \citenamefont {Antón}, \citenamefont {Somaschi}, \citenamefont {Hilaire}, \citenamefont {Ollivier}, \citenamefont {Esmann}, \citenamefont {Cohen}, \citenamefont {Vidro}, \citenamefont {Millet}, \citenamefont {Lemaître}, \citenamefont {Sagnes}, \citenamefont {Harouri}, \citenamefont {Lanco}, \citenamefont {Senellart},\ and\ \citenamefont {Eisenberg}}]{istratiSequentialGenerationLinear2020a}%
  \BibitemOpen
  \bibfield  {author} {\bibinfo {author} {\bibfnamefont {D.}~\bibnamefont {Istrati}}, \bibinfo {author} {\bibfnamefont {Y.}~\bibnamefont {Pilnyak}}, \bibinfo {author} {\bibfnamefont {J.~C.}\ \bibnamefont {Loredo}}, \bibinfo {author} {\bibfnamefont {C.}~\bibnamefont {Antón}}, \bibinfo {author} {\bibfnamefont {N.}~\bibnamefont {Somaschi}}, \bibinfo {author} {\bibfnamefont {P.}~\bibnamefont {Hilaire}}, \bibinfo {author} {\bibfnamefont {H.}~\bibnamefont {Ollivier}}, \bibinfo {author} {\bibfnamefont {M.}~\bibnamefont {Esmann}}, \bibinfo {author} {\bibfnamefont {L.}~\bibnamefont {Cohen}}, \bibinfo {author} {\bibfnamefont {L.}~\bibnamefont {Vidro}}, \bibinfo {author} {\bibfnamefont {C.}~\bibnamefont {Millet}}, \bibinfo {author} {\bibfnamefont {A.}~\bibnamefont {Lemaître}}, \bibinfo {author} {\bibfnamefont {I.}~\bibnamefont {Sagnes}}, \bibinfo {author} {\bibfnamefont {A.}~\bibnamefont {Harouri}}, \bibinfo {author} {\bibfnamefont {L.}~\bibnamefont {Lanco}}, \bibinfo {author} {\bibfnamefont {P.}~\bibnamefont
  {Senellart}},\ and\ \bibinfo {author} {\bibfnamefont {H.~S.}\ \bibnamefont {Eisenberg}},\ }\bibfield  {title} {\bibinfo {title} {Sequential generation of linear cluster states from a single photon emitter},\ }\href {https://doi.org/10.1038/s41467-020-19341-4} {\bibfield  {journal} {\bibinfo  {journal} {Nature Communications}\ }\textbf {\bibinfo {volume} {11}},\ \bibinfo {pages} {5501} (\bibinfo {year} {2020})}\BibitemShut {NoStop}%
\bibitem [{\citenamefont {Pont}\ \emph {et~al.}(2024)\citenamefont {Pont}, \citenamefont {Corrielli}, \citenamefont {Fyrillas}, \citenamefont {Agresti}, \citenamefont {Carvacho}, \citenamefont {Maring}, \citenamefont {Emeriau}, \citenamefont {Ceccarelli}, \citenamefont {Albiero}, \citenamefont {{Dias Ferreira}}, \citenamefont {Somaschi}, \citenamefont {Senellart}, \citenamefont {Sagnes}, \citenamefont {Morassi}, \citenamefont {Lema{\^{i}}tre}, \citenamefont {Senellart}, \citenamefont {Sciarrino}, \citenamefont {Liscidini}, \citenamefont {Belabas},\ and\ \citenamefont {Osellame}}]{Pont2024}%
  \BibitemOpen
  \bibfield  {author} {\bibinfo {author} {\bibfnamefont {M.}~\bibnamefont {Pont}}, \bibinfo {author} {\bibfnamefont {G.}~\bibnamefont {Corrielli}}, \bibinfo {author} {\bibfnamefont {A.}~\bibnamefont {Fyrillas}}, \bibinfo {author} {\bibfnamefont {I.}~\bibnamefont {Agresti}}, \bibinfo {author} {\bibfnamefont {G.}~\bibnamefont {Carvacho}}, \bibinfo {author} {\bibfnamefont {N.}~\bibnamefont {Maring}}, \bibinfo {author} {\bibfnamefont {P.~E.}\ \bibnamefont {Emeriau}}, \bibinfo {author} {\bibfnamefont {F.}~\bibnamefont {Ceccarelli}}, \bibinfo {author} {\bibfnamefont {R.}~\bibnamefont {Albiero}}, \bibinfo {author} {\bibfnamefont {P.~H.}\ \bibnamefont {{Dias Ferreira}}}, \bibinfo {author} {\bibfnamefont {N.}~\bibnamefont {Somaschi}}, \bibinfo {author} {\bibfnamefont {J.}~\bibnamefont {Senellart}}, \bibinfo {author} {\bibfnamefont {I.}~\bibnamefont {Sagnes}}, \bibinfo {author} {\bibfnamefont {M.}~\bibnamefont {Morassi}}, \bibinfo {author} {\bibfnamefont {A.}~\bibnamefont {Lema{\^{i}}tre}}, \bibinfo {author}
  {\bibfnamefont {P.}~\bibnamefont {Senellart}}, \bibinfo {author} {\bibfnamefont {F.}~\bibnamefont {Sciarrino}}, \bibinfo {author} {\bibfnamefont {M.}~\bibnamefont {Liscidini}}, \bibinfo {author} {\bibfnamefont {N.}~\bibnamefont {Belabas}},\ and\ \bibinfo {author} {\bibfnamefont {R.}~\bibnamefont {Osellame}},\ }\bibfield  {title} {\bibinfo {title} {{High-fidelity four-photon GHZ states on chip}},\ }\href {https://doi.org/10.1038/s41534-024-00830-z} {\bibfield  {journal} {\bibinfo  {journal} {npj Quantum Information}\ }\textbf {\bibinfo {volume} {10}},\ \bibinfo {pages} {1} (\bibinfo {year} {2024})}\BibitemShut {NoStop}%
\bibitem [{\citenamefont {Chen}\ \emph {et~al.}(2024)\citenamefont {Chen}, \citenamefont {Peng}, \citenamefont {Guo}, \citenamefont {Gu}, \citenamefont {Ding}, \citenamefont {Liu}, \citenamefont {Zhao}, \citenamefont {You}, \citenamefont {Qin}, \citenamefont {Wang}, \citenamefont {He}, \citenamefont {Renema}, \citenamefont {Huo}, \citenamefont {Wang}, \citenamefont {Lu},\ and\ \citenamefont {Pan}}]{Chen2024}%
  \BibitemOpen
  \bibfield  {author} {\bibinfo {author} {\bibfnamefont {S.}~\bibnamefont {Chen}}, \bibinfo {author} {\bibfnamefont {L.~C.}\ \bibnamefont {Peng}}, \bibinfo {author} {\bibfnamefont {Y.~P.}\ \bibnamefont {Guo}}, \bibinfo {author} {\bibfnamefont {X.~M.}\ \bibnamefont {Gu}}, \bibinfo {author} {\bibfnamefont {X.}~\bibnamefont {Ding}}, \bibinfo {author} {\bibfnamefont {R.~Z.}\ \bibnamefont {Liu}}, \bibinfo {author} {\bibfnamefont {J.~Y.}\ \bibnamefont {Zhao}}, \bibinfo {author} {\bibfnamefont {X.}~\bibnamefont {You}}, \bibinfo {author} {\bibfnamefont {J.}~\bibnamefont {Qin}}, \bibinfo {author} {\bibfnamefont {Y.~F.}\ \bibnamefont {Wang}}, \bibinfo {author} {\bibfnamefont {Y.~M.}\ \bibnamefont {He}}, \bibinfo {author} {\bibfnamefont {J.~J.}\ \bibnamefont {Renema}}, \bibinfo {author} {\bibfnamefont {Y.~H.}\ \bibnamefont {Huo}}, \bibinfo {author} {\bibfnamefont {H.}~\bibnamefont {Wang}}, \bibinfo {author} {\bibfnamefont {C.~Y.}\ \bibnamefont {Lu}},\ and\ \bibinfo {author} {\bibfnamefont {J.~W.}\ \bibnamefont {Pan}},\
  }\bibfield  {title} {\bibinfo {title} {{Heralded Three-Photon Entanglement from a Single-Photon Source on a Photonic Chip}},\ }\href {https://doi.org/10.1103/PhysRevLett.132.130603} {\bibfield  {journal} {\bibinfo  {journal} {Physical Review Letters}\ }\textbf {\bibinfo {volume} {132}},\ \bibinfo {pages} {1} (\bibinfo {year} {2024})}\BibitemShut {NoStop}%
\bibitem [{\citenamefont {Reiserer}\ and\ \citenamefont {Rempe}(2015)}]{reisererCavitybasedQuantumNetworks2015}%
  \BibitemOpen
  \bibfield  {author} {\bibinfo {author} {\bibfnamefont {A.}~\bibnamefont {Reiserer}}\ and\ \bibinfo {author} {\bibfnamefont {G.}~\bibnamefont {Rempe}},\ }\bibfield  {title} {\bibinfo {title} {Cavity-based quantum networks with single atoms and optical photons},\ }\href {https://doi.org/10.1103/RevModPhys.87.1379} {\bibfield  {journal} {\bibinfo  {journal} {Reviews of Modern Physics}\ }\textbf {\bibinfo {volume} {87}},\ \bibinfo {pages} {1379} (\bibinfo {year} {2015})}\BibitemShut {NoStop}%
\bibitem [{\citenamefont {Lindner}\ and\ \citenamefont {Rudolph}(2009)}]{lindnerPhotonicClusterState2009}%
  \BibitemOpen
  \bibfield  {author} {\bibinfo {author} {\bibfnamefont {N.~H.}\ \bibnamefont {Lindner}}\ and\ \bibinfo {author} {\bibfnamefont {T.}~\bibnamefont {Rudolph}},\ }\bibfield  {title} {\bibinfo {title} {A photonic cluster state machine gun},\ }\href {https://doi.org/10.1103/PhysRevLett.103.113602} {\bibfield  {journal} {\bibinfo  {journal} {Physical Review Letters}\ }\textbf {\bibinfo {volume} {103}},\ \bibinfo {pages} {113602} (\bibinfo {year} {2009})}\BibitemShut {NoStop}%
\bibitem [{\citenamefont {Yang}\ \emph {et~al.}(2022)\citenamefont {Yang}, \citenamefont {Yu}, \citenamefont {Li}, \citenamefont {Jing}, \citenamefont {Bao},\ and\ \citenamefont {Pan}}]{yangSequentialGenerationMultiphoton2022}%
  \BibitemOpen
  \bibfield  {author} {\bibinfo {author} {\bibfnamefont {C.-W.}\ \bibnamefont {Yang}}, \bibinfo {author} {\bibfnamefont {Y.}~\bibnamefont {Yu}}, \bibinfo {author} {\bibfnamefont {J.}~\bibnamefont {Li}}, \bibinfo {author} {\bibfnamefont {B.}~\bibnamefont {Jing}}, \bibinfo {author} {\bibfnamefont {X.-H.}\ \bibnamefont {Bao}},\ and\ \bibinfo {author} {\bibfnamefont {J.-W.}\ \bibnamefont {Pan}},\ }\bibfield  {title} {\bibinfo {title} {Sequential generation of multiphoton entanglement with a {{Rydberg}} superatom},\ }\href {https://doi.org/10.1038/s41566-022-01054-3} {\bibfield  {journal} {\bibinfo  {journal} {Nature Photonics}\ }\textbf {\bibinfo {volume} {16}},\ \bibinfo {pages} {658} (\bibinfo {year} {2022})}\BibitemShut {NoStop}%
\bibitem [{\citenamefont {Thomas}\ \emph {et~al.}(2022)\citenamefont {Thomas}, \citenamefont {Ruscio}, \citenamefont {Morin},\ and\ \citenamefont {Rempe}}]{thomasEfficientGenerationEntangled2022}%
  \BibitemOpen
  \bibfield  {author} {\bibinfo {author} {\bibfnamefont {P.}~\bibnamefont {Thomas}}, \bibinfo {author} {\bibfnamefont {L.}~\bibnamefont {Ruscio}}, \bibinfo {author} {\bibfnamefont {O.}~\bibnamefont {Morin}},\ and\ \bibinfo {author} {\bibfnamefont {G.}~\bibnamefont {Rempe}},\ }\bibfield  {title} {\bibinfo {title} {Efficient generation of entangled multiphoton graph states from a single atom},\ }\href {https://doi.org/10.1038/s41586-022-04987-5} {\bibfield  {journal} {\bibinfo  {journal} {Nature}\ }\textbf {\bibinfo {volume} {608}},\ \bibinfo {pages} {677} (\bibinfo {year} {2022})}\BibitemShut {NoStop}%
\bibitem [{\citenamefont {Thomas}\ \emph {et~al.}(2024)\citenamefont {Thomas}, \citenamefont {Ruscio}, \citenamefont {Morin},\ and\ \citenamefont {Rempe}}]{thomasFusionDeterministicallyGenerated2024b}%
  \BibitemOpen
  \bibfield  {author} {\bibinfo {author} {\bibfnamefont {P.}~\bibnamefont {Thomas}}, \bibinfo {author} {\bibfnamefont {L.}~\bibnamefont {Ruscio}}, \bibinfo {author} {\bibfnamefont {O.}~\bibnamefont {Morin}},\ and\ \bibinfo {author} {\bibfnamefont {G.}~\bibnamefont {Rempe}},\ }\bibfield  {title} {\bibinfo {title} {Fusion of deterministically generated photonic graph states},\ }\href {https://doi.org/10.1038/s41586-024-07357-5} {\bibfield  {journal} {\bibinfo  {journal} {Nature}\ }\textbf {\bibinfo {volume} {629}},\ \bibinfo {pages} {567} (\bibinfo {year} {2024})}\BibitemShut {NoStop}%
\bibitem [{\citenamefont {Cogan}\ \emph {et~al.}(2023)\citenamefont {Cogan}, \citenamefont {Su}, \citenamefont {Kenneth},\ and\ \citenamefont {Gershoni}}]{coganDeterministicGenerationIndistinguishable2023}%
  \BibitemOpen
  \bibfield  {author} {\bibinfo {author} {\bibfnamefont {D.}~\bibnamefont {Cogan}}, \bibinfo {author} {\bibfnamefont {Z.-E.}\ \bibnamefont {Su}}, \bibinfo {author} {\bibfnamefont {O.}~\bibnamefont {Kenneth}},\ and\ \bibinfo {author} {\bibfnamefont {D.}~\bibnamefont {Gershoni}},\ }\bibfield  {title} {\bibinfo {title} {Deterministic generation of indistinguishable photons in a cluster state},\ }\href {https://doi.org/10.1038/s41566-022-01152-2} {\bibfield  {journal} {\bibinfo  {journal} {Nature Photonics}\ }\textbf {\bibinfo {volume} {17}},\ \bibinfo {pages} {324} (\bibinfo {year} {2023})}\BibitemShut {NoStop}%
\bibitem [{\citenamefont {Coste}\ \emph {et~al.}(2023{\natexlab{a}})\citenamefont {Coste}, \citenamefont {Fioretto}, \citenamefont {Belabas}, \citenamefont {Wein}, \citenamefont {Hilaire}, \citenamefont {Frantzeskakis}, \citenamefont {Gundin}, \citenamefont {Goes}, \citenamefont {Somaschi}, \citenamefont {Morassi}, \citenamefont {Lemaître}, \citenamefont {Sagnes}, \citenamefont {Harouri}, \citenamefont {Economou}, \citenamefont {Auffeves}, \citenamefont {Krebs}, \citenamefont {Lanco},\ and\ \citenamefont {Senellart}}]{costeHighrateEntanglementSemiconductor2023a}%
  \BibitemOpen
  \bibfield  {author} {\bibinfo {author} {\bibfnamefont {N.}~\bibnamefont {Coste}}, \bibinfo {author} {\bibfnamefont {D.~A.}\ \bibnamefont {Fioretto}}, \bibinfo {author} {\bibfnamefont {N.}~\bibnamefont {Belabas}}, \bibinfo {author} {\bibfnamefont {S.~C.}\ \bibnamefont {Wein}}, \bibinfo {author} {\bibfnamefont {P.}~\bibnamefont {Hilaire}}, \bibinfo {author} {\bibfnamefont {R.}~\bibnamefont {Frantzeskakis}}, \bibinfo {author} {\bibfnamefont {M.}~\bibnamefont {Gundin}}, \bibinfo {author} {\bibfnamefont {B.}~\bibnamefont {Goes}}, \bibinfo {author} {\bibfnamefont {N.}~\bibnamefont {Somaschi}}, \bibinfo {author} {\bibfnamefont {M.}~\bibnamefont {Morassi}}, \bibinfo {author} {\bibfnamefont {A.}~\bibnamefont {Lemaître}}, \bibinfo {author} {\bibfnamefont {I.}~\bibnamefont {Sagnes}}, \bibinfo {author} {\bibfnamefont {A.}~\bibnamefont {Harouri}}, \bibinfo {author} {\bibfnamefont {S.~E.}\ \bibnamefont {Economou}}, \bibinfo {author} {\bibfnamefont {A.}~\bibnamefont {Auffeves}}, \bibinfo {author} {\bibfnamefont
  {O.}~\bibnamefont {Krebs}}, \bibinfo {author} {\bibfnamefont {L.}~\bibnamefont {Lanco}},\ and\ \bibinfo {author} {\bibfnamefont {P.}~\bibnamefont {Senellart}},\ }\bibfield  {title} {\bibinfo {title} {High-rate entanglement between a semiconductor spin and indistinguishable photons},\ }\href {https://doi.org/10.1038/s41566-023-01186-0} {\bibfield  {journal} {\bibinfo  {journal} {Nature Photonics}\ }\textbf {\bibinfo {volume} {17}},\ \bibinfo {pages} {582} (\bibinfo {year} {2023}{\natexlab{a}})}\BibitemShut {NoStop}%
\bibitem [{\citenamefont {Su}\ \emph {et~al.}(2024)\citenamefont {Su}, \citenamefont {Taitler}, \citenamefont {Schwartz}, \citenamefont {Cogan}, \citenamefont {Nassar}, \citenamefont {Kenneth}, \citenamefont {Lindner},\ and\ \citenamefont {Gershoni}}]{suContinuousDeterministicAllphotonic2024a}%
  \BibitemOpen
  \bibfield  {author} {\bibinfo {author} {\bibfnamefont {Z.-E.}\ \bibnamefont {Su}}, \bibinfo {author} {\bibfnamefont {B.}~\bibnamefont {Taitler}}, \bibinfo {author} {\bibfnamefont {I.}~\bibnamefont {Schwartz}}, \bibinfo {author} {\bibfnamefont {D.}~\bibnamefont {Cogan}}, \bibinfo {author} {\bibfnamefont {I.}~\bibnamefont {Nassar}}, \bibinfo {author} {\bibfnamefont {O.}~\bibnamefont {Kenneth}}, \bibinfo {author} {\bibfnamefont {N.~H.}\ \bibnamefont {Lindner}},\ and\ \bibinfo {author} {\bibfnamefont {D.}~\bibnamefont {Gershoni}},\ }\bibfield  {title} {\bibinfo {title} {Continuous and deterministic all-photonic cluster state of indistinguishable photons},\ }\href {https://doi.org/10.1088/1361-6633/ad4c93} {\bibfield  {journal} {\bibinfo  {journal} {Reports on Progress in Physics}\ }\textbf {\bibinfo {volume} {87}},\ \bibinfo {pages} {077601} (\bibinfo {year} {2024})}\BibitemShut {NoStop}%
\bibitem [{\citenamefont {Meng}\ \emph {et~al.}(2024)\citenamefont {Meng}, \citenamefont {Chan}, \citenamefont {Nielsen}, \citenamefont {Appel}, \citenamefont {Liu}, \citenamefont {Wang}, \citenamefont {Bart}, \citenamefont {Wieck}, \citenamefont {Ludwig}, \citenamefont {Midolo}, \citenamefont {Tiranov}, \citenamefont {S{\o}rensen},\ and\ \citenamefont {Lodahl}}]{mengDeterministicPhotonSource2024}%
  \BibitemOpen
  \bibfield  {author} {\bibinfo {author} {\bibfnamefont {Y.}~\bibnamefont {Meng}}, \bibinfo {author} {\bibfnamefont {M.~L.}\ \bibnamefont {Chan}}, \bibinfo {author} {\bibfnamefont {R.~B.}\ \bibnamefont {Nielsen}}, \bibinfo {author} {\bibfnamefont {M.~H.}\ \bibnamefont {Appel}}, \bibinfo {author} {\bibfnamefont {Z.}~\bibnamefont {Liu}}, \bibinfo {author} {\bibfnamefont {Y.}~\bibnamefont {Wang}}, \bibinfo {author} {\bibfnamefont {N.}~\bibnamefont {Bart}}, \bibinfo {author} {\bibfnamefont {A.~D.}\ \bibnamefont {Wieck}}, \bibinfo {author} {\bibfnamefont {A.}~\bibnamefont {Ludwig}}, \bibinfo {author} {\bibfnamefont {L.}~\bibnamefont {Midolo}}, \bibinfo {author} {\bibfnamefont {A.}~\bibnamefont {Tiranov}}, \bibinfo {author} {\bibfnamefont {A.~S.}\ \bibnamefont {S{\o}rensen}},\ and\ \bibinfo {author} {\bibfnamefont {P.}~\bibnamefont {Lodahl}},\ }\bibfield  {title} {\bibinfo {title} {Deterministic photon source of genuine three-qubit entanglement},\ }\href {https://doi.org/10.1038/s41467-024-52086-y} {\bibfield
  {journal} {\bibinfo  {journal} {Nature Communications}\ }\textbf {\bibinfo {volume} {15}},\ \bibinfo {pages} {7774} (\bibinfo {year} {2024})}\BibitemShut {NoStop}%
\bibitem [{\citenamefont {Ferreira}\ \emph {et~al.}(2024)\citenamefont {Ferreira}, \citenamefont {Kim}, \citenamefont {Butler}, \citenamefont {Pichler},\ and\ \citenamefont {Painter}}]{Ferreira2024}%
  \BibitemOpen
  \bibfield  {author} {\bibinfo {author} {\bibfnamefont {V.~S.}\ \bibnamefont {Ferreira}}, \bibinfo {author} {\bibfnamefont {G.}~\bibnamefont {Kim}}, \bibinfo {author} {\bibfnamefont {A.}~\bibnamefont {Butler}}, \bibinfo {author} {\bibfnamefont {H.}~\bibnamefont {Pichler}},\ and\ \bibinfo {author} {\bibfnamefont {O.}~\bibnamefont {Painter}},\ }\bibfield  {title} {\bibinfo {title} {{Deterministic generation of multidimensional photonic cluster states with a single quantum emitter}},\ }\href {https://doi.org/10.1038/s41567-024-02408-0} {\bibfield  {journal} {\bibinfo  {journal} {Nature Physics}\ }\textbf {\bibinfo {volume} {20}},\ \bibinfo {pages} {865} (\bibinfo {year} {2024})}\BibitemShut {NoStop}%
\bibitem [{\citenamefont {O'Sullivan}\ \emph {et~al.}(2024)\citenamefont {O'Sullivan}, \citenamefont {Reuer}, \citenamefont {Grigorev}, \citenamefont {Dai}, \citenamefont {{Hern{\'a}ndez-Ant{\'o}n}}, \citenamefont {{Mu{\~n}oz-Arias}}, \citenamefont {Hellings}, \citenamefont {Flasby}, \citenamefont {Zanuz}, \citenamefont {Besse}, \citenamefont {Blais}, \citenamefont {Malz}, \citenamefont {Eichler},\ and\ \citenamefont {Wallraff}}]{OSullivan2024}%
  \BibitemOpen
  \bibfield  {author} {\bibinfo {author} {\bibfnamefont {J.}~\bibnamefont {O'Sullivan}}, \bibinfo {author} {\bibfnamefont {K.}~\bibnamefont {Reuer}}, \bibinfo {author} {\bibfnamefont {A.}~\bibnamefont {Grigorev}}, \bibinfo {author} {\bibfnamefont {X.}~\bibnamefont {Dai}}, \bibinfo {author} {\bibfnamefont {A.}~\bibnamefont {{Hern{\'a}ndez-Ant{\'o}n}}}, \bibinfo {author} {\bibfnamefont {M.~H.}\ \bibnamefont {{Mu{\~n}oz-Arias}}}, \bibinfo {author} {\bibfnamefont {C.}~\bibnamefont {Hellings}}, \bibinfo {author} {\bibfnamefont {A.}~\bibnamefont {Flasby}}, \bibinfo {author} {\bibfnamefont {D.~C.}\ \bibnamefont {Zanuz}}, \bibinfo {author} {\bibfnamefont {J.-C.}\ \bibnamefont {Besse}}, \bibinfo {author} {\bibfnamefont {A.}~\bibnamefont {Blais}}, \bibinfo {author} {\bibfnamefont {D.}~\bibnamefont {Malz}}, \bibinfo {author} {\bibfnamefont {C.}~\bibnamefont {Eichler}},\ and\ \bibinfo {author} {\bibfnamefont {A.}~\bibnamefont {Wallraff}},\ }\href@noop {} {\bibinfo {title} {Deterministic generation of a 20-qubit
  two-dimensional photonic cluster state}} (\bibinfo {year} {2024}),\ \Eprint {https://arxiv.org/abs/2409.06623} {arXiv:2409.06623} \BibitemShut {NoStop}%
\bibitem [{\citenamefont {Somaschi}\ \emph {et~al.}(2016)\citenamefont {Somaschi}, \citenamefont {Giesz}, \citenamefont {{De Santis}}, \citenamefont {Loredo}, \citenamefont {Almeida}, \citenamefont {Hornecker}, \citenamefont {Portalupi}, \citenamefont {Grange}, \citenamefont {Ant{\'{o}}n}, \citenamefont {Demory}, \citenamefont {G{\'{o}}mez}, \citenamefont {Sagnes}, \citenamefont {Lanzillotti-Kimura}, \citenamefont {Lema{\'{i}}tre}, \citenamefont {Auffeves}, \citenamefont {White}, \citenamefont {Lanco},\ and\ \citenamefont {Senellart}}]{Somaschi2016}%
  \BibitemOpen
  \bibfield  {author} {\bibinfo {author} {\bibfnamefont {N.}~\bibnamefont {Somaschi}}, \bibinfo {author} {\bibfnamefont {V.}~\bibnamefont {Giesz}}, \bibinfo {author} {\bibfnamefont {L.}~\bibnamefont {{De Santis}}}, \bibinfo {author} {\bibfnamefont {J.~C.}\ \bibnamefont {Loredo}}, \bibinfo {author} {\bibfnamefont {M.~P.}\ \bibnamefont {Almeida}}, \bibinfo {author} {\bibfnamefont {G.}~\bibnamefont {Hornecker}}, \bibinfo {author} {\bibfnamefont {S.~L.}\ \bibnamefont {Portalupi}}, \bibinfo {author} {\bibfnamefont {T.}~\bibnamefont {Grange}}, \bibinfo {author} {\bibfnamefont {C.}~\bibnamefont {Ant{\'{o}}n}}, \bibinfo {author} {\bibfnamefont {J.}~\bibnamefont {Demory}}, \bibinfo {author} {\bibfnamefont {C.}~\bibnamefont {G{\'{o}}mez}}, \bibinfo {author} {\bibfnamefont {I.}~\bibnamefont {Sagnes}}, \bibinfo {author} {\bibfnamefont {N.~D.}\ \bibnamefont {Lanzillotti-Kimura}}, \bibinfo {author} {\bibfnamefont {A.}~\bibnamefont {Lema{\'{i}}tre}}, \bibinfo {author} {\bibfnamefont {A.}~\bibnamefont {Auffeves}}, \bibinfo
  {author} {\bibfnamefont {A.~G.}\ \bibnamefont {White}}, \bibinfo {author} {\bibfnamefont {L.}~\bibnamefont {Lanco}},\ and\ \bibinfo {author} {\bibfnamefont {P.}~\bibnamefont {Senellart}},\ }\bibfield  {title} {\bibinfo {title} {{Near-optimal single-photon sources in the solid state}},\ }\href {https://doi.org/10.1038/nphoton.2016.23} {\bibfield  {journal} {\bibinfo  {journal} {Nature Photonics}\ }\textbf {\bibinfo {volume} {10}},\ \bibinfo {pages} {340} (\bibinfo {year} {2016})}\BibitemShut {NoStop}%
\bibitem [{\citenamefont {Tomm}\ \emph {et~al.}(2021)\citenamefont {Tomm}, \citenamefont {Javadi}, \citenamefont {Antoniadis}, \citenamefont {Najer}, \citenamefont {L{\"{o}}bl}, \citenamefont {Korsch}, \citenamefont {Schott}, \citenamefont {Valentin}, \citenamefont {Wieck}, \citenamefont {Ludwig},\ and\ \citenamefont {Warburton}}]{Tomm2021}%
  \BibitemOpen
  \bibfield  {author} {\bibinfo {author} {\bibfnamefont {N.}~\bibnamefont {Tomm}}, \bibinfo {author} {\bibfnamefont {A.}~\bibnamefont {Javadi}}, \bibinfo {author} {\bibfnamefont {N.~O.}\ \bibnamefont {Antoniadis}}, \bibinfo {author} {\bibfnamefont {D.}~\bibnamefont {Najer}}, \bibinfo {author} {\bibfnamefont {M.~C.}\ \bibnamefont {L{\"{o}}bl}}, \bibinfo {author} {\bibfnamefont {A.~R.}\ \bibnamefont {Korsch}}, \bibinfo {author} {\bibfnamefont {R.}~\bibnamefont {Schott}}, \bibinfo {author} {\bibfnamefont {S.~R.}\ \bibnamefont {Valentin}}, \bibinfo {author} {\bibfnamefont {A.~D.}\ \bibnamefont {Wieck}}, \bibinfo {author} {\bibfnamefont {A.}~\bibnamefont {Ludwig}},\ and\ \bibinfo {author} {\bibfnamefont {R.~J.}\ \bibnamefont {Warburton}},\ }\bibfield  {title} {\bibinfo {title} {{A bright and fast source of coherent single photons}},\ }\href {https://doi.org/10.1038/s41565-020-00831-x} {\bibfield  {journal} {\bibinfo  {journal} {Nature Nanotechnology}\ }\textbf {\bibinfo {volume} {16}},\ \bibinfo {pages} {399}
  (\bibinfo {year} {2021})},\ \Eprint {https://arxiv.org/abs/2007.12654} {2007.12654} \BibitemShut {NoStop}%
\bibitem [{\citenamefont {Ding}\ \emph {et~al.}(2025)\citenamefont {Ding}, \citenamefont {Guo}, \citenamefont {Xu}, \citenamefont {Liu}, \citenamefont {Zou}, \citenamefont {Zhao}, \citenamefont {Ge}, \citenamefont {Zhang}, \citenamefont {Liu}, \citenamefont {Wang}, \citenamefont {Chen}, \citenamefont {Wang}, \citenamefont {He}, \citenamefont {Huo}, \citenamefont {Lu},\ and\ \citenamefont {Pan}}]{dingHighefficiencySinglephotonSource2025}%
  \BibitemOpen
  \bibfield  {author} {\bibinfo {author} {\bibfnamefont {X.}~\bibnamefont {Ding}}, \bibinfo {author} {\bibfnamefont {Y.-P.}\ \bibnamefont {Guo}}, \bibinfo {author} {\bibfnamefont {M.-C.}\ \bibnamefont {Xu}}, \bibinfo {author} {\bibfnamefont {R.-Z.}\ \bibnamefont {Liu}}, \bibinfo {author} {\bibfnamefont {G.-Y.}\ \bibnamefont {Zou}}, \bibinfo {author} {\bibfnamefont {J.-Y.}\ \bibnamefont {Zhao}}, \bibinfo {author} {\bibfnamefont {Z.-X.}\ \bibnamefont {Ge}}, \bibinfo {author} {\bibfnamefont {Q.-H.}\ \bibnamefont {Zhang}}, \bibinfo {author} {\bibfnamefont {H.-L.}\ \bibnamefont {Liu}}, \bibinfo {author} {\bibfnamefont {L.-J.}\ \bibnamefont {Wang}}, \bibinfo {author} {\bibfnamefont {M.-C.}\ \bibnamefont {Chen}}, \bibinfo {author} {\bibfnamefont {H.}~\bibnamefont {Wang}}, \bibinfo {author} {\bibfnamefont {Y.-M.}\ \bibnamefont {He}}, \bibinfo {author} {\bibfnamefont {Y.-H.}\ \bibnamefont {Huo}}, \bibinfo {author} {\bibfnamefont {C.-Y.}\ \bibnamefont {Lu}},\ and\ \bibinfo {author} {\bibfnamefont {J.-W.}\ \bibnamefont
  {Pan}},\ }\bibfield  {title} {\bibinfo {title} {High-efficiency single-photon source above the loss-tolerant threshold for efficient linear optical quantum computing},\ }\href {https://doi.org/10.1038/s41566-025-01639-8} {\bibfield  {journal} {\bibinfo  {journal} {Nature Photonics}\ ,\ \bibinfo {pages} {1}} (\bibinfo {year} {2025})}\BibitemShut {NoStop}%
\bibitem [{\citenamefont {Pettersson}\ \emph {et~al.}(2024)\citenamefont {Pettersson}, \citenamefont {S{\o}rensen},\ and\ \citenamefont {Paesani}}]{petterssonDeterministicGenerationConcatenated2024}%
  \BibitemOpen
  \bibfield  {author} {\bibinfo {author} {\bibfnamefont {L.~A.}\ \bibnamefont {Pettersson}}, \bibinfo {author} {\bibfnamefont {A.~S.}\ \bibnamefont {S{\o}rensen}},\ and\ \bibinfo {author} {\bibfnamefont {S.}~\bibnamefont {Paesani}},\ }\href@noop {} {\bibinfo {title} {Deterministic generation of concatenated graph codes from quantum emitters}} (\bibinfo {year} {2024}),\ \Eprint {https://arxiv.org/abs/2406.16684} {arXiv:2406.16684} \BibitemShut {NoStop}%
\bibitem [{\citenamefont {Hilaire}\ \emph {et~al.}(2023)\citenamefont {Hilaire}, \citenamefont {Vidro}, \citenamefont {Eisenberg},\ and\ \citenamefont {Economou}}]{hilaireNeardeterministicHybridGeneration2023}%
  \BibitemOpen
  \bibfield  {author} {\bibinfo {author} {\bibfnamefont {P.}~\bibnamefont {Hilaire}}, \bibinfo {author} {\bibfnamefont {L.}~\bibnamefont {Vidro}}, \bibinfo {author} {\bibfnamefont {H.~S.}\ \bibnamefont {Eisenberg}},\ and\ \bibinfo {author} {\bibfnamefont {S.~E.}\ \bibnamefont {Economou}},\ }\bibfield  {title} {\bibinfo {title} {Near-deterministic hybrid generation of arbitrary photonic graph states using a single quantum emitter and linear optics},\ }\href {https://doi.org/10.22331/q-2023-04-27-992} {\bibfield  {journal} {\bibinfo  {journal} {Quantum}\ }\textbf {\bibinfo {volume} {7}},\ \bibinfo {pages} {992} (\bibinfo {year} {2023})}\BibitemShut {NoStop}%
\bibitem [{\citenamefont {Dousse}\ \emph {et~al.}(2008)\citenamefont {Dousse}, \citenamefont {Lanco}, \citenamefont {Suffczy{\'{n}}ski}, \citenamefont {Semenova}, \citenamefont {Miard}, \citenamefont {Lema{\^{i}}tre}, \citenamefont {Sagnes}, \citenamefont {Roblin}, \citenamefont {Bloch},\ and\ \citenamefont {Senellart}}]{Dousse2008}%
  \BibitemOpen
  \bibfield  {author} {\bibinfo {author} {\bibfnamefont {A.}~\bibnamefont {Dousse}}, \bibinfo {author} {\bibfnamefont {L.}~\bibnamefont {Lanco}}, \bibinfo {author} {\bibfnamefont {J.}~\bibnamefont {Suffczy{\'{n}}ski}}, \bibinfo {author} {\bibfnamefont {E.}~\bibnamefont {Semenova}}, \bibinfo {author} {\bibfnamefont {A.}~\bibnamefont {Miard}}, \bibinfo {author} {\bibfnamefont {A.}~\bibnamefont {Lema{\^{i}}tre}}, \bibinfo {author} {\bibfnamefont {I.}~\bibnamefont {Sagnes}}, \bibinfo {author} {\bibfnamefont {C.}~\bibnamefont {Roblin}}, \bibinfo {author} {\bibfnamefont {J.}~\bibnamefont {Bloch}},\ and\ \bibinfo {author} {\bibfnamefont {P.}~\bibnamefont {Senellart}},\ }\bibfield  {title} {\bibinfo {title} {{Controlled light-matter coupling for a single quantum dot embedded in a pillar microcavity using far-field optical lithography}},\ }\href {https://doi.org/10.1103/PhysRevLett.101.267404} {\bibfield  {journal} {\bibinfo  {journal} {Physical Review Letters}\ }\textbf {\bibinfo {volume} {101}},\ \bibinfo {pages}
  {30} (\bibinfo {year} {2008})}\BibitemShut {NoStop}%
\bibitem [{\citenamefont {Thomas}\ \emph {et~al.}(2021)\citenamefont {Thomas}, \citenamefont {Billard}, \citenamefont {Coste}, \citenamefont {Wein}, \citenamefont {Priya}, \citenamefont {Ollivier}, \citenamefont {Krebs}, \citenamefont {Taza{\"{i}}rt}, \citenamefont {Harouri}, \citenamefont {Lemaitre}, \citenamefont {Sagnes}, \citenamefont {Anton}, \citenamefont {Lanco}, \citenamefont {Somaschi}, \citenamefont {Loredo},\ and\ \citenamefont {Senellart}}]{Thomas2021}%
  \BibitemOpen
  \bibfield  {author} {\bibinfo {author} {\bibfnamefont {S.~E.}\ \bibnamefont {Thomas}}, \bibinfo {author} {\bibfnamefont {M.}~\bibnamefont {Billard}}, \bibinfo {author} {\bibfnamefont {N.}~\bibnamefont {Coste}}, \bibinfo {author} {\bibfnamefont {S.~C.}\ \bibnamefont {Wein}}, \bibinfo {author} {\bibnamefont {Priya}}, \bibinfo {author} {\bibfnamefont {H.}~\bibnamefont {Ollivier}}, \bibinfo {author} {\bibfnamefont {O.}~\bibnamefont {Krebs}}, \bibinfo {author} {\bibfnamefont {L.}~\bibnamefont {Taza{\"{i}}rt}}, \bibinfo {author} {\bibfnamefont {A.}~\bibnamefont {Harouri}}, \bibinfo {author} {\bibfnamefont {A.}~\bibnamefont {Lemaitre}}, \bibinfo {author} {\bibfnamefont {I.}~\bibnamefont {Sagnes}}, \bibinfo {author} {\bibfnamefont {C.}~\bibnamefont {Anton}}, \bibinfo {author} {\bibfnamefont {L.}~\bibnamefont {Lanco}}, \bibinfo {author} {\bibfnamefont {N.}~\bibnamefont {Somaschi}}, \bibinfo {author} {\bibfnamefont {J.~C.}\ \bibnamefont {Loredo}},\ and\ \bibinfo {author} {\bibfnamefont {P.}~\bibnamefont
  {Senellart}},\ }\bibfield  {title} {\bibinfo {title} {{Bright Polarized Single-Photon Source Based on a Linear Dipole}},\ }\href {https://doi.org/10.1103/PhysRevLett.126.233601} {\bibfield  {journal} {\bibinfo  {journal} {Physical Review Letters}\ }\textbf {\bibinfo {volume} {126}},\ \bibinfo {pages} {1} (\bibinfo {year} {2021})}\BibitemShut {NoStop}%
\bibitem [{\citenamefont {Coste}\ \emph {et~al.}(2023{\natexlab{b}})\citenamefont {Coste}, \citenamefont {Gundin}, \citenamefont {Fioretto}, \citenamefont {Thomas}, \citenamefont {Millet}, \citenamefont {Mehdi}, \citenamefont {Somaschi}, \citenamefont {Morassi}, \citenamefont {Pont}, \citenamefont {Lema{\^i}tre}, \citenamefont {Belabas}, \citenamefont {Krebs}, \citenamefont {Lanco},\ and\ \citenamefont {Senellart}}]{costeProbingDynamicsCoherence2023}%
  \BibitemOpen
  \bibfield  {author} {\bibinfo {author} {\bibfnamefont {N.}~\bibnamefont {Coste}}, \bibinfo {author} {\bibfnamefont {M.}~\bibnamefont {Gundin}}, \bibinfo {author} {\bibfnamefont {D.~A.}\ \bibnamefont {Fioretto}}, \bibinfo {author} {\bibfnamefont {S.~E.}\ \bibnamefont {Thomas}}, \bibinfo {author} {\bibfnamefont {C.}~\bibnamefont {Millet}}, \bibinfo {author} {\bibfnamefont {E.}~\bibnamefont {Mehdi}}, \bibinfo {author} {\bibfnamefont {N.}~\bibnamefont {Somaschi}}, \bibinfo {author} {\bibfnamefont {M.}~\bibnamefont {Morassi}}, \bibinfo {author} {\bibfnamefont {M.}~\bibnamefont {Pont}}, \bibinfo {author} {\bibfnamefont {A.}~\bibnamefont {Lema{\^i}tre}}, \bibinfo {author} {\bibfnamefont {N.}~\bibnamefont {Belabas}}, \bibinfo {author} {\bibfnamefont {O.}~\bibnamefont {Krebs}}, \bibinfo {author} {\bibfnamefont {L.}~\bibnamefont {Lanco}},\ and\ \bibinfo {author} {\bibfnamefont {P.}~\bibnamefont {Senellart}},\ }\bibfield  {title} {\bibinfo {title} {Probing the dynamics and coherence of a semiconductor hole spin via
  acoustic phonon-assisted excitation},\ }\href {https://doi.org/10.1088/2058-9565/acbd6a} {\bibfield  {journal} {\bibinfo  {journal} {Quantum Science and Technology}\ }\textbf {\bibinfo {volume} {8}},\ \bibinfo {pages} {025021} (\bibinfo {year} {2023}{\natexlab{b}})}\BibitemShut {NoStop}%
\bibitem [{\citenamefont {Greilich}\ \emph {et~al.}(2009)\citenamefont {Greilich}, \citenamefont {Economou}, \citenamefont {Spatzek}, \citenamefont {Yakovlev}, \citenamefont {Reuter}, \citenamefont {Wieck}, \citenamefont {Reinecke},\ and\ \citenamefont {Bayer}}]{greilichUltrafastOpticalRotations2009}%
  \BibitemOpen
  \bibfield  {author} {\bibinfo {author} {\bibfnamefont {A.}~\bibnamefont {Greilich}}, \bibinfo {author} {\bibfnamefont {S.~E.}\ \bibnamefont {Economou}}, \bibinfo {author} {\bibfnamefont {S.}~\bibnamefont {Spatzek}}, \bibinfo {author} {\bibfnamefont {D.~R.}\ \bibnamefont {Yakovlev}}, \bibinfo {author} {\bibfnamefont {D.}~\bibnamefont {Reuter}}, \bibinfo {author} {\bibfnamefont {A.~D.}\ \bibnamefont {Wieck}}, \bibinfo {author} {\bibfnamefont {T.~L.}\ \bibnamefont {Reinecke}},\ and\ \bibinfo {author} {\bibfnamefont {M.}~\bibnamefont {Bayer}},\ }\bibfield  {title} {\bibinfo {title} {Ultrafast optical rotations of electron spins in quantum dots},\ }\href {https://doi.org/10.1038/nphys1226} {\bibfield  {journal} {\bibinfo  {journal} {Nature Physics}\ }\textbf {\bibinfo {volume} {5}},\ \bibinfo {pages} {262} (\bibinfo {year} {2009})}\BibitemShut {NoStop}%
\bibitem [{\citenamefont {Press}\ \emph {et~al.}(2008)\citenamefont {Press}, \citenamefont {Ladd}, \citenamefont {Zhang},\ and\ \citenamefont {Yamamoto}}]{Press2008}%
  \BibitemOpen
  \bibfield  {author} {\bibinfo {author} {\bibfnamefont {D.}~\bibnamefont {Press}}, \bibinfo {author} {\bibfnamefont {T.~D.}\ \bibnamefont {Ladd}}, \bibinfo {author} {\bibfnamefont {B.}~\bibnamefont {Zhang}},\ and\ \bibinfo {author} {\bibfnamefont {Y.}~\bibnamefont {Yamamoto}},\ }\bibfield  {title} {\bibinfo {title} {{Complete quantum control of a single quantum dot spin using ultrafast optical pulses}},\ }\href {https://doi.org/10.1038/nature07530} {\bibfield  {journal} {\bibinfo  {journal} {Nature}\ }\textbf {\bibinfo {volume} {456}},\ \bibinfo {pages} {218} (\bibinfo {year} {2008})}\BibitemShut {NoStop}%
\bibitem [{\citenamefont {Berezovsky}\ \emph {et~al.}(2008)\citenamefont {Berezovsky}, \citenamefont {Mikkelsen}, \citenamefont {Stoltz}, \citenamefont {Coldren},\ and\ \citenamefont {Awschalom}}]{Berezovsky2008}%
  \BibitemOpen
  \bibfield  {author} {\bibinfo {author} {\bibfnamefont {J.}~\bibnamefont {Berezovsky}}, \bibinfo {author} {\bibfnamefont {M.~H.}\ \bibnamefont {Mikkelsen}}, \bibinfo {author} {\bibfnamefont {N.~G.}\ \bibnamefont {Stoltz}}, \bibinfo {author} {\bibfnamefont {L.~A.}\ \bibnamefont {Coldren}},\ and\ \bibinfo {author} {\bibfnamefont {D.~D.}\ \bibnamefont {Awschalom}},\ }\bibfield  {title} {\bibinfo {title} {{Picosecond coherent optical manipulation of a single electron spin in a quantum dot}},\ }\href {https://doi.org/10.1126/science.1154798} {\bibfield  {journal} {\bibinfo  {journal} {Science}\ }\textbf {\bibinfo {volume} {320}},\ \bibinfo {pages} {349} (\bibinfo {year} {2008})}\BibitemShut {NoStop}%
\bibitem [{\citenamefont {Stockill}\ \emph {et~al.}(2016)\citenamefont {Stockill}, \citenamefont {{Le Gall}}, \citenamefont {Matthiesen}, \citenamefont {Huthmacher}, \citenamefont {Clarke}, \citenamefont {Hugues},\ and\ \citenamefont {Atat{\"{u}}re}}]{Stockill2016}%
  \BibitemOpen
  \bibfield  {author} {\bibinfo {author} {\bibfnamefont {R.}~\bibnamefont {Stockill}}, \bibinfo {author} {\bibfnamefont {C.}~\bibnamefont {{Le Gall}}}, \bibinfo {author} {\bibfnamefont {C.}~\bibnamefont {Matthiesen}}, \bibinfo {author} {\bibfnamefont {L.}~\bibnamefont {Huthmacher}}, \bibinfo {author} {\bibfnamefont {E.}~\bibnamefont {Clarke}}, \bibinfo {author} {\bibfnamefont {M.}~\bibnamefont {Hugues}},\ and\ \bibinfo {author} {\bibfnamefont {M.}~\bibnamefont {Atat{\"{u}}re}},\ }\bibfield  {title} {\bibinfo {title} {{Quantum dot spin coherence governed by a strained nuclear environment}},\ }\href {https://doi.org/10.1038/ncomms12745} {\bibfield  {journal} {\bibinfo  {journal} {Nature Communications}\ }\textbf {\bibinfo {volume} {7}},\ \bibinfo {pages} {1} (\bibinfo {year} {2016})}\BibitemShut {NoStop}%
\bibitem [{\citenamefont {Ramesh}\ \emph {et~al.}(2025)\citenamefont {Ramesh}, \citenamefont {Annoni}, \citenamefont {Margaria}, \citenamefont {Fioretto}, \citenamefont {Pishchagin}, \citenamefont {Morassi}, \citenamefont {Lema{\^i}tre}, \citenamefont {Doty}, \citenamefont {Senellart}, \citenamefont {Lanco}, \citenamefont {Belabas}, \citenamefont {Wein},\ and\ \citenamefont {Krebs}}]{rameshImpactHole$g$factor2025}%
  \BibitemOpen
  \bibfield  {author} {\bibinfo {author} {\bibfnamefont {P.~R.}\ \bibnamefont {Ramesh}}, \bibinfo {author} {\bibfnamefont {E.}~\bibnamefont {Annoni}}, \bibinfo {author} {\bibfnamefont {N.}~\bibnamefont {Margaria}}, \bibinfo {author} {\bibfnamefont {D.~A.}\ \bibnamefont {Fioretto}}, \bibinfo {author} {\bibfnamefont {A.}~\bibnamefont {Pishchagin}}, \bibinfo {author} {\bibfnamefont {M.}~\bibnamefont {Morassi}}, \bibinfo {author} {\bibfnamefont {A.}~\bibnamefont {Lema{\^i}tre}}, \bibinfo {author} {\bibfnamefont {M.~F.}\ \bibnamefont {Doty}}, \bibinfo {author} {\bibfnamefont {P.}~\bibnamefont {Senellart}}, \bibinfo {author} {\bibfnamefont {L.}~\bibnamefont {Lanco}}, \bibinfo {author} {\bibfnamefont {N.}~\bibnamefont {Belabas}}, \bibinfo {author} {\bibfnamefont {S.~C.}\ \bibnamefont {Wein}},\ and\ \bibinfo {author} {\bibfnamefont {O.}~\bibnamefont {Krebs}},\ }\href@noop {} {\bibinfo {title} {The impact of hole $g$-factor anisotropy on spin-photon entanglement generation with {{InGaAs}} quantum dots}} (\bibinfo
  {year} {2025}),\ \Eprint {https://arxiv.org/abs/2502.07627} {arXiv:2502.07627} \BibitemShut {NoStop}%
\bibitem [{\citenamefont {Wein}(2024)}]{weinSimulatingPhotonCounting2024}%
  \BibitemOpen
  \bibfield  {author} {\bibinfo {author} {\bibfnamefont {S.~C.}\ \bibnamefont {Wein}},\ }\bibfield  {title} {\bibinfo {title} {Simulating photon counting from dynamic quantum emitters by exploiting zero-photon measurements},\ }\href {https://doi.org/10.1103/PhysRevA.109.023713} {\bibfield  {journal} {\bibinfo  {journal} {Physical Review A}\ }\textbf {\bibinfo {volume} {109}},\ \bibinfo {pages} {023713} (\bibinfo {year} {2024})}\BibitemShut {NoStop}%
\bibitem [{\citenamefont {Browne}\ and\ \citenamefont {Rudolph}(2005)}]{browneResourceEfficientLinearOptical2005}%
  \BibitemOpen
  \bibfield  {author} {\bibinfo {author} {\bibfnamefont {D.~E.}\ \bibnamefont {Browne}}\ and\ \bibinfo {author} {\bibfnamefont {T.}~\bibnamefont {Rudolph}},\ }\bibfield  {title} {\bibinfo {title} {Resource-{{Efficient Linear Optical Quantum Computation}}},\ }\href {https://doi.org/10.1103/PhysRevLett.95.010501} {\bibfield  {journal} {\bibinfo  {journal} {Physical Review Letters}\ }\textbf {\bibinfo {volume} {95}},\ \bibinfo {pages} {010501} (\bibinfo {year} {2005})}\BibitemShut {NoStop}%
\bibitem [{\citenamefont {{Herrera-Mart{\'i}}}\ \emph {et~al.}(2010)\citenamefont {{Herrera-Mart{\'i}}}, \citenamefont {Fowler}, \citenamefont {Jennings},\ and\ \citenamefont {Rudolph}}]{herrera-martiPhotonicImplementationTopological2010}%
  \BibitemOpen
  \bibfield  {author} {\bibinfo {author} {\bibfnamefont {D.~A.}\ \bibnamefont {{Herrera-Mart{\'i}}}}, \bibinfo {author} {\bibfnamefont {A.~G.}\ \bibnamefont {Fowler}}, \bibinfo {author} {\bibfnamefont {D.}~\bibnamefont {Jennings}},\ and\ \bibinfo {author} {\bibfnamefont {T.}~\bibnamefont {Rudolph}},\ }\bibfield  {title} {\bibinfo {title} {Photonic implementation for the topological cluster-state quantum computer},\ }\href {https://doi.org/10.1103/PhysRevA.82.032332} {\bibfield  {journal} {\bibinfo  {journal} {Physical Review A}\ }\textbf {\bibinfo {volume} {82}},\ \bibinfo {pages} {032332} (\bibinfo {year} {2010})}\BibitemShut {NoStop}%
\bibitem [{\citenamefont {Chan}\ \emph {et~al.}(2024)\citenamefont {Chan}, \citenamefont {Bell}, \citenamefont {Pettersson}, \citenamefont {Chen}, \citenamefont {Yard}, \citenamefont {S{\o}rensen},\ and\ \citenamefont {Paesani}}]{chanTailoringFusionbasedPhotonic2024}%
  \BibitemOpen
  \bibfield  {author} {\bibinfo {author} {\bibfnamefont {M.~L.}\ \bibnamefont {Chan}}, \bibinfo {author} {\bibfnamefont {T.~J.}\ \bibnamefont {Bell}}, \bibinfo {author} {\bibfnamefont {L.~A.}\ \bibnamefont {Pettersson}}, \bibinfo {author} {\bibfnamefont {S.~X.}\ \bibnamefont {Chen}}, \bibinfo {author} {\bibfnamefont {P.}~\bibnamefont {Yard}}, \bibinfo {author} {\bibfnamefont {A.~S.}\ \bibnamefont {S{\o}rensen}},\ and\ \bibinfo {author} {\bibfnamefont {S.}~\bibnamefont {Paesani}},\ }\href@noop {} {\bibinfo {title} {Tailoring fusion-based photonic quantum computing schemes to quantum emitters}} (\bibinfo {year} {2024}),\ \Eprint {https://arxiv.org/abs/2410.06784} {arXiv:2410.06784} \BibitemShut {NoStop}%
\bibitem [{\citenamefont {Margaria}\ \emph {et~al.}(2024)\citenamefont {Margaria}, \citenamefont {Pastier}, \citenamefont {Bennour}, \citenamefont {Billard}, \citenamefont {Ivanov}, \citenamefont {Hease}, \citenamefont {Stepanov}, \citenamefont {Adiyatullin}, \citenamefont {Singla}, \citenamefont {Pont}, \citenamefont {Descampeaux}, \citenamefont {Bernard}, \citenamefont {Pishchagin}, \citenamefont {Morassi}, \citenamefont {Lema{\^i}tre}, \citenamefont {Volz}, \citenamefont {Giesz}, \citenamefont {Somaschi}, \citenamefont {Maring}, \citenamefont {Boissier}, \citenamefont {Au},\ and\ \citenamefont {Senellart}}]{margariaEfficientFiberpigtailedSource2024}%
  \BibitemOpen
  \bibfield  {author} {\bibinfo {author} {\bibfnamefont {N.}~\bibnamefont {Margaria}}, \bibinfo {author} {\bibfnamefont {F.}~\bibnamefont {Pastier}}, \bibinfo {author} {\bibfnamefont {T.}~\bibnamefont {Bennour}}, \bibinfo {author} {\bibfnamefont {M.}~\bibnamefont {Billard}}, \bibinfo {author} {\bibfnamefont {E.}~\bibnamefont {Ivanov}}, \bibinfo {author} {\bibfnamefont {W.}~\bibnamefont {Hease}}, \bibinfo {author} {\bibfnamefont {P.}~\bibnamefont {Stepanov}}, \bibinfo {author} {\bibfnamefont {A.~F.}\ \bibnamefont {Adiyatullin}}, \bibinfo {author} {\bibfnamefont {R.}~\bibnamefont {Singla}}, \bibinfo {author} {\bibfnamefont {M.}~\bibnamefont {Pont}}, \bibinfo {author} {\bibfnamefont {M.}~\bibnamefont {Descampeaux}}, \bibinfo {author} {\bibfnamefont {A.}~\bibnamefont {Bernard}}, \bibinfo {author} {\bibfnamefont {A.}~\bibnamefont {Pishchagin}}, \bibinfo {author} {\bibfnamefont {M.}~\bibnamefont {Morassi}}, \bibinfo {author} {\bibfnamefont {A.}~\bibnamefont {Lema{\^i}tre}}, \bibinfo {author} {\bibfnamefont
  {T.}~\bibnamefont {Volz}}, \bibinfo {author} {\bibfnamefont {V.}~\bibnamefont {Giesz}}, \bibinfo {author} {\bibfnamefont {N.}~\bibnamefont {Somaschi}}, \bibinfo {author} {\bibfnamefont {N.}~\bibnamefont {Maring}}, \bibinfo {author} {\bibfnamefont {S.}~\bibnamefont {Boissier}}, \bibinfo {author} {\bibfnamefont {T.~H.}\ \bibnamefont {Au}},\ and\ \bibinfo {author} {\bibfnamefont {P.}~\bibnamefont {Senellart}},\ }\href@noop {} {\bibinfo {title} {Efficient fiber-pigtailed source of indistinguishable single photons}} (\bibinfo {year} {2024}),\ \Eprint {https://arxiv.org/abs/2410.07760} {arXiv:2410.07760} \BibitemShut {NoStop}%
\bibitem [{\citenamefont {Maring}\ \emph {et~al.}(2024)\citenamefont {Maring}, \citenamefont {Fyrillas}, \citenamefont {Pont}, \citenamefont {Ivanov}, \citenamefont {Stepanov}, \citenamefont {Margaria}, \citenamefont {Hease}, \citenamefont {Pishchagin}, \citenamefont {Lema{\^{i}}tre}, \citenamefont {Sagnes}, \citenamefont {Au}, \citenamefont {Boissier}, \citenamefont {Bertasi}, \citenamefont {Baert}, \citenamefont {Valdivia}, \citenamefont {Billard}, \citenamefont {Acar}, \citenamefont {Brieussel}, \citenamefont {Mezher}, \citenamefont {Wein}, \citenamefont {Salavrakos}, \citenamefont {Sinnott}, \citenamefont {Fioretto}, \citenamefont {Emeriau}, \citenamefont {Belabas}, \citenamefont {Mansfield}, \citenamefont {Senellart}, \citenamefont {Senellart},\ and\ \citenamefont {Somaschi}}]{Maring2024}%
  \BibitemOpen
  \bibfield  {author} {\bibinfo {author} {\bibfnamefont {N.}~\bibnamefont {Maring}}, \bibinfo {author} {\bibfnamefont {A.}~\bibnamefont {Fyrillas}}, \bibinfo {author} {\bibfnamefont {M.}~\bibnamefont {Pont}}, \bibinfo {author} {\bibfnamefont {E.}~\bibnamefont {Ivanov}}, \bibinfo {author} {\bibfnamefont {P.}~\bibnamefont {Stepanov}}, \bibinfo {author} {\bibfnamefont {N.}~\bibnamefont {Margaria}}, \bibinfo {author} {\bibfnamefont {W.}~\bibnamefont {Hease}}, \bibinfo {author} {\bibfnamefont {A.}~\bibnamefont {Pishchagin}}, \bibinfo {author} {\bibfnamefont {A.}~\bibnamefont {Lema{\^{i}}tre}}, \bibinfo {author} {\bibfnamefont {I.}~\bibnamefont {Sagnes}}, \bibinfo {author} {\bibfnamefont {T.~H.}\ \bibnamefont {Au}}, \bibinfo {author} {\bibfnamefont {S.}~\bibnamefont {Boissier}}, \bibinfo {author} {\bibfnamefont {E.}~\bibnamefont {Bertasi}}, \bibinfo {author} {\bibfnamefont {A.}~\bibnamefont {Baert}}, \bibinfo {author} {\bibfnamefont {M.}~\bibnamefont {Valdivia}}, \bibinfo {author} {\bibfnamefont {M.}~\bibnamefont
  {Billard}}, \bibinfo {author} {\bibfnamefont {O.}~\bibnamefont {Acar}}, \bibinfo {author} {\bibfnamefont {A.}~\bibnamefont {Brieussel}}, \bibinfo {author} {\bibfnamefont {R.}~\bibnamefont {Mezher}}, \bibinfo {author} {\bibfnamefont {S.~C.}\ \bibnamefont {Wein}}, \bibinfo {author} {\bibfnamefont {A.}~\bibnamefont {Salavrakos}}, \bibinfo {author} {\bibfnamefont {P.}~\bibnamefont {Sinnott}}, \bibinfo {author} {\bibfnamefont {D.~A.}\ \bibnamefont {Fioretto}}, \bibinfo {author} {\bibfnamefont {P.~E.}\ \bibnamefont {Emeriau}}, \bibinfo {author} {\bibfnamefont {N.}~\bibnamefont {Belabas}}, \bibinfo {author} {\bibfnamefont {S.}~\bibnamefont {Mansfield}}, \bibinfo {author} {\bibfnamefont {P.}~\bibnamefont {Senellart}}, \bibinfo {author} {\bibfnamefont {J.}~\bibnamefont {Senellart}},\ and\ \bibinfo {author} {\bibfnamefont {N.}~\bibnamefont {Somaschi}},\ }\bibfield  {title} {\bibinfo {title} {{A versatile single-photon-based quantum computing platform}},\ }\href {https://doi.org/10.1038/s41566-024-01403-4} {\bibfield
  {journal} {\bibinfo  {journal} {Nature Photonics}\ }\textbf {\bibinfo {volume} {18}},\ \bibinfo {pages} {603} (\bibinfo {year} {2024})}\BibitemShut {NoStop}%
\bibitem [{\citenamefont {Steindl}\ \emph {et~al.}(2023)\citenamefont {Steindl}, \citenamefont {{van der Ent}}, \citenamefont {{van der Meer}}, \citenamefont {Frey}, \citenamefont {Norman}, \citenamefont {Bowers}, \citenamefont {Bouwmeester},\ and\ \citenamefont {L{\"o}ffler}}]{steindlResonantTwoLaserSpinState2023}%
  \BibitemOpen
  \bibfield  {author} {\bibinfo {author} {\bibfnamefont {P.}~\bibnamefont {Steindl}}, \bibinfo {author} {\bibfnamefont {T.}~\bibnamefont {{van der Ent}}}, \bibinfo {author} {\bibfnamefont {H.}~\bibnamefont {{van der Meer}}}, \bibinfo {author} {\bibfnamefont {J.}~\bibnamefont {Frey}}, \bibinfo {author} {\bibfnamefont {J.}~\bibnamefont {Norman}}, \bibinfo {author} {\bibfnamefont {J.}~\bibnamefont {Bowers}}, \bibinfo {author} {\bibfnamefont {D.}~\bibnamefont {Bouwmeester}},\ and\ \bibinfo {author} {\bibfnamefont {W.}~\bibnamefont {L{\"o}ffler}},\ }\bibfield  {title} {\bibinfo {title} {Resonant {{Two-Laser Spin-State Spectroscopy}} of a {{Negatively Charged Quantum-Dot--Microcavity System}} with a {{Cold Permanent Magnet}}},\ }\href {https://doi.org/10.1103/PhysRevApplied.20.014026} {\bibfield  {journal} {\bibinfo  {journal} {Physical Review Applied}\ }\textbf {\bibinfo {volume} {20}},\ \bibinfo {pages} {014026} (\bibinfo {year} {2023})}\BibitemShut {NoStop}%
\bibitem [{\citenamefont {Greilich}\ \emph {et~al.}(2007)\citenamefont {Greilich}, \citenamefont {Shabaev}, \citenamefont {Yakovlev}, \citenamefont {Efros}, \citenamefont {Yugova}, \citenamefont {Reuter}, \citenamefont {Wieck},\ and\ \citenamefont {Bayer}}]{Greilich2007}%
  \BibitemOpen
  \bibfield  {author} {\bibinfo {author} {\bibfnamefont {A.}~\bibnamefont {Greilich}}, \bibinfo {author} {\bibfnamefont {A.}~\bibnamefont {Shabaev}}, \bibinfo {author} {\bibfnamefont {D.~R.}\ \bibnamefont {Yakovlev}}, \bibinfo {author} {\bibfnamefont {A.~L.}\ \bibnamefont {Efros}}, \bibinfo {author} {\bibfnamefont {I.~A.}\ \bibnamefont {Yugova}}, \bibinfo {author} {\bibfnamefont {D.}~\bibnamefont {Reuter}}, \bibinfo {author} {\bibfnamefont {A.~D.}\ \bibnamefont {Wieck}},\ and\ \bibinfo {author} {\bibfnamefont {M.}~\bibnamefont {Bayer}},\ }\bibfield  {title} {\bibinfo {title} {{Nuclei-Induced Frequency Focusing of Electron Spin Coherence}},\ }\href {https://doi.org/10.1126/science.1146850} {\bibfield  {journal} {\bibinfo  {journal} {Science}\ }\textbf {\bibinfo {volume} {317}},\ \bibinfo {pages} {1896} (\bibinfo {year} {2007})}\BibitemShut {NoStop}%
\bibitem [{\citenamefont {{\'{E}}thier-Majcher}\ \emph {et~al.}(2017)\citenamefont {{\'{E}}thier-Majcher}, \citenamefont {Gangloff}, \citenamefont {Stockill}, \citenamefont {Clarke}, \citenamefont {Hugues}, \citenamefont {{Le Gall}},\ and\ \citenamefont {Atat{\"{u}}re}}]{Ethier-Majcher2017a}%
  \BibitemOpen
  \bibfield  {author} {\bibinfo {author} {\bibfnamefont {G.}~\bibnamefont {{\'{E}}thier-Majcher}}, \bibinfo {author} {\bibfnamefont {D.}~\bibnamefont {Gangloff}}, \bibinfo {author} {\bibfnamefont {R.}~\bibnamefont {Stockill}}, \bibinfo {author} {\bibfnamefont {E.}~\bibnamefont {Clarke}}, \bibinfo {author} {\bibfnamefont {M.}~\bibnamefont {Hugues}}, \bibinfo {author} {\bibfnamefont {C.}~\bibnamefont {{Le Gall}}},\ and\ \bibinfo {author} {\bibfnamefont {M.}~\bibnamefont {Atat{\"{u}}re}},\ }\bibfield  {title} {\bibinfo {title} {{Improving a Solid-State Qubit through an Engineered Mesoscopic Environment}},\ }\href {https://doi.org/10.1103/PhysRevLett.119.130503} {\bibfield  {journal} {\bibinfo  {journal} {Physical Review Letters}\ }\textbf {\bibinfo {volume} {119}},\ \bibinfo {pages} {1} (\bibinfo {year} {2017})}\BibitemShut {NoStop}%
\bibitem [{\citenamefont {Gangloff}\ \emph {et~al.}(2019)\citenamefont {Gangloff}, \citenamefont {{\'{E}}thier-Majcher}, \citenamefont {Lang}, \citenamefont {Denning}, \citenamefont {Bodey}, \citenamefont {Jackson}, \citenamefont {Clarke}, \citenamefont {Hugues}, \citenamefont {{Le Gall}},\ and\ \citenamefont {Atat{\"{u}}re}}]{Gangloff2019}%
  \BibitemOpen
  \bibfield  {author} {\bibinfo {author} {\bibfnamefont {D.~A.}\ \bibnamefont {Gangloff}}, \bibinfo {author} {\bibfnamefont {G.}~\bibnamefont {{\'{E}}thier-Majcher}}, \bibinfo {author} {\bibfnamefont {C.}~\bibnamefont {Lang}}, \bibinfo {author} {\bibfnamefont {E.~V.}\ \bibnamefont {Denning}}, \bibinfo {author} {\bibfnamefont {J.~H.}\ \bibnamefont {Bodey}}, \bibinfo {author} {\bibfnamefont {D.~M.}\ \bibnamefont {Jackson}}, \bibinfo {author} {\bibfnamefont {E.}~\bibnamefont {Clarke}}, \bibinfo {author} {\bibfnamefont {M.}~\bibnamefont {Hugues}}, \bibinfo {author} {\bibfnamefont {C.}~\bibnamefont {{Le Gall}}},\ and\ \bibinfo {author} {\bibfnamefont {M.}~\bibnamefont {Atat{\"{u}}re}},\ }\bibfield  {title} {\bibinfo {title} {{Quantum interface of an electron and a nuclear ensemble}},\ }\href {https://doi.org/10.1126/science.aaw2906} {\bibfield  {journal} {\bibinfo  {journal} {Science}\ }\textbf {\bibinfo {volume} {364}},\ \bibinfo {pages} {62} (\bibinfo {year} {2019})}\BibitemShut {NoStop}%
\bibitem [{\citenamefont {Prechtel}\ \emph {et~al.}(2016)\citenamefont {Prechtel}, \citenamefont {Kuhlmann}, \citenamefont {Houel}, \citenamefont {Ludwig}, \citenamefont {Valentin}, \citenamefont {Wieck},\ and\ \citenamefont {Warburton}}]{Prechtel2016a}%
  \BibitemOpen
  \bibfield  {author} {\bibinfo {author} {\bibfnamefont {J.~H.}\ \bibnamefont {Prechtel}}, \bibinfo {author} {\bibfnamefont {A.~V.}\ \bibnamefont {Kuhlmann}}, \bibinfo {author} {\bibfnamefont {J.}~\bibnamefont {Houel}}, \bibinfo {author} {\bibfnamefont {A.}~\bibnamefont {Ludwig}}, \bibinfo {author} {\bibfnamefont {S.~R.}\ \bibnamefont {Valentin}}, \bibinfo {author} {\bibfnamefont {A.~D.}\ \bibnamefont {Wieck}},\ and\ \bibinfo {author} {\bibfnamefont {R.~J.}\ \bibnamefont {Warburton}},\ }\bibfield  {title} {\bibinfo {title} {{Decoupling a hole spin qubit from the nuclear spins}},\ }\href {https://doi.org/10.1038/nmat4704} {\bibfield  {journal} {\bibinfo  {journal} {Nature Materials}\ }\textbf {\bibinfo {volume} {15}},\ \bibinfo {pages} {981} (\bibinfo {year} {2016})}\BibitemShut {NoStop}%
\bibitem [{\citenamefont {Hahn}(1950)}]{Hahn1950}%
  \BibitemOpen
  \bibfield  {author} {\bibinfo {author} {\bibfnamefont {E.~L.}\ \bibnamefont {Hahn}},\ }\bibfield  {title} {\bibinfo {title} {{Spin echoes}},\ }\href {https://doi.org/10.1103/PhysRev.80.580} {\bibfield  {journal} {\bibinfo  {journal} {Physical Review}\ }\textbf {\bibinfo {volume} {80}},\ \bibinfo {pages} {580} (\bibinfo {year} {1950})}\BibitemShut {NoStop}%
\bibitem [{\citenamefont {Press}\ \emph {et~al.}(2010)\citenamefont {Press}, \citenamefont {{De Greve}}, \citenamefont {McMahon}, \citenamefont {Ladd}, \citenamefont {Friess}, \citenamefont {Schneider}, \citenamefont {Kamp}, \citenamefont {H{\"{o}}fling}, \citenamefont {Forchel},\ and\ \citenamefont {Yamamoto}}]{Press2010}%
  \BibitemOpen
  \bibfield  {author} {\bibinfo {author} {\bibfnamefont {D.}~\bibnamefont {Press}}, \bibinfo {author} {\bibfnamefont {K.}~\bibnamefont {{De Greve}}}, \bibinfo {author} {\bibfnamefont {P.~L.}\ \bibnamefont {McMahon}}, \bibinfo {author} {\bibfnamefont {T.~D.}\ \bibnamefont {Ladd}}, \bibinfo {author} {\bibfnamefont {B.}~\bibnamefont {Friess}}, \bibinfo {author} {\bibfnamefont {C.}~\bibnamefont {Schneider}}, \bibinfo {author} {\bibfnamefont {M.}~\bibnamefont {Kamp}}, \bibinfo {author} {\bibfnamefont {S.}~\bibnamefont {H{\"{o}}fling}}, \bibinfo {author} {\bibfnamefont {A.}~\bibnamefont {Forchel}},\ and\ \bibinfo {author} {\bibfnamefont {Y.}~\bibnamefont {Yamamoto}},\ }\bibfield  {title} {\bibinfo {title} {{Ultrafast optical spin echo in a single quantum dot}},\ }\href {https://doi.org/10.1038/nphoton.2010.83} {\bibfield  {journal} {\bibinfo  {journal} {Nature Photonics}\ }\textbf {\bibinfo {volume} {4}},\ \bibinfo {pages} {367} (\bibinfo {year} {2010})}\BibitemShut {NoStop}%
\bibitem [{\citenamefont {Wein}\ \emph {et~al.}(2024)\citenamefont {Wein}, \citenamefont {de~Brugi{\`e}re}, \citenamefont {Music}, \citenamefont {Senellart}, \citenamefont {Bourdoncle},\ and\ \citenamefont {Mansfield}}]{weinMinimizingResourceOverhead2024}%
  \BibitemOpen
  \bibfield  {author} {\bibinfo {author} {\bibfnamefont {S.~C.}\ \bibnamefont {Wein}}, \bibinfo {author} {\bibfnamefont {T.~G.}\ \bibnamefont {de~Brugi{\`e}re}}, \bibinfo {author} {\bibfnamefont {L.}~\bibnamefont {Music}}, \bibinfo {author} {\bibfnamefont {P.}~\bibnamefont {Senellart}}, \bibinfo {author} {\bibfnamefont {B.}~\bibnamefont {Bourdoncle}},\ and\ \bibinfo {author} {\bibfnamefont {S.}~\bibnamefont {Mansfield}},\ }\href@noop {} {\bibinfo {title} {Minimizing resource overhead in fusion-based quantum computation using hybrid spin-photon devices}} (\bibinfo {year} {2024}),\ \Eprint {https://arxiv.org/abs/2412.08611} {arXiv:2412.08611} \BibitemShut {NoStop}%
\bibitem [{\citenamefont {de~Gliniasty}\ \emph {et~al.}(2024)\citenamefont {de~Gliniasty}, \citenamefont {Hilaire}, \citenamefont {Emeriau}, \citenamefont {Wein}, \citenamefont {Salavrakos},\ and\ \citenamefont {Mansfield}}]{gliniastySpinOpticalQuantumComputing2024}%
  \BibitemOpen
  \bibfield  {author} {\bibinfo {author} {\bibfnamefont {G.}~\bibnamefont {de~Gliniasty}}, \bibinfo {author} {\bibfnamefont {P.}~\bibnamefont {Hilaire}}, \bibinfo {author} {\bibfnamefont {P.-E.}\ \bibnamefont {Emeriau}}, \bibinfo {author} {\bibfnamefont {S.~C.}\ \bibnamefont {Wein}}, \bibinfo {author} {\bibfnamefont {A.}~\bibnamefont {Salavrakos}},\ and\ \bibinfo {author} {\bibfnamefont {S.}~\bibnamefont {Mansfield}},\ }\bibfield  {title} {\bibinfo {title} {A {{Spin-Optical Quantum Computing Architecture}}},\ }\href {https://doi.org/10.22331/q-2024-07-24-1423} {\bibfield  {journal} {\bibinfo  {journal} {Quantum}\ }\textbf {\bibinfo {volume} {8}},\ \bibinfo {pages} {1423} (\bibinfo {year} {2024})}\BibitemShut {NoStop}%
\bibitem [{\citenamefont {Hilaire}\ \emph {et~al.}(2024)\citenamefont {Hilaire}, \citenamefont {Dessertaine}, \citenamefont {Bourdoncle}, \citenamefont {Denys}, \citenamefont {de~Gliniasty}, \citenamefont {{Valent{\'i}-Rojas}},\ and\ \citenamefont {Mansfield}}]{hilaireEnhancedFaulttolerancePhotonic2024}%
  \BibitemOpen
  \bibfield  {author} {\bibinfo {author} {\bibfnamefont {P.}~\bibnamefont {Hilaire}}, \bibinfo {author} {\bibfnamefont {T.}~\bibnamefont {Dessertaine}}, \bibinfo {author} {\bibfnamefont {B.}~\bibnamefont {Bourdoncle}}, \bibinfo {author} {\bibfnamefont {A.}~\bibnamefont {Denys}}, \bibinfo {author} {\bibfnamefont {G.}~\bibnamefont {de~Gliniasty}}, \bibinfo {author} {\bibfnamefont {G.}~\bibnamefont {{Valent{\'i}-Rojas}}},\ and\ \bibinfo {author} {\bibfnamefont {S.}~\bibnamefont {Mansfield}},\ }\href@noop {} {\bibinfo {title} {Enhanced {{Fault-tolerance}} in {{Photonic Quantum Computing}}: {{Floquet Code Outperforms Surface Code}} in {{Tailored Architecture}}}} (\bibinfo {year} {2024}),\ \Eprint {https://arxiv.org/abs/2410.07065} {arXiv:2410.07065} \BibitemShut {NoStop}%
\bibitem [{\citenamefont {Bauch}\ \emph {et~al.}(2024)\citenamefont {Bauch}, \citenamefont {K{\"o}cher}, \citenamefont {Heinisch},\ and\ \citenamefont {Schumacher}}]{bauchTimebinEntanglementDeterministic2024}%
  \BibitemOpen
  \bibfield  {author} {\bibinfo {author} {\bibfnamefont {D.}~\bibnamefont {Bauch}}, \bibinfo {author} {\bibfnamefont {N.}~\bibnamefont {K{\"o}cher}}, \bibinfo {author} {\bibfnamefont {N.}~\bibnamefont {Heinisch}},\ and\ \bibinfo {author} {\bibfnamefont {S.}~\bibnamefont {Schumacher}},\ }\bibfield  {title} {\bibinfo {title} {Time-bin entanglement in the deterministic generation of linear photonic cluster states},\ }\href {https://doi.org/10.1063/5.0214197} {\bibfield  {journal} {\bibinfo  {journal} {APL Quantum}\ }\textbf {\bibinfo {volume} {1}},\ \bibinfo {pages} {036110} (\bibinfo {year} {2024})}\BibitemShut {NoStop}%
\bibitem [{\citenamefont {Wein}\ \emph {et~al.}(2020)\citenamefont {Wein}, \citenamefont {Ji}, \citenamefont {Wu}, \citenamefont {Kimiaee~Asadi}, \citenamefont {Ghobadi},\ and\ \citenamefont {Simon}}]{weinAnalyzingPhotoncountHeralded2020}%
  \BibitemOpen
  \bibfield  {author} {\bibinfo {author} {\bibfnamefont {S.~C.}\ \bibnamefont {Wein}}, \bibinfo {author} {\bibfnamefont {J.-W.}\ \bibnamefont {Ji}}, \bibinfo {author} {\bibfnamefont {Y.-F.}\ \bibnamefont {Wu}}, \bibinfo {author} {\bibfnamefont {F.}~\bibnamefont {Kimiaee~Asadi}}, \bibinfo {author} {\bibfnamefont {R.}~\bibnamefont {Ghobadi}},\ and\ \bibinfo {author} {\bibfnamefont {C.}~\bibnamefont {Simon}},\ }\bibfield  {title} {\bibinfo {title} {Analyzing photon-count heralded entanglement generation between solid-state spin qubits by decomposing the master-equation dynamics},\ }\href {https://doi.org/10.1103/PhysRevA.102.033701} {\bibfield  {journal} {\bibinfo  {journal} {Physical Review A}\ }\textbf {\bibinfo {volume} {102}},\ \bibinfo {pages} {033701} (\bibinfo {year} {2020})}\BibitemShut {NoStop}%
\bibitem [{\citenamefont {Johansson}\ \emph {et~al.}(2012)\citenamefont {Johansson}, \citenamefont {Nation},\ and\ \citenamefont {Nori}}]{johanssonQuTiPOpensourcePython2012}%
  \BibitemOpen
  \bibfield  {author} {\bibinfo {author} {\bibfnamefont {J.~R.}\ \bibnamefont {Johansson}}, \bibinfo {author} {\bibfnamefont {P.~D.}\ \bibnamefont {Nation}},\ and\ \bibinfo {author} {\bibfnamefont {F.}~\bibnamefont {Nori}},\ }\bibfield  {title} {\bibinfo {title} {{{QuTiP}}: {{An}} open-source {{Python}} framework for the dynamics of open quantum systems},\ }\href {https://doi.org/10.1016/j.cpc.2012.02.021} {\bibfield  {journal} {\bibinfo  {journal} {Computer Physics Communications}\ }\textbf {\bibinfo {volume} {183}},\ \bibinfo {pages} {1760} (\bibinfo {year} {2012})}\BibitemShut {NoStop}%
\end{thebibliography}%

\clearpage 

\begin{widetext}

\section*{Supplementary information}
\setcounter{figure}{0}
\renewcommand{\thefigure}{S\arabic{figure}}

\subsection{Spin-photon state evolution}

We detail in this section the spin-photon state evolution in the ideal case (no decoherence and instantaneous photon emission) for various configurations of the experimental protocol, following spin $\ket{\downarrow}$ (only for the linear cluster configuration) or $\ket{\uparrow}$  state initialization. We use $E_S$ to denote the photon emission process conditioned on the spin state, whereas $R_y(\pi/2)$ and $Z$ are defined in the main text. \\

\textbf{State evolution 4-partite linear cluster state:}\\

\begin{fleqn}
\begin{gather}
\nonumber \ket{\psi_1}=\ket{\uparrow}\\ 
\nonumber \xrightarrow{R_y(\pi/2)}\dfrac{1} {\sqrt2}\left(\ket{\uparrow}+\ket{\downarrow}\right)\\
\nonumber \xrightarrow{E_s}\ket{\psi_2}= \dfrac{1}{\sqrt2}\left(\ket{R_2}\ket{\uparrow}+\ket{L_2}\ket{\downarrow}\right)\\ \nonumber \xrightarrow{R_y(\pi/2)} \dfrac{1} {\sqrt2}\left(\ket{R_2}\dfrac{1} {\sqrt2}(\ket{\uparrow}+\ket{\downarrow})+\ket{L_2}\dfrac{1} {\sqrt2}(-\ket{\uparrow}+\ket{\downarrow})\right)\\
\nonumber \xrightarrow{E_s}\ket{\psi_3}= \dfrac{1} {\sqrt2}\left(\ket{R_2}\dfrac{1} {\sqrt2}(\ket{R_3}\ket{\uparrow}+\ket{L_3}\ket{\downarrow})+\ket{L_2}\dfrac{1} {\sqrt2}(-\ket{R_3}\ket{\uparrow}+\ket{L_3}\ket{\downarrow})\right)\\
\nonumber  = \dfrac{1}{\sqrt2}\left( \dfrac{1}{\sqrt2}(\ket{R_2}-\ket{L_2})\ket{R_3}\ket{\uparrow} + \dfrac{1}{\sqrt2}(\ket{R_2}+\ket{L_2})\ket{L_3}\ket{\downarrow} \right) = \dfrac{1}{\sqrt2}\left(\ket{-_2,R_{3}}\ket{\uparrow}+\ket{+_2,L_{3}}\ket{\downarrow}\right)\\
\nonumber \xrightarrow{R_y(\pi/2)} \dfrac{1}{\sqrt2}\left(\ket{-_2,R_3}\dfrac{1} {\sqrt2}(\ket{\uparrow}+\ket{\downarrow})+\ket{+_2,L_3}\dfrac{1} {\sqrt2}(-\ket{\uparrow}+\ket{\downarrow})\right)\\
\nonumber \xrightarrow{E_s}\ket{\psi_4}= \dfrac{1}{2}\left(\ket{-_2,R_3}(\ket{R_4}\ket{\uparrow}+\ket{L_4}\ket{\downarrow})+\ket{+_2,L_3}(-\ket{R_4}\ket{\uparrow}+\ket{L_4}\ket{\downarrow})\right)\\
 \nonumber =\dfrac{1}{2}( \left(\ket{-_2,R_3}-\ket{+_2,L_3}\right)\ket{R_4}\ket{\uparrow} +\left(\ket{-_2,R_3}+\ket{+_2,L_3}\right)\ket{L_4}\ket{\downarrow})\\\nonumber 
\end{gather}
\end{fleqn}\\

\begin{fleqn}
\begin{gather}
\nonumber \ket{\psi_1}=\ket{\downarrow}\\ 
\nonumber \xrightarrow{R_y(\pi/2)}\dfrac{1} {\sqrt2}\left(-\ket{\uparrow}+\ket{\downarrow}\right)\\
\nonumber \xrightarrow{E_s}\ket{\psi_2}= \dfrac{1}{\sqrt2}\left(-\ket{R_2}\ket{\uparrow}+\ket{L_2}\ket{\downarrow}\right)\\ \nonumber \xrightarrow{R_y(\pi/2)} \dfrac{1} {\sqrt2}\left(-\ket{R_2}\dfrac{1} {\sqrt2}(\ket{\uparrow}+\ket{\downarrow})+\ket{L_2}\dfrac{1} {\sqrt2}(-\ket{\uparrow}+\ket{\downarrow})\right)\\
\nonumber \xrightarrow{E_s}\ket{\psi_3}= \dfrac{1} {\sqrt2}\left(-\ket{R_2}\dfrac{1} {\sqrt2}(\ket{R_3}\ket{\uparrow}+\ket{L_3}\ket{\downarrow})+\ket{L_2}\dfrac{1} {\sqrt2}(-\ket{R_3}\ket{\uparrow}+\ket{L_3}\ket{\downarrow})\right)\\
\nonumber  = \dfrac{-1}{\sqrt2}\left( \dfrac{1}{\sqrt2}(\ket{R_2}+\ket{L_2})\ket{R_3}\ket{\uparrow} + \dfrac{1}{\sqrt2}(\ket{R_2}-\ket{L_2})\ket{L_3}\ket{\downarrow} \right) = \dfrac{-1}{\sqrt2}\left(\ket{+_2,R_{3}}\ket{\uparrow}+\ket{-_2,L_{3}}\ket{\downarrow}\right)\\
\nonumber \xrightarrow{R_y(\pi/2)} \dfrac{-1}{\sqrt2}\left(\ket{+_2,R_3}\dfrac{1} {\sqrt2}(\ket{\uparrow}+\ket{\downarrow})+\ket{-_2,L_3}\dfrac{1} {\sqrt2}(-\ket{\uparrow}+\ket{\downarrow})\right)\\
\nonumber \xrightarrow{E_s}\ket{\psi_4}= \dfrac{-1}{2}\left(\ket{+_2,R_3}(\ket{R_4}\ket{\uparrow}+\ket{L_4}\ket{\downarrow})+\ket{-_2,L_3}(-\ket{R_4}\ket{\uparrow}+\ket{L_4}\ket{\downarrow})\right)\\
 \nonumber =\dfrac{-1}{2}( \left(\ket{+_2,R_3}-\ket{-_2,L_3}\right)\ket{R_4}\ket{\uparrow} +\left(\ket{+_2,R_3}+\ket{-_2,L_3}\right)\ket{L_4}\ket{\downarrow})\\\nonumber 
\end{gather}
\end{fleqn}\\

\textbf{State evolution 4-partite GHZ state:}

\begin{fleqn}
\begin{gather}
\nonumber \ket{\psi_1}=\ket{\uparrow}\\
\nonumber \xrightarrow{R_y(\pi/2)}\dfrac{1} {\sqrt2}\left(\ket{\uparrow}+\ket{\downarrow}\right)\\
\nonumber \xrightarrow{E_s}\ket{\psi_2}= \dfrac{1}{\sqrt2}\left(\ket{R_2}\ket{\uparrow}+\ket{L_2}\ket{\downarrow}\right)\\
\nonumber \xrightarrow{Z} \dfrac{1}{\sqrt2}\left(-\ket{R_2}\ket{\uparrow}+\ket{L_2}\ket{\downarrow}\right)\\
\nonumber \xrightarrow{E_s}\ket{\psi_3}= \dfrac{1}{\sqrt2}\left(-\ket{R_2,R_3}\ket{\uparrow}+\ket{L_2,L_3}\ket{\downarrow}\right)\\
\nonumber \xrightarrow{Z} \dfrac{1}{\sqrt2}\left(\ket{R_2,R_3}\ket{\uparrow}+\ket{L_2,L_3}\ket{\downarrow}\right)\\
\nonumber \xrightarrow{E_s}\ket{\psi_4}= \dfrac{1}{\sqrt2}\left(\ket{R_2,R_3}\ket{R_4}\ket{\uparrow}+\ket{L_2,L_3}\ket{L_4}\ket{\downarrow}\right)\\\nonumber 
\end{gather}
\end{fleqn}\\

\textbf{State evolution redundantly encoded 4-partite linear cluster state \#1: }

\begin{fleqn}
\begin{gather}
\nonumber \ket{\psi_1}=\ket{\uparrow}\\
\nonumber \xrightarrow{R_y(\pi/2)}\dfrac{1} {\sqrt2}\left(\ket{\uparrow}+\ket{\downarrow}\right)\\
\nonumber \xrightarrow{E_s}\ket{\psi_2}=\dfrac{1}{\sqrt2}\left(\ket{R_2}\ket{\uparrow}+\ket{L_2}\ket{\downarrow}\right)\\
\nonumber \xrightarrow{Z} \dfrac{1}{\sqrt2}\left(-\ket{R_2}\ket{\uparrow}+\ket{L_2}\ket{\downarrow}\right)\\
\nonumber \xrightarrow{E_s} \ket{\psi_3}=\dfrac{1}{\sqrt2}\left(-\ket{R_2,R_3}\ket{\uparrow}+\ket{L_2,L_3}\ket{\downarrow}\right)\\
\nonumber \xrightarrow{R_y(\pi/2)} \dfrac{1}{\sqrt2}\left(-\ket{R_2,R_3}\dfrac{1} {\sqrt2}\left(\ket{\uparrow}+\ket{\downarrow}\right)+\ket{L_2,L_3}\dfrac{1} {\sqrt2}\left(-\ket{\uparrow}+\ket{\downarrow}\right)\right)\\
\nonumber \xrightarrow{E_s}\ket{\psi_4}= \dfrac{1}{\sqrt2}\left(-\ket{R_2,R_3}\dfrac{1} {\sqrt2}\left(\ket{R_4}\ket{\uparrow}+\ket{L_4}\ket{\downarrow}\right)+\ket{L_2,L_3}\dfrac{1} {\sqrt2}\left(-\ket{R_4}\ket{\uparrow}+\ket{L_4}\ket{\downarrow}\right)\right)\\
 \nonumber = \dfrac{-1}{2}\left(\ket{R_2,R_3} +  \ket{L_2,L_3}\ket{R_4}\ket{\uparrow} + 
\ket{R_2,R_3} - \ket{L_2,L_3}\ket{L_4}\ket{\downarrow}\right)\\\nonumber 
\end{gather}
\end{fleqn}\\

\clearpage

\textbf{State evolution redundantly encoded 4-partite linear cluster state \#2:}

\begin{fleqn}
\begin{gather}
\nonumber \ket{\psi_1}=\ket{\uparrow}\\
\nonumber \xrightarrow{R_y(\pi/2)}\dfrac{1} {\sqrt2}\left(\ket{\uparrow}+\ket{\downarrow}\right)\\
\nonumber \xrightarrow{E_s}\ket{\psi_2}= \dfrac{1}{\sqrt2}\left(\ket{R_2}\ket{\uparrow}+\ket{L_2}\ket{\downarrow}\right)\\ \nonumber \xrightarrow{R_y(\pi/2)} \dfrac{1} {\sqrt2}\left(\ket{R_2}\dfrac{1} {\sqrt2}(\ket{\uparrow}+\ket{\downarrow})+\ket{L_2}\dfrac{1} {\sqrt2}(-\ket{\uparrow}+\ket{\downarrow})\right)\\
\nonumber \xrightarrow{E_s}\ket{\psi_3}= \dfrac{1} {\sqrt2}\left(\ket{R_2}\dfrac{1} {\sqrt2}(\ket{R_3}\ket{\uparrow}+\ket{L_3}\ket{\downarrow})+\ket{L_2}\dfrac{1} {\sqrt2}(-\ket{R_3}\ket{\uparrow}+\ket{L_3}\ket{\downarrow})\right)\\
 \nonumber = \dfrac{1}{\sqrt2}\left( \dfrac{1}{\sqrt2}(\ket{R_2}-\ket{L_2})\ket{R_3}\ket{\uparrow} + \dfrac{1}{\sqrt2}(\ket{R_2}+\ket{L_2})\ket{L_3}\ket{\downarrow} \right) = \dfrac{1}{\sqrt2}\left(\ket{-_2,R_{3}}\ket{\uparrow}+\ket{+_2,L_{3}}\ket{\downarrow}\right)\\
 \nonumber \xrightarrow{Z} \dfrac{1}{\sqrt2}\left(-\ket{-_2,R_{3}}\ket{\uparrow}+\ket{+_2,L_{3}}\ket{\downarrow}\right)\\
\nonumber \xrightarrow{E_s}\ket{\psi_4}= \dfrac{1}{\sqrt2}\left(-\ket{-_2,R_{3}}\ket{R_4}\ket{\uparrow}+\ket{+_2,L_{3}}\ket{L_4}\ket{\downarrow}\right)\\\nonumber 
\end{gather}
\end{fleqn}\\

\clearpage

\subsection{Caterpillar state generation with a single quantum emitter.}  

Here we explain the local equivalence between caterpillar graph states and the graph states generated in this work. \\

At any time the state of a caterpillar graph can be represented as,
$$
\ket{G} = \ket{\uparrow}\ket{G\backslash \{s\}} + \ket{\downarrow} \ket{\overline{G\backslash \{s\}}},
$$
(omitting normalization factors) where $\ket{G\backslash \{s\}}$ is the subgraph state without the spin qubit node and
$$
\ket{\overline{G\backslash \{s\}}} = \prod_{i \in \mathcal{N}_s}Z_i \ket{G\backslash \{s\}},
$$
where we apply a $Z$ rotation on all the (photonic) qubits connected by an edge to the spin node, i.e.~the neighboring set $\mathcal N_s$ of the spin node, in the graph $G$.

As was shown in Ref.~\cite{hilaireNeardeterministicHybridGeneration2023}, it is possible to construct caterpillar states from four operations on the spin of a quantum emitter: initialization in the $\ket{+}=\ket{\uparrow + \downarrow}$ state, photon emission $E_s = \ket{R,\uparrow}\bra{\uparrow}+\ket{L,\downarrow}\bra{\downarrow}$, Hadamard gate $H$  and spin measurement in either the $\ket{\pm}$ basis or the $\ket{\uparrow / \downarrow}$ basis. In this work, we use a spin rotation $R_y(\pi/2)$ gate instead of a $H$ gate. Additionally, an OSRP (with $\varphi=\pi$) positioned exactly in between two consecutive photon emissions effectively perform a $Z$ gate. Finally, spin initialization in $\ket{\pm}$ is achieved by heralding the spin in the $\ket{\uparrow}$ or $\ket{\downarrow}$ state, followed by a $R_y(\pi/2)$ spin rotation. We thus need to ensure that these gates are sufficient to produce caterpillar graph states up to single qubit rotation on the photonic states. To do so, we show that the general state after a sequence $R_y(\pi/2)E_s$ (read from right to left) is local-Clifford equivalent to the one after $H E_s$. Similarly, the state resulting from the sequence $Z E_s$ is local-Clifford equivalent to the one after $E_s$.\\

\paragraph{Local-Clifford equivalence between the $E_s$ and $Z E_s$ operations.}\mbox{}\\

After a photon emission $E_s$ we should obtain the state
$$
E_s \ket{G} = \ket{\uparrow}\ket{R}_p\ket{G\backslash \{s\}} + \ket{\downarrow} \ket{L}_p \ket{\overline{G\backslash \{s\}}},
$$
which is close to the one obtained after a photon emission and an OSRP:
$$
Z E_s \ket{G} = \ket{\uparrow}\ket{R}_p\ket{G\backslash \{s\}} - \ket{\downarrow} \ket{L}_p \ket{\overline{G\backslash \{s\}}},
$$
and can be corrected by a $Z_p = \ket{R}\bra{R} - \ket{L}\bra{L}$ gate on the newly emitted photon $p$:
$$
Z_p Z E_s \ket{G} = E_s \ket{G},
$$

\paragraph{Local-Clifford equivalence between the $H E_s$ and $R_y(\pi/2) E_s$ operations.}\mbox{}\\

Following a photon emission and a Hadamard gate we obtain:
$$
H E_s \ket{G} = \ket{+}\ket{R}_p\ket{G\backslash \{s\}} + \ket{-} \ket{L}_p \ket{\overline{G\backslash \{s\}}},
$$
which is also close to the state obtained after a photon emission and a $R_y(\pi/2)$ rotation:
$$
R_y(\pi/2) E_s \ket{G} = \ket{+}\ket{R}_p\ket{G\backslash \{s\}} - \ket{-} \ket{L}_p \ket{\overline{G\backslash \{s\}}},
$$
and can be corrected similarly:
$$
Z_p  R_y(\pi/2) E_s \ket{G} = H E_s \ket{G},
$$
We thus see that by applying $Z$ corrections on the newly emitted photons, we can mimic the effect of the operations required for the generation of caterpillar graph states. 

\clearpage

\subsection{Extended data}

Fig.~\ref{fig:measured cluster imag} presents the measured imaginary part of the two-photon density matrix for all the states defined in the main text, complementing the real part of the density matrices shown in Fig.~\ref{fig:cluster}.

\begin{figure}[h]
\includegraphics[scale=0.36]{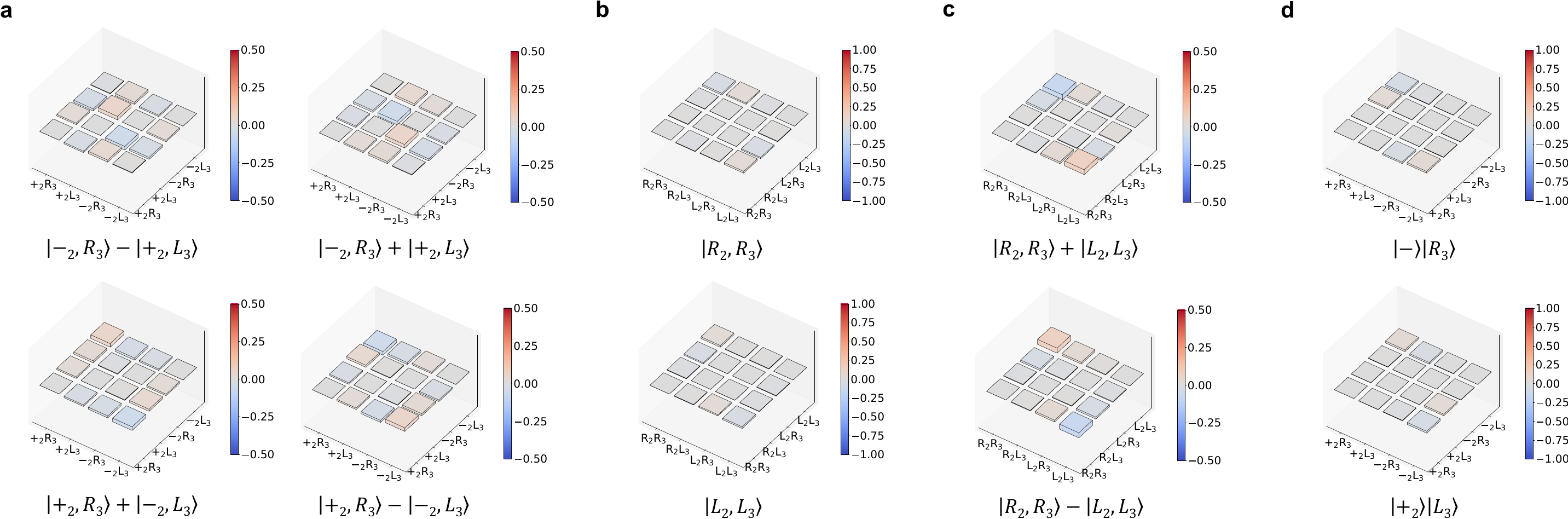}
\caption{\textbf{Measured imaginary part two-photon density matrices} \textbf{a-d}, Measured imaginary part of the two-photon density matrix for different 4-partite states generated : linear cluster (\textbf{a}), GHZ (\textbf{b}), and two redundantly encoded linear cluster (\textbf{c,e}), corresponding to the experimental results shown in Fig.~\ref{fig:cluster}\textbf{b-e}.}
\label{fig:measured cluster imag}
\end{figure}

Fig.~\ref{fig:HOM} presents measurements of the second-order correlation function, $g^{(2)}(0)$ , and Hong-Ou-Mandel (HOM) visibility to characterize the single-photon purity and indistinguishability, respectively. Both measurements are performed using a Mach-Zehnder interferometer with a 12.3 ns delay in one arm, matching the 81 MHz repetition rate of the pulsed laser. Input polarization is set using a linear polarizer, and a set of waveplates in each arm is used for polarization control. One arm is blocked when measuring $g^{(2)}(0)$. The HOM visibility is measured for two different experimental configurations: no magnetic field without OSRP and with a 60 mT magnetic field amplitude with OSRP.

\begin{figure}[h!]
\includegraphics[scale=0.55]{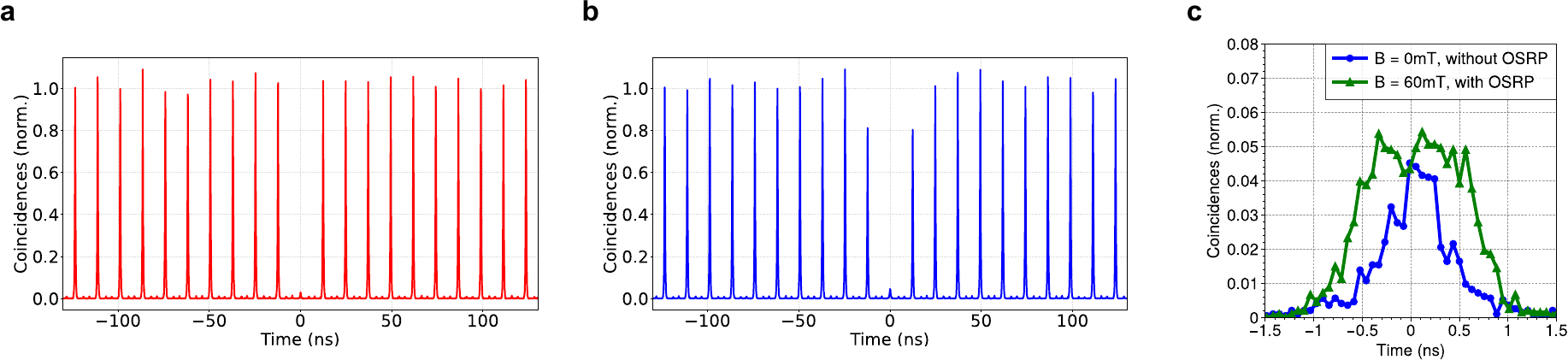}
\caption{\textbf{Photon purity and indistinguishability.} \textbf{a}, Second-order auto-correlation histogram showing a single photon purity $g^{(2)}(0) = 0.033 \pm 0.002$. 
\textbf{b}, Correlation histogram showing the indistinguishability of consecutively emitted photons at $B = 0$ mT and without applying OSRP. We extract a wave packet overlap of $M = 0.927 \pm 0.005$, corrected for the non-zero measured $g^{(2)}(0)$. \textbf{c}, Zero delay peak of the correlation histogram for two experimental configurations: no magnetic field without OSRP (blue circles) and for a 60mT magnetic field amplitude with OSRP (green triangles). The extracting wave packet overlaps are $M = 0.927 \pm 0.005$, and $0.827 \pm 0.007$ respectively, showing a high photon indistinguishability, albeit reduced in the presence of magnetic field and OSRP}.
\label{fig:HOM}
\end{figure}

\clearpage

\subsection{Simulated data}

\begin{figure}[h]
\includegraphics[scale=0.36]{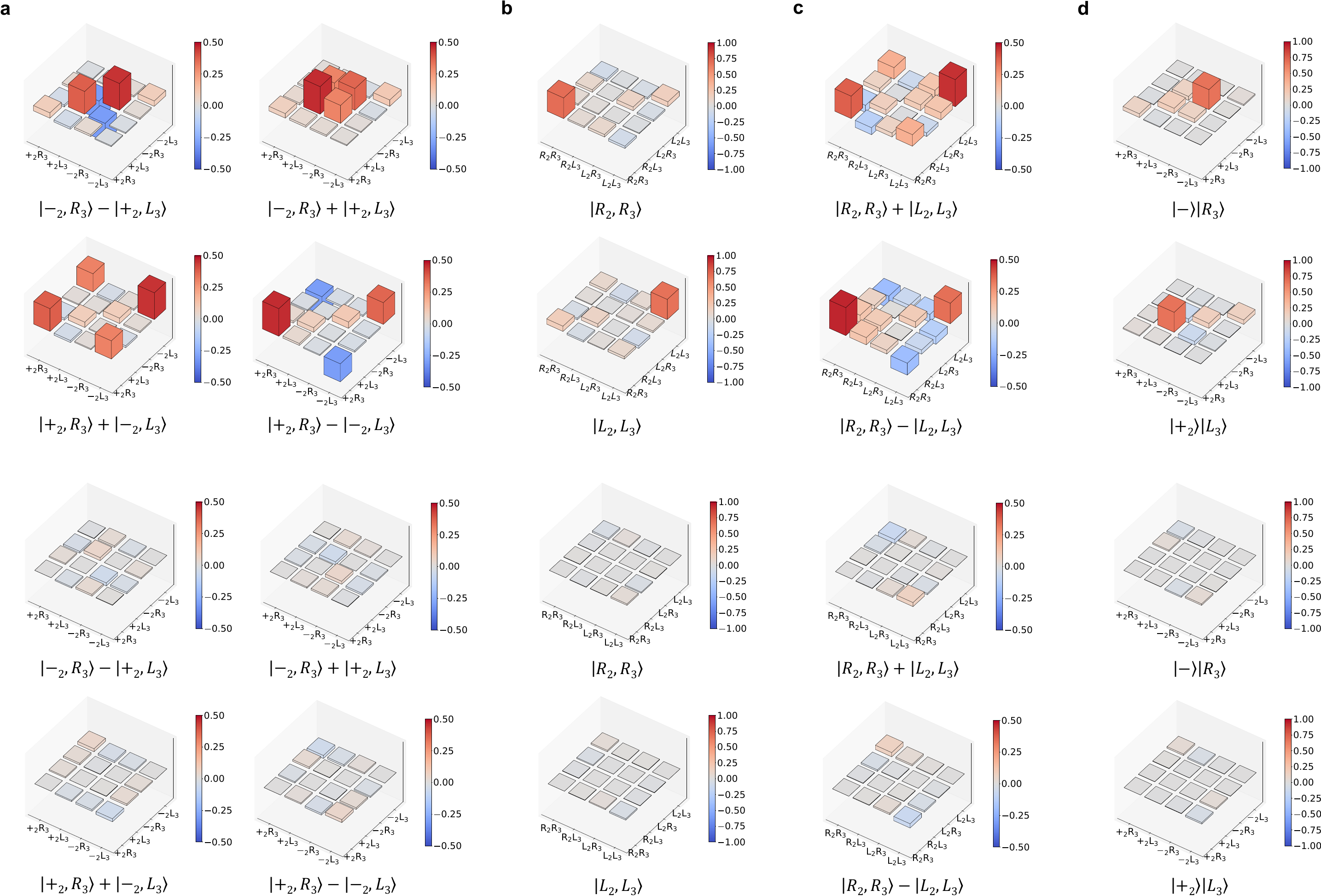}
\caption{\textbf{Simulated two-photon density matrices} \textbf{a-d}, Simulated real (top) and imaginary (bottom) part of the two-photon density matrix for different 4-partite states generated : linear cluster (\textbf{a}), GHZ (\textbf{b}), and two redundantly encoded linear cluster (\textbf{c,e}), corresponding to the experimental results shown in Fig.~\ref{fig:cluster}\textbf{b-e}.}
\label{fig:simulated all}
\end{figure}

In order to explain the experimental results and estimate the 4-partite entanglement fidelity, we use a physically motivated model to simulate the evolution of the quantum light-matter state. The parameters of the model are then determined by minimizing the error (maximizing the state fidelity) of the simulated data with respect to the experimental data.

We use the same basic trion model as detailed in the Supplementary information of Ref. \cite{costeHighrateEntanglementSemiconductor2023a}, which we summarise here. The Hamiltonian for this model is
\begin{equation*}
    H = \frac{\Delta_e}{2}\sigma_y^{(e)}+\frac{\Delta_h}{2}\sigma_y^{(h)}
\end{equation*}
where $\sigma_y^{(e)}=i(\ket{\downarrow}\bra{\uparrow}-\ket{\uparrow}\bra{\downarrow})$ and $\sigma_y^{(h)}=i(\ket{\downarrow\uparrow\Downarrow}\bra{\downarrow\uparrow\Uparrow}-\ket{\downarrow\uparrow\Uparrow}\bra{\downarrow\uparrow\Downarrow})$ are the electron and hole Pauli $y$ operators, respectively, that capture the impact of the magnetic field. The Zeeman splittings $\Delta_e=\mu_Bg_eB$ and $\Delta_h=\mu_Bg_hB$ are described by an isotropic coupling to a static magnetic field $B$, where $g_e$ ($g_h$) is the effective electron (hole) g factor and $\mu_B$ is the Bohr magneton. As in Ref. \cite{costeHighrateEntanglementSemiconductor2023a}, we also model a fluctuating Overhauser field ($B_{OH}$) impacting the electronic state through an additional Zeeman Hamiltonian
\begin{equation*}
    H_O =\frac{1}{2}g_e\mu_B\vec{B}_{OH}\cdot\vec{\sigma}^{(e)},
\end{equation*}
where $\vec{\sigma}^{(e)}=(\sigma_x^{(e)}, \sigma_y^{(e)}, \sigma_z^{(e)})$ is the vector of electronic Pauli operators and  $\vec{B}_{OH}$ is assumed to be isotropic and normally distributed. The impact of spontaneous emission is captured using a Markovian master equation of the form
\begin{equation*}
    \frac{d}{dt}\rho(t) = -\frac{i}{\hbar}\left[H+H_O,\rho(t)\right]+\gamma\mathcal{D}_{\sigma_R}\rho(t)+\gamma\mathcal{D}_{\sigma_L}\rho(t) = \mathcal{L}\rho(t)
\end{equation*}
where $\mathcal{L}$ is the system Liouvillian, $\gamma=1/T_1$ is the Purcell-enhanced decay rate of the trion state, $\mathcal{D}_\sigma\rho=\sigma\rho\sigma^\dagger - \sigma^\dagger\sigma\rho/2-\rho\sigma^\dagger\sigma/2$ is the action of the Lindblad dissipator superoperator and $\sigma_{R}=\ket{\uparrow}\bra{\downarrow\uparrow\Uparrow}$ ($\sigma_{L}=\ket{\downarrow}\bra{\downarrow\uparrow\Downarrow}$) is the optical lowering operator coupled to right (left) circularly-polarized light.

We model the optical excitation pulses as being instantaneous and linearly polarized, corresponding to the application of the unitary rotation
\begin{equation*}
    R_\text{ex}(\varphi)=\exp\left(-i\pi(\cos(\varphi)\sigma_{y,H}+\sin(\varphi)\sigma_{y,V})/2\right),
\end{equation*}
where $\sigma_{y,H}=-i(\sigma_H-\sigma_H^\dagger)$, $\sigma_{y,V}=-i(\sigma_V-\sigma_V^\dagger)$ for $\sigma_H=(\sigma_L+\sigma_R)/\sqrt{2}$ and $\sigma_V=-i(\sigma_L-\sigma_R)/\sqrt{2}$. Similarly, we model the optical spin rotation as a unitary phase rotation
\begin{equation*}
    R_\text{osrp}(\theta) = \exp\left(-i\theta\sigma^{(e)}_z/2\right),
\end{equation*}
where $\sigma^{(e)}_z=\ket{\uparrow}\bra{\uparrow}-\ket{\downarrow}\bra{\downarrow}$. In addition, we assume that each excitation or OSRP may cause a decoherence effect due to multiphoton emission or phonon interactions, which we model by a pure dephasing channel 
\begin{equation*}
    \mathcal{C}_{\text{deph}}^{(j)}(\lambda) = \exp\left[-\frac{1}{2}\log(\lambda)\mathcal{D}_{\sigma^{(j)}_z}\right]
\end{equation*}
applied immediately after the pulse. For the excitation pulse, $j=h$ so that the hole spin is dephased prior to decay, where $\sigma^{(h)}_z=\ket{\downarrow\uparrow\Uparrow}\bra{\downarrow\uparrow\Uparrow}-\ket{\downarrow\uparrow\Downarrow}\bra{\downarrow\uparrow\Downarrow}$. For the OSRP, $j=e$ so that the electron spin is dephased prior to continued precession.

To compute the light-matter entangled state in the spin-polarization basis, and to predict polarization correlations, we must use the model to compute the system light-matter process map $\mathcal{C}_\text{emit}(t)$ that maps a single-qubit electron spin density matrix to a two-qubit spin-photon density matrix. The primary challenge in computing $\mathcal{C}_\text{emit}(t)$ is that the polarization qubits are encoded onto photons occupying pulse modes, which are inherently multi-mode states. This means that the photonic qubit is a time-evolving object, and this time evolution can impact the photonic qubit measurement results. In other words, the time-integration performed by photocounting over the lifetime of the photon has a direct impact on the effective spin-photon state. To capture this effect, it is necessary to compute the spin state conditioned on the time-integrated polarization state.

The standard approach to handle this time-integration problem is to explicitly compute instantaneous spin-photon field correlation functions and then subsequently integrate those over time to mimic the blurring induced by the photodetection process~\cite{bauchTimebinEntanglementDeterministic2024}. While perfectly valid, this approach has the numerical limitation of exponential scaling in the size of the field correlation, restricting analysis to at most three-particle correlations. In addition, field correlation functions only accurately capture the physics of photodetection in the low-efficiency regime where the detector has a linear response to the intensity of light. For photonic quantum computing, we are primarily interested in the regime of high-efficiency detection where detectors either resolve the photon number directly (photon-number resolved detectors), or they respond identically regardless of the photon number (threshold detectors).

To solve both of these problems, we use the so-called zero-photon generator (ZPG) method~\cite{weinSimulatingPhotonCounting2024} to compute the state of the spin qubit conditioned on the time-integrated detection polarization. The ZPG is the generator of the photon-number decomposition of the system master equation~\cite{weinAnalyzingPhotoncountHeralded2020}, and this decomposition correctly captures spin-photon correlation functions regardless of the efficiency regime. The ZPG method is a technique to solve this decomposition in a way that circumvents multi-dimensional time integration, since it instead extracts the integrated statistics of the field by solving an inversion problem.

To compute $\mathcal{C}_\text{emit}(t)$ using the ZPG method, we first assume that we are using two detectors to measure polarization after passing the photonic qubit through a polarizing beam-splitter. The ZPG equation for this scenario is:
\begin{equation}
    \frac{d}{dt}\rho^{(\mathbf{0})}_{\eta,\overline{\eta}}(t|\mathbf{p}) = \left(\mathcal{L} - \eta\mathcal{J}_\mathbf{p} - \overline{\eta}\mathcal{J}_{\overline{\mathbf{p}}}\right)\rho^{(\mathbf{0})}_{\eta,\overline{\eta}}(t|\mathbf{p}),
\end{equation}
where $\mathcal{J}_i\rho = \sigma_i\rho\sigma_i^\dagger$ is the jump superoperator, $\sigma_\mathbf{p} = \cos(\theta)\sigma_H + \sin(\theta)e^{i\phi}\sigma_V$ is the dipole operator of polarization $\mathbf{p} = (\cos(\theta), \sin(\theta)e^{i\phi})$, and $\overline{\mathbf{p}}$ is the orthogonal polarization. Solving this equation provides the state $\rho^{(0)}_{\eta,\overline{\eta}}(t|\mathbf{p})$ at time $t$, conditioned on having not detected any light at both detectors between time $t_0$ and $t$ for some initial state $\rho(t_0)$, given the first detector measures polarization $\mathbf{p}$ with efficiency $\eta$, the second measures the orthogonal polarization $\overline{\mathbf{p}}$ with efficiency $\overline{\eta}$.

By choosing $\eta$ and $\overline{\eta}$ to take values of either 0 or 1, for a total of 4 configurations, we can reconstruct the desired threshold-detection conditioned states:
\begin{equation}
\begin{aligned}
    \rho^{(0, 0)} &= \rho^{(\mathbf{0})}_{1, 1}\\
    \rho^{(1, 0)} &= \rho^{(\mathbf{0})}_{0, 1} - \rho^{(\mathbf{0})}_{1, 1}\\
    \rho^{(0, 1)} &= \rho^{(\mathbf{0})}_{1, 0} - \rho^{(\mathbf{0})}_{1, 1}\\
    \rho^{(1, 1)} &= \rho^{(\mathbf{0})}_{0, 0} - \rho^{(\mathbf{0})}_{1, 0} - \rho^{(\mathbf{0})}_{0, 1}
     + \rho^{(\mathbf{0})}_{1, 1},
\end{aligned}
\end{equation}
where we have dropped the argument $(t|\mathbf{p})$ for simplicity. The above relationship between the zero-photon conditioned states and the threshold-detection conditioned states can be formally derived (see the appendix of Ref.~\cite{weinSimulatingPhotonCounting2024}). However, the intuition is that the state $\rho^{(\mathbf{0})}_{0, 1}$ is the state conditioned on the absence of light detected of polarization $\overline{\mathbf{p}}$ but no restriction on $\mathbf{p}$. The state $\rho^{(\mathbf{0})}_{1, 1}$ is the state conditioned on the absence of any light detected. Hence, by subtracting the two, we are left with the state $\rho^{(0, 1)}$ conditioned on the detection of at least one photon of polarization $\mathbf{p}$ and none of $\overline{\mathbf{p}}$. Importantly, $\rho^{(1, 0)}(t|\mathbf{p})$ and $\rho^{(0, 1)}(t|\mathbf{p})$ are the conditional states of the system at time $t$ given that we measured an emitted polarization qubit in the basis defined by $\mathbf{p}$.

With this method, we gain direct access to the spin-photon polarization correlations where the photonic part has been time-integrated but where the spin part still depends on time. From this, by choosing $\mathbf{p}$ to be the 6 polarizations of the Poincar\'e sphere, and $\rho(t_0)$ to be four unique Pauli eigenstates, we obtain enough information to fully reconstruct the effective $4\times16$ spin-photon process map $\mathcal{C}_\text{emit}(t)$. In particular, we solve 64 correlation functions:
\begin{equation}
\begin{aligned}
    \langle \sigma_i^{(e)}(t)\sigma_I^{(p)}(t)\rangle &= \frac{\langle\sigma_i^{(e)}(t|L)\rangle+\langle\sigma_i^{(e)}(t|R)\rangle}{P_L(t)+P_R(t)}\\
    \langle \sigma_i^{(e)}(t)\sigma_x^{(p)}(t)\rangle &= \frac{\langle\sigma_i^{(e)}(t|H)\rangle-\langle\sigma_i^{(e)}(t|V)\rangle}{P_H(t)+P_V(t)}\\
    \langle \sigma_i^{(e)}(t)\sigma_y^{(p)}(t)\rangle &= \frac{\langle\sigma_i^{(e)}(t|D)\rangle-\langle\sigma_i^{(e)}(t|A)\rangle}{P_D(t)+P_A(t)}\\
    \langle \sigma_i^{(e)}(t)\sigma_z^{(p)}(t)\rangle &= \frac{\langle\sigma_i^{(e)}(t|L)\rangle-\langle\sigma_i^{(e)}(t|R)\rangle}{P_L(t)+P_R(t)},
\end{aligned}
\end{equation}
for $i\in\{I, x, y, z\}$ and for four different initial states spin: $\ket{\psi_k}\in \{\ket{0}$, $\ket{1}$, $\ket{+}$, $\ket{i}$\}. Here, $P_\mathbf{p}(t)=\text{Tr}(\rho^{(1, 0)}(t|\mathbf{p}))$ is the probability of detecting at least one photon of polarization $\mathbf{p}$ and exactly zero of polarization $\overline{\mathbf{p}}$, and $\sigma_i^{p}$ for $i\in\{I, x, y, z\}$ are the Pauli operators for the photonic qubit. With the above 64 correlation functions, we can linearly invert the relation $\text{Tr}\{\sigma^{(e)}_i(t)\sigma^{(p)}_j(t) \mathcal{C}_\text{emit}(t)(\ketbra{\psi_k}{\psi_k})\}=\langle \sigma_i^{(e)}(t)\sigma_j^{(p)}(t)\rangle_k$ to reconstruct $\mathcal{C}_\text{emit}(t)$ at any chosen time $t$. Note that, the method we use to construct $\mathcal{C}_\text{emit}$, assumes we have detected at least one photon, and rejects events where both detectors clicked. Hence, it accurately represents the same process map that would be reconstructed if a process tomography was carried out experimentally using threshold detectors and post-selection.

Direct numerical simulation of $\mathcal{C}_\text{emit}$ is very fast using the ZPG method, taking about 1 second on a standard laptop to reach $>6$ digits of precision for a given single parameter set. However, this is still generally too slow to numerically fit the multiple data sets produced by our experiments while varying six parameters and averaging the Overhauser field. To circumvent this, we first implement the above ZPG method in \emph{Wolfram Mathematica} to produce a closed-form analytical expression of the full process map, which is far too intractable to gain any physical insight from. We export this expression into C code and compile it into a shared library that can be accessed through a Python interface. This approach can evaluate $\mathcal{C}_\text{emit}(t)$ in less than 10 ms for a given parameter set. With this compiled process map solution, we use QuTiP \cite{johanssonQuTiPOpensourcePython2012} to apply the process map as a Qobj object to an initially mixed spin state, take measurements and partial traces to prepare and readout the spin, and to apply the optical spin rotation operations that produce various photonic graph states, and average the fluctuating Overhauser field. This rapid simulation technique allows us to easily construct large spin-photon or purely photonic density matrices for a given set of model parameters and pulse configurations, and then use standard SciPy optimization functions to maximize the fidelity of the simulated density matrices to the experimentally reconstructed density matrices while accounting for phase averaging due to 100-1000 orientations and magnitudes of the Overhauser field.

Since the uncertainty in the fit is primarily due to deviations of the model from the exact experimental configuration rather than due to the optimization procedure, we estimate the parameter uncertainty by performing multiple optimization runs on various combinations of the experimental data. In this way, the mean value represents the parameter set that globally minimizes the average simulation error, while the standard deviation represents the range of parameters needed to minimize the error for each individual density matrix. Thus, a small uncertainty indicates that the model captures the observations well.

\begin{table}[h] 
\centering
\hspace{-90mm}\textbf{a}\hspace{78mm} \textbf{b}
\vspace{2mm}

\rowcolors{1}{gray!20}{white}
\begin{tabular}{|c|c|c|}
\hline
 Symbol & Description & Fixed Value\\\hline
 $T_1$& Emitter lifetime & 200 ps\\
 $B$ & Static magnetic field & 60 mT\\
 $\tau_\text{ex}$ & Excitation pulse period & 600 ps\\
 $\tau_\text{osrp}$ & OSRP timing (after excitation) & 300 ps\\\hline
\end{tabular}
\hphantom{x}
\rowcolors{1}{gray!20}{white}
\begin{tabular}{|c|c|c|c|}
\hline
 Symbol & Description & Fitted Value & Ideal Value \\\hline
 $B_{OH}$& OH field standard deviation & 9.0(5) mT & 0\\
 $g_e$ & Electron g factor & 0.60(4) &N/A\\
 $g_h$ & Hole g factor & 0.30(6)&0\\
 $\lambda_\text{ex}$ & Excitation purity & 0.94(6) & 1\\
 $\varphi_\text{ex}$ & Excitation polarization angle & 0.02(2) $\pi$& 0\\
 $\lambda_\text{osrp}$ & OSRP purity & 0.74(9) & 1\\
 $\theta_\text{osrp}$ & OSRP rotation &1.03(5) $\pi$& $\pi$\\\hline
\end{tabular}
\caption{\textbf{Model parameters. a,} Parameter symbols, description, and values fixed based on the known experimental configuration. \textbf{b}, Parameters determined by fitting the physical model to experimental data sets. The last column indicates the ideal values that produce the target state (along with $T_1\rightarrow 0$). $B_{OH}$ indicates the effective magnitude of the magnetic field fluctuations seen by the electron spin, which determines the electron spin coherence time. The fitted value and uncertainty represent the mean and standard deviation, respectively, determined by fitting subsets of data.}
\label{tab:parameter table} 
\end{table}

Once the model parameters are determined, we simulate the two-photon density matrices produced by the four pulse configurations studied in Fig.~\ref{fig:cluster} of the main text. They are given in Fig.~\ref{fig:simulated all}. Using the simulated density matrices, we evaluate the mean and standard deviation of the two-photon state fidelity with respect to the ideal state for all four pulse configurations and for each combination of the measured polarization of the first and fourth photons. The results of these simulations are given in Table \ref{tab:parameter table} along with the summary of the experimentally measured values. We find that the measured and simulated fidelity of all configurations have overlapping uncertainty and, in most cases, agree well within the uncertainty. The average absolute difference between the measured and simulated fidelity is just $3\pm 2\%$.

\begin{table}[h] 
\centering 
\rowcolors{1}{gray!20}{white}
\begin{tabular}{|c|c|c|c|c|c|c|c|c|c|c|c|c|c|}
\hline
 & \multicolumn{4}{|c|}{Linear cluster} & \multicolumn{2}{|c|}{GHZ} & \multicolumn{2}{|c|}{Redundant linear cluster \#1} & \multicolumn{2}{|c|}{Redundant linear cluster \#2} \\ \hline
 &\makebox[1.2cm]{$R_1R_4$}&\makebox[1.2cm]{$R_1L_4$}&\makebox[1.2cm]{$L_1R_4$} &\makebox[1.2cm]{$L_1L_4$}&\makebox[1.2cm]{$R_1R_4$} & \makebox[1.2cm]{$R_1L_4$} & \makebox[1.2cm]{$R_1R_4$}&\makebox[1.2cm]{$R_1L_4$}&\makebox[1.2cm]{$R_1R_4$}&\makebox[1.2cm]{$R_1L_4$}\\
$F^{meas}_{2}$ & $0.69(2)$ & $0.78(4)  $ & $0.73(1)  $ &  $0.80(4)  $ & $0.71(2)  $ &  $0.68(2)  $ & $0.58(3)  $ & $0.61(2)  $ & $0.67(2)  $ &$0.66(2)  $ \\ \
$F^{sim}_{2}$ & $0.70(4)  $ & $0.71(4)  $ & $0.71(4)  $ &  $0.72(4)  $ & $0.67(4)  $ &  $0.68(7)  $ & $0.58(4)  $ & $0.57(5)  $ & $0.68(4)  $ &$0.68(3)  $ \\ 
$F^{sim}_{4}$ & \multicolumn{4}{|c|}{$0.66(5)$}   & \multicolumn{2}{|c|}{$0.45(5)$}   & \multicolumn{2}{|c|}{$0.53(5)$} &  \multicolumn{2}{|c|}{$0.53(5)$}  \\ \hline
\end{tabular}
\caption{\textbf{Fidelity to ideal states.} Measured two-photon fidelity ($F_{2\_meas}$), simulated two-photon fidelity ($F_{2\_sim}$), and simulated 4-partite fidelity ($F_{4\_sim}$) for various spin-photon entangled graph states, specified by the polarization state of the first and last photon measured ($R$  or $L$).}
\label{tab:fidelity table} 
\end{table}

\begin{figure}[h]
\includegraphics[scale=0.3]{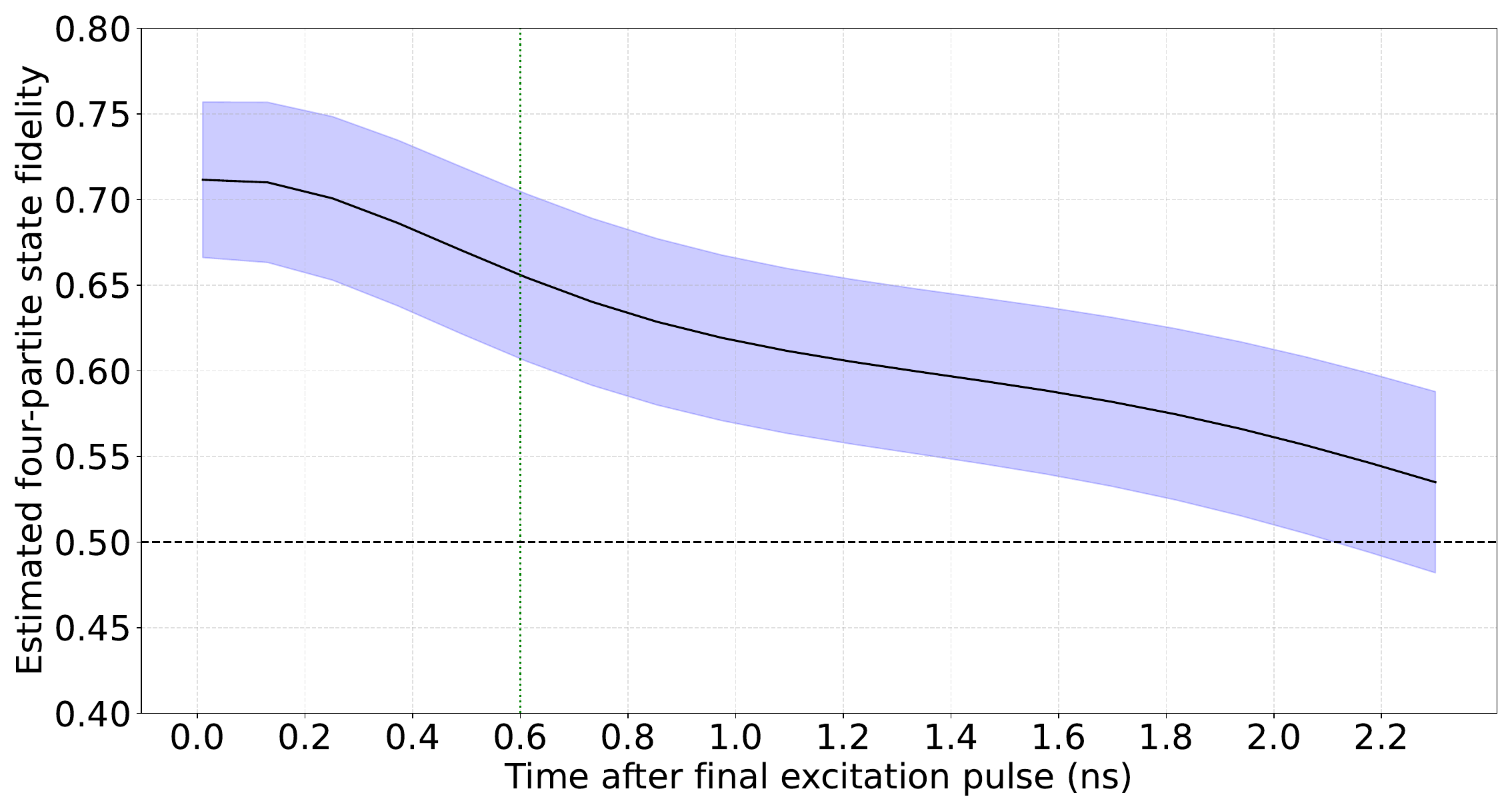}
\caption{\textbf{4-partite fidelity.} Estimated 4-partite linear cluster state fidelity as a function of the time after the final excitation pulse. The black solid line and shaded purple area represents the mean and standard deviation of the fidelity, respectively, obtained by sampling 100 sets of model parameters from the fitted mean and standard deviations given in Table \ref{tab:parameter table}. For each of the 100 sampled parameter sets, the four-partite density matrix was simulated for 100 sampled values of the Overhauser field and averaged before evaluating the fidelity with respect to the target state. The black horizontal dashed line represents the 50\% fidelity threshold, while the green dotted line marks the 600 ps $\pi/2$ spin rotation time, which is the time at which the 4-partite fidelity is evaluated in the main text.}
\label{fig:simulated fourpartite}
\end{figure}

Using the same simulation parameters, we also gain access to an estimate of the full 4-partite spin-photon-photon-photon density matrix following the detection of an $R$-polarized first photon. Since this state is entangled with the spin, it necessarily depends on time as the spin continuously decoheres due to the solid-state environment. The spin also precesses, meaning that the target state itself evolves in time. Importantly, emission time jitter will cause a shift in the effective spin precession time, which can easily be compensated in the experiment by altering the pulse timings. To account for this effect, we compute the maximum fidelity of the simulated state up to some time $t$ after the final excitation pulse with respect to a target state that is evolved to within $t\pm T_1$. That is, we shift the evolution of the target state by up to 200 ps to better reproduce experimental observation. From this, we find that (for our simulation parameters) the emission time jitter delays the effective precession of the spin by up to 100 ps by the time spontaneous emission has concluded ($t\gg T_1$).

The results of these simulations are given in Table \ref{tab:fidelity table}, where the 4-partite fidelity is taken at $t=600$ ps following the last excitation pulse. We also show the estimated 4-partite fidelity and uncertainty for the linear cluster state pulse configuration as a function of the time $t$ after the final excitation in Fig.~\ref{fig:simulated fourpartite}. This indicates that genuine 4-partite entanglement survives well beyond the 600 ps $\pi/2$ rotation time and that our protocol can already be extended to more photons while still producing genuine multipartite entanglement.\\

Finally, we now adapt the simulation parameters based on realistic near-term improvements to the source. Specifically, we consider a positive trion, which is proven to have a coherence time ten times longer than  negative trions in InGaAs QD~\cite{costeProbingDynamicsCoherence2023}, due to the reduced hyperfine interaction between the hole in the ground state and the surrounding nuclei. With a photon radiative lifetime of 100 ps and improved spin gate fidelity of 0.995 (which can be realistically achieved by increasing the duration and detuning of the OSRP), we estimate (see all parameters in Fig.~\ref{fig:simulated fidelity prediction table}\textbf{b}) that generating linear cluster and GHZ state with over 50\% fidelity is achievable for up to 20 and 30 photons, respectively, as shown in Fig.~\ref{fig:simulated fidelity prediction table}\textbf{a}. Additionally, our simulations also show that an arbitrary 10-photon caterpillar state (represented in Fig.~\ref{fig:caterpillar}) can be produced with $80\pm1\%$ fidelity, making this a realistic near-term goal.   

\begin{figure}[h]
\includegraphics[scale=0.65]{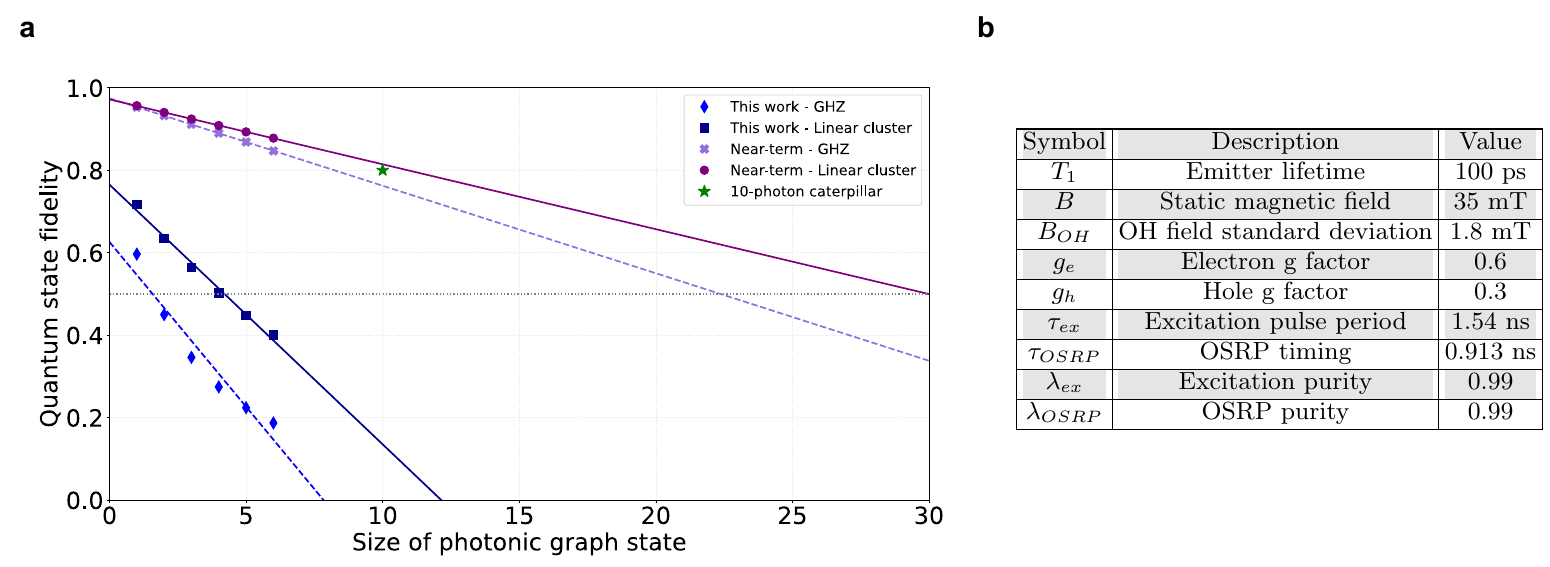}
\caption{\textbf{Current work and near-term simulated fidelity.} \textbf{a}, Simulated quantum state fidelity of GHZ and linear cluster states as a function of the number of photons for both the negative trion source used in this work (diamond and square markers) and a realistic near-term positive trion source (cross and circle markers) whose parameters are detailed in \textbf{b}. Dashed and solid lines represent linear regressions of the simulated data points and the horizontal dotted line marks the fidelity threshold of 0.5. The green star indicates the simulated fidelity for the caterpillar graph state shown in Fig.~\ref{fig:caterpillar}. \textbf{b}, Realistic near-term parameters used to simulate the photonic graph states displayed in \textbf{a}.}
\label{fig:simulated fidelity prediction table}.
\end{figure}

\clearpage

\clearpage
\subsection{Experimental setup}

\begin{figure}[h!]
\includegraphics[scale=0.6]{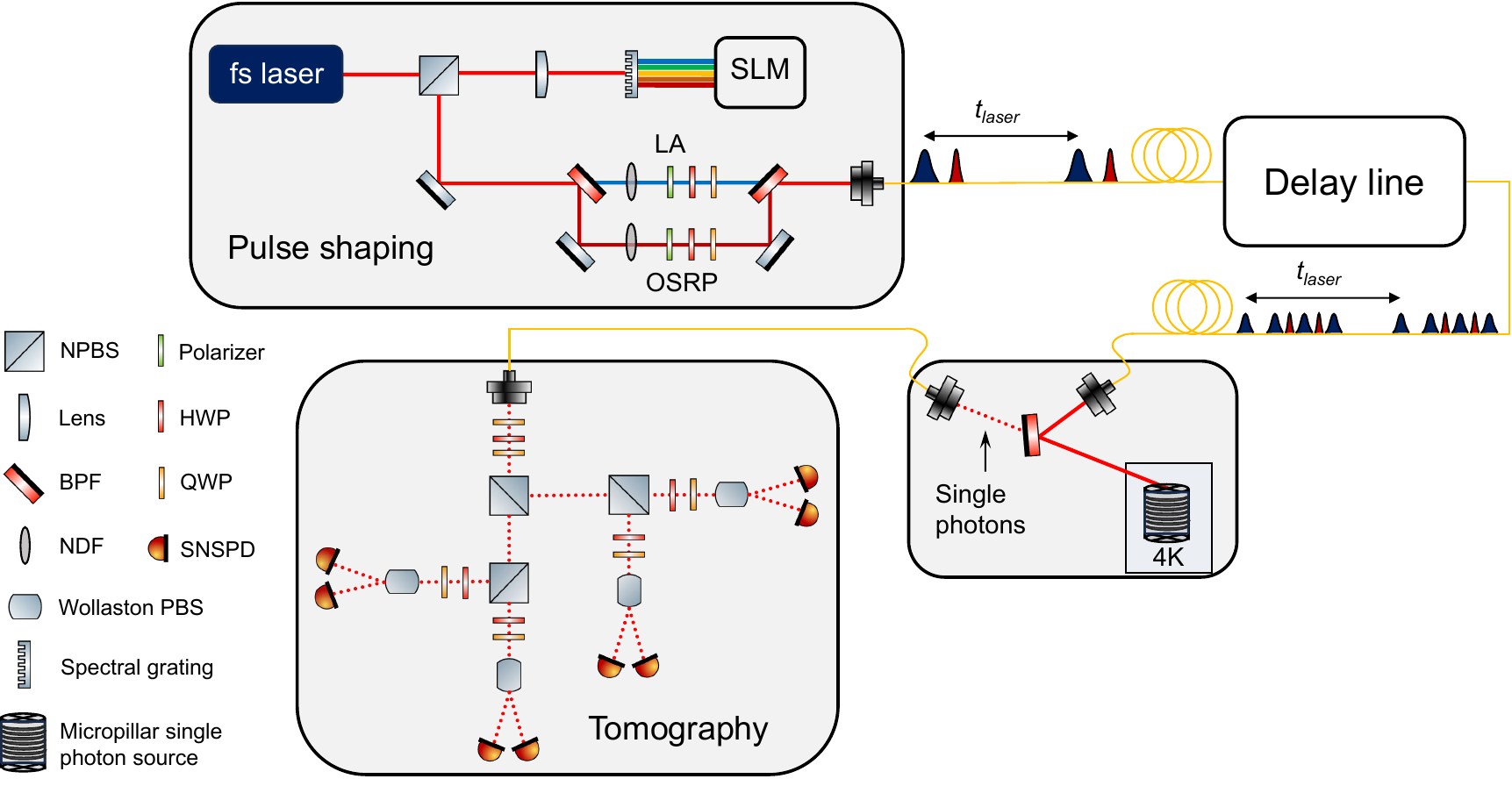}
\caption{\textbf{Schematic of the experimental setup used for reconfigurable graph state generation} Experimental setup consists of four main parts. The first is pulse shaping, where the longitudinal-acoustic phonon-assisted excitation pulses (LA) and the optical spin rotation pulses (OSRP) are carved from the same femtosecond laser using a spatial light modulator (SLM) and are then separated using band pass filters (BPF). The power in each path is controlled by a neutral density filter (NDF), and polarization is set by a polarizer, half waveplate (HWP) and quarter waveplate (QWP). Next, a delay line is used to set arbitrary delays between the 4 excitation pulses and up to 2 OSRPs in each pulse sequence. Each OSRP can be turned on or off by spectral filtering. Each pulse sequence is separated by the repetition rate of the laser ($t_{laser}$). The third part is the delivery of the pulses to the QD-cavity device (kept in a closed-cycle cryostat at 4K) and collection of the emitted single photons. Finally, the single photons are sent to a demultiplexed tomography setup. The tomography setup is divided into four paths using non-polarizing beam splitters (NPBS). Each path consists of a HWP, QWP, Wollaston polarizing beam splitter (Wollaston PBS) and 2 superconducting nanowire single photon detectors (SNSPD).}
\label{fig:setup}
\end{figure}

\end{widetext}
\end{document}